\documentclass{aa}
\usepackage{graphicx}
\usepackage{txfonts}
\usepackage{natbib}
\usepackage[caption=false]{subfig}
\usepackage[linktocpage]{hyperref}
\usepackage{xcolor}
\usepackage{threeparttable}
\usepackage{etoolbox}


\begin{document}

   \title{The role of AGN feedback on the structure, kinematics and evolution of ETGs in the Horizon simulations}
          
    \author{M. S. Rosito \inst{1}, S. E. Pedrosa \inst{2,3}, P. B. Tissera \inst{4,5}, N. E. Chisari \inst{6}, R. Dom\'{i}nguez-Tenreiro \inst{2,7}, Y. Dubois\inst{8}, S. Peirani\inst{8,9}, J. Devriendt\inst{10}, C. Pichon\inst{8,11}, A. Slyz\inst{10}.
 } 

   \institute{Departamento de Matem\'{a}tica. Facultad de Ciencias Exactas y Naturales. Universidad de Buenos Aires. Pabell\'{o}n I. Ciudad Universitaria C1428EGA Buenos Aires, Argentina. \and
   Departamento F\'{i}sica Te\'{o}rica, Universidad Aut\'{o}noma de Madrid, E-28049 Cantoblanco, Madrid, Spain. \and
  Instituto de Astronom\'{i}a y F\'{i}sica del Espacio, CONICET-UBA. 1428. Buenos Aires, Argentina. \and
  Instituto de Astrof\'{i}sica, Pontificia Universidad Cat\'olica, Av. Vicuña Mackenna 4860, Santiago, Chile.
 \and Centro de Astro-Ingenier\'ia, Pontificia Universidad Cat\'olica de Chile, Av. Vicu\~na Mackenna 4860, Santiago, Chile.
  \and Institute for Theoretical Physics, Utrecht University, The Netherlands, Princetonplein 5, 3584 CC Utrecht, The Netherlands.
  \and Centro de Investigaci\'{o}n Avanzada en F\'{i}sica Fundamental, Universidad Aut\'{o}noma de Madrid, E-28049 Cantoblanco, Madrid, Spain.
  \and Institut d'Astrophysique de Paris, CNRS \& Sorbonne Universit\'e, UMR 7095, 98 bis Boulevard Arago, 75014, Paris, France.
  \and Universit\'e C\^ote d’Azur, Observatoire de la C\^ote d'Azur, CNRS, Laboratoire Lagrange, Bd. de l'observatoire, 06304 Nice, France.
  \and Department of Physics, University of Oxford, Keble Road, Oxford OX1 3RH, United Kingdom.
  \and Korea Institute for Advanced Study, 85 Hoegiro, Dongdaemun-gu, Seoul, 02455, Republic of Korea.}


\abstract
   {Feedback processes play a fundamental role in the regulation of the star formation (SF) activity in galaxies and, in particular, in the quenching of early-type galaxies (ETGs) as has been inferred by observational and numerical studies of $\Lambda$-CDM models. At $z=0$, ETGs exhibit well-known fundamental scaling relations, but the connection between them and the physical processes shaping ETG evolution remains unknown.}
   {This work aims at studying the impact of the energetic feedback due to active galactic nuclei (AGN) on the formation and evolution of ETGs. We focus on assessing the impact of AGN feedback on the evolution of the mass--plane and the fundamental plane (FP, defined by using mass surface density)  as well as on morphology, kinematics, and stellar age across the FP. }
   { The Horizon-AGN and Horizon-noAGN cosmological hydrodynamical simulations were performed with identical initial conditions and including the same physical processes except for the activation of the AGN feedback in the former. We select a sample of central ETGs from both simulations using the same criteria  and exhaustively study their SF activity, kinematics, and scaling relations for $z \le 3$.}
   {  We find that Horizon-AGN ETGs identified at $z=0$ follow the observed fundamental scaling relations (mass--plane, FP,  mass--size relation) and qualitatively reproduce kinematic features albeit conserving a rotational inner component with a mass fraction regulated by the AGN feedback. AGN feedback  seems to be required to reproduce the bimodality in the spin parameter distribution reported by observational works  
   and the mass--size relation (with more massive galaxies having older stellar populations (SPs), larger sizes, and being slower rotators). 
   We study the evolution of the fundamental relations with redshift, finding a mild evolution of the mass--plane of  Horizon-AGN ETGs for $z<1$ whereas a  stronger change is detected for $z > 1$.
   ETGs in Horizon-noAGN show a strong systematic redshift evolution of the mass--plane. The FP of Horizon-AGN ETGs is in agreement with observations at $z=0$. At all analysed redshifts, when AGN feedback is switched off,  a fraction of galaxies departs from the expected FP due to  the presence of a few extended galaxies with an excess of stellar surface density.
   We find that AGN feedback regulates the star formation activity as a function of stellar mass and redshift, being able to reproduce the observed relations. Our results show  the impact of AGN feedback on the $M/L$ ratio and its relation with the tilt of the Luminosity FP (L-FP: defined by using averaged surface brightness).
   Overall, AGN feedback 
   has an impact on the regulation of the star formation activity,  size, stellar surface density, stellar ages, rotation, and masses of ETGs that is reflected on the fundamental relations, particularly on the FP.
   We detect a dependency of the FP on stellar age and galaxy morphology  which evolves with redshfit. The characteristics of the galaxy distribution on the FP according to these properties change drastically by $z\sim 1$ in Horizon-AGN and hence  this feature  could provide further insight into the action of AGN feedback.}
   {}

  {}


  \authorrunning{M. S. Rosito et al.}
  \maketitle
   
\section{Introduction}

In the current cosmological paradigm a number of complex processes, such as major and/or minor mergers, secular evolution, and other interactions or environmental effects, contribute to shaping the morphology of galaxies \citep[e.g.][and references therein]{tissera2012, AvilaResse2014, Somerville2015, RG2016, Dubois2016}, and particularly ellipticals or ETGs.
The general picture for ETG formation considers different scenarios to explain their observed variety of properties \citep[e.g.][]{Kormendy2016}.
Some analytical models \citep{Solar2020}, as well as N-body simulations
\citep{Wechsler2002, Zhao2003}, have shown that two different phases can be
distinguished along halo mass assembly: first, a violent, fast phase, with high-mass
aggregation rates (i.e., merger rates); and later on, a slow phase, where the mass
aggregation rates are much lower. Subsequent numerical works agree with this scenario and its implications for
the properties of massive galactic objects at low $z$ \citep[e.g.][]{DTenreiro2006,Oser+2010, Cook2009, DT2011}.
Instead,  \cite{Naab2013} proposes two phases:  an in-situ SF at high redshift, followed by minor mergers that allow the accretion of stars from other galaxies at later stages. More recently, \cite{Clauwens2018} distinguish three different formation phases as a function of stellar mass in galaxies from EAGLE simulation \citep{Crain2015, Schaye2015}.
These authors conclude that low-mass galaxies grow due to in-situ SF caused partially by mergers and present random stellar motion.  Intermediate-mass galaxies are mostly disc-dominated. Finally, in high-mass galaxies, in-situ SF becomes less important and they are more spheroidal.
Feedback processes are essential to regulate the SF, and therefore to explain the quenching of ETGs \citep[e.g.][]{Burkert2004, Hayward2016, Dubois2016}.
In particular, the importance of AGN feedback for regulating the growth and the SF for massive galaxies has been extensively discussed in numerical simulations \citep{Crain2009, Schaye2010, Haas2013, Somerville2015, Rosas2016}, which implement more sophisticated modelling that leads to more realistic quenched massive elliptical galaxies \citep[e.g.][]{Pillepich2018, Habouzit2019}. 

A remarkable characteristic of ETGs is that they determine clear scaling relations.
The fundamental plane \citep[FP][]{Faber1987, Dressler1987, Davis1987} links size, velocity dispersion, and luminosity 
\citep[e.g.][]{Mo2010}. Under some assumptions, such as homology and constant mass-to-light ratio ($M/L$),  the virial theorem can provide a physical interpretation for the FP \citep[e.g.][]{Binney1987}. However, the observed parameters of this plane are tilted from the virial predictions \citep[e.g.][]{Bernardi2003, Hyde2009, Cappellari2013, Saulder2013}. This tilt has been a matter of controversy, which has fostered different explanations such as the variation of the $M/L$ \citep{Faber1987}, the non-homology in the surface brightness distribution \citep[e.g.][]{Prugniel1997,Graham1997, Bertin2002, Trujillo2004}, or the variation in the fraction of dark matter \citep[e.g.][]{Renzini1993,Ciotti1996,Borriello2003}. \cite{cappellarireview2016} reports that the expected virial relation is recoved when the surface brightness is replaced by the dynamical mass. 
Based on hydrodynamical simulations in a cosmological context, \citet{Onorbe2005,Onorbe2006}  analyse the FP of massive galaxies finding a tilt from the virial parameters at galactic scale whereas, at halo scale, the virial relations hold. 
These authors claim that the tilt originates at high redshift \citep{Onorbe2011} and that it only shows a mild evolution after $z \sim 1.5$ \citep{DTenreiro2006}, presumably due to a spatial homology breaking. This is associated with a systematic decrease with increasing galaxy mass of the relative amount of dissipation experienced by the baryonic mass component during mass assembly
\citep[see also][]{Taranu2015}.

Recently \citet{Lu2020} analyse the fundamental relations for ETGs in The Next Generation Illustris Simulations \citep[IllustrisTNG;][]{Marinacci2018,Naiman2018, Nelson2018,Pillepich2018,Springel2018}, finding a FP with a mild evolution and a mass--plane relation in place, with small scatter ($\sim 0.08$ dex), since $z=2$.

Regarding the mass--size relation,  \cite{Li2018} report that the evolution of the stellar populations (SPs) results in variations of properties such as mean age and metallicity across this plane in the MaNGA survey. 
\cite{Rosito2019b} study this relation for a sample of galaxies with a large variety of morphologies selected from the EAGLE project and the variation of some important galaxy properties across the mass--size relation. They also analyse the angular momentum distribution across the mass--size relation, finding that, for a fixed value of the spin parameter, more massive galaxies are more extended. In particular, they report a bimodality in the $\lambda$-$\varepsilon$ plane with galaxies in the first peak populated by fast rotators with younger SPs and in the second one, by slow rotators with older SPs, in agreement with observational works \citep{Graham2018}.

Feedback mechanisms also play an important role in producing realistic galaxy sizes, affecting the mass--size relation.
\cite{Crain2015} emphasise that galaxy sizes have to be taken into account when calibrating EAGLE project simulations since the absence of feedback produces more compact galaxies than those observed as a result of the excessive gas dissipation. Moreover, they remark that the variation of subgrid parameters that allows black holes to achieve high accretion rates in low-mass haloes ($< 10^{11.5}$ M$_{\odot}$) enables earlier AGN feedback effect leading to the production of more extended massive galaxies \citep{Rosas2015}. 
\cite{Habouzit2019} report a good agreement of IllustrisTNG galaxies with recent observations regarding the mass--size relation. This simulation includes an improved model of AGN feedback in comparison to that implemented in Illustris \citep{Genel2014}.

Although ETGs are supported by velocity dispersion,  
the amount of rotation among them varies, with fast rotators being notably more frequent than slow rotators \citep{Emsellem2011}. A more complete ETGs kinematic classification is summarised in \cite{cappellarireview2016}, who shows that the rotational component plays a non-negligible role in the ETG structure.
The importance of rotation in elliptical galaxies can also be seen in cosmological simulations as shown in \cite{Onorbe2007} who find clear rotation curves in some of them. 
\cite{Rosito2018} and \cite{Rosito2019a} analyse the presence of a disc component in the inner region of ETGs, coexisting with the spheroidal-dominated component (i.e., bulge), identified in simulations of the Fenix project \citep{Pedrosa2015} and the EAGLE project \citep{Crain2015, Schaye2015}, respectively.

The main goal of the present work is to unveil the role of AGN feedback on the fundamental scaling relations and their evolution with time.
For this purpose, we analyse two simulations from the Horizon project \citep{Dubois2014,Kaviraj2017} which share the same initial conditions and subgrid physics except for the  AGN feedback: Horizon-AGN and Horizon-noAGN \citep[][i.e. no AGN feedback is modeled]{Peirani2017}. Central ETGs are identified by applying the same selection criteria in both simulations and thus, a cross-identify procedure is not applied in our work. We also analysed the star formation rates, kinematics, and the ages of the SPs in order to interpret the results.

Several results have been already reported using the Horizon project which are relevant to our work.
\cite{Dubois2016} remark on the importance of simulating AGN feedback to reproduce properly the morphology of galaxies, especially of the massive ones.
Furthermore, they conclude that the assembly of ETGs is mostly driven by mergers, having higher fractions of ex-situ stars than disc-dominated galaxies, and that the quenching, regulated by AGN feedback, freezes the resultant morphology triggered by those mergers. 
Additionally, by isolating the effect of AGN feedback, \cite{Beckmann2017} compare the stellar mass ratios of twin galaxies from these simulations, i.e., they cross-identify individual galaxies from both simulations, analysing thus the impact of AGN feedback on quenching as a function of stellar mass: smaller galaxies quench at higher redshifts than massive ones ($M_* > 10^{11}$ M$_{\odot}$)  and  more massive ones  are more affected by AGN feedback. They highlight the impact of AGN feedback in regulating the SF in galaxies of different morphologies as a result of both the decrease of total gas due to outflows and the difficulty to recover the gas from the halo.
Massive ETGs are also studied by \cite{Peirani2019}, who find that AGN feedback plays a fundamental role to achieve an agreement with observational trends, in particular, they focus on the mass-weighted density slope \citep{Dutton2014} and its correlation with other galaxy features, such as mass and effective radius. 
Recently \citet{vdSande2019} find that  Horizon-AGN galaxies are larger at a given stellar mass than those observed in SAMI galaxy survey \citep{SAMI2012, Bryant2015}, ATLAS$^{\rm 3D}$ project \citep{AtlasI}, CALIFA survey \citep{CALIFA2012} and MASSIVE survey \citep{Ma2014}. 

This paper is organised as follows. In Section \ref{sec:meth}, we present the methodology adopted to perform  our research: the descriptions of the simulations, the definitions of the main properties to be studied, and the criteria for selecting and classifying  the samples of central ETGs that are analysed exhaustively throughout this work. In Section \ref{sec:char}, the characteristics of the simulated samples at $z=0$ are analysed. The evolution of the fundamental relations is studied in Section \ref{sec:empfp} and the results are discussed in Section \ref{sec:disc}. Section \ref{sec:conclusions} summarises our main findings. 

\section{Methods}
\label{sec:meth}

\subsection{Horizon-AGN and Horizon-noAGN simulations}
\label{sec:simu}

The Horizon simulations are cosmological hydrodynamical simulations consistent with a $\Lambda$-CDM universe with $\Omega_m=0.272$, $\Omega_{\Lambda}=0.728$, $\Omega_b=0.045$, $H_0=100 \ h$ km s$^{-1}$ Mpc$^{-1}$ being $h=0.704$, and the normalisation of the power spectrum $\sigma_8 =0.81$ \citep{Dubois2014}. These parameters are consistent with a WMAP cosmology \citep{Komatsu2011}. The simulated box size is 100 $h^{-1}$ Mpc per side with 1024$^3$ dark matter particles of $8 \times 10^7$ M$_{\odot}$ \citep{Dubois2014}.

These simulations are run with the RAMSES code \citep{Teyssier2002} which is based on the adaptative mesh refinement technique (AMR). The initial mesh is refined down to $\Delta x$ = 1 kpc. This refinement is developed by means of a quasi-Lagrangian method: if the number of dark matter particles in a cell exceeds eight, or the total baryonic matter inside the cell is eight times greater than the dark matter resolution, a new refinement is performed. Furthermore, the minimum size of the cells in physical units (taking into account the scale factor) remains approximately constant.
The initial conditions are computed using the MPGRAFIC code \citep{Prunet2008}.

Gas cooling processes (including hydrogen, helium, and some metals) are described following \cite{Sutherland1993} model. Cooling down to 10$^4$ K is allowed. On the other hand, gas heating occurs after reionisation ($z \sim 10$) and is modeled with a UV background following \cite{Haardt1996}. 
SF takes place when the gas density is greater than 0.1 H cm$^{-3}$ and is produced randomly \citep{Rasera2006, Dubois2008}. Such process follows a Schmidt law: $\dot{\rho_*} = \epsilon_* \rho_{\rm g} / t_{\rm ff}$, where $\dot{\rho_*}$ is the star formation rate (SFR) density, $\rho_{\rm g}$ is the gas density, $t_{\rm ff}$ the gas local free-fall time and $\epsilon_* = 0.02$ is a constant \citep{Kenni98, Krumholz2007}. The stellar mass resolution is $2 \times 10^6$ M$_{\odot}$ and a Salpeter initial mass function (IMF) is assumed \citep{Salpeter55}. The simulations also include Type Ia (SNIa) and Type II (SNII) supernovae (SN) feedback \citep{Dubois2014}.
Gas metallicity is modified by galactic winds and the enriched gas is ejected during a SN event. The following elements are modelled: O, Fe, C, N, Mg, and Si \citep{Dubois2014}.

Individual galactic structures are identified using the AdaptaHOP code \citep{Aubert2004} updated by \cite{Tweed2009}. In the original galaxy catalog, only objects resolved with more than 50 stellar particles are considered, yielding 150000 galaxies a $z=0$.
In the present work, we apply a more restrictive minimum particle number by requesting simulated galaxies to have more than 5000 stellar particles in order to be able to properly resolve galaxies and the evolution with redshift of their fundamental relations.

In the case of Horizon-AGN, black holes are created in regions with gas density higher than a fixed threshold and have an initial mass of $10^5$ M$_{\odot}$. Black holes are not allowed to form within small distances from each other (less than 50 kpc). The accretion rate ($\dot{M}_{\rm BH}$) and Eddington accretion rate ($\dot{M}_{\rm Edd}$) are calculated as in \cite{Dubois2012}. There are two AGN feedback modes implemented according to the quotient $\chi=\dot{M}_{\rm BH}/\dot{M}_{\rm Edd}$. If $\chi < 0.01$, the situation corresponds to a "radio" mode, otherwise, to a "quasar" mode \citep{Merloni2008}. The fraction of energy released by the black hole is assumed to be a free parameter that depends on the mode.
Despite the calibration of Horizon-AGN enforcing the $M_{\rm BH}-\sigma$ relation, there is evidence that the gas fractions inside haloes are higher than those observed \citep{Chisari2018}. Thus, the results of our work can be assumed to be conservative in gauging the impact of AGN feedback.

These models reproduce a number of black holes observational constraints, such as the black hole mass function and relations between these objects and galaxy properties \citep{Volonteri2016}. There are, however, some differences between these mock objects with the observed ones. In particular, the faint end of the luminosity function is overestimated in the presence of massive black holes at $z \sim 0.5-2$, which is a common problem in cosmological simulations \citep{Sijacki2015}. A complete analysis of Horizon-AGN black holes can be found in \cite{Volonteri2016} where the authors propose a stronger SN feedback to improve the reproduction of the observational trends. On the other hand, \cite{Beckmann2017} find a relation between the mass of the smallest galaxy affected by AGN feedback with redshift for Horizon-AGN galaxies that is in agreement with observations \citep{Baldry2004,Peng+2010full}.

\subsection{Measured properties of simulated galaxies}
\label{sec:defi}

For the purpose of developing a comprehensive analysis of the simulated galaxies from Horizon-AGN and Horizon-noAGN and comparing thus their properties, we compute a number of relevant parameters that will be used in this work as follows:

\begin{itemize}
    \item Stellar optical radius \citep[$R_{\rm opt}$, ][]{tissera2012}: 3D radius that encloses $\sim 80$ per-cent of the stellar mass.  
    \item  Total stellar mass ($M_*$): sum of masses of individual stellar particles within $R_{\rm opt}$. 
    \item Stellar half-mass radius ($R_{\rm hm}$): 3D radius which encloses 50 per-cent of the total stellar mass. 
    \item  Stellar $V/\sigma$: mean rotational velocity (in cylindrical coordinates) over the average velocity dispersion considering the three cylindrical components of velocity. These parameters are measured using all stellar particles belonging to the same galactic structure identified by the AdaptaHOP code \citep[e.g.][]{Dubois2014, Chisari2015}.
    \item $F_{\rm rot}$: ratio between the mass of the stellar disc spatially coexisting with the bulge, i.e., within $0.5 R_{\rm hm}$, and the bulge stellar mass. See Appendix \ref{app:morp} for details about the definition of the galaxy components.
    \item Star formation rate (SFR): quotient between the mass of stars younger than $\Delta t$ and $\Delta t$, being $\Delta t = 0.5$   Gyr\footnote{We did not use an ``instantaneous" measure of the SF activity to avoid numerical noise due to the low number of very young particles (see also \cite{Rosito2018} and \cite{Rosito2019a}).} within the $R_{\rm opt}$.
    \item Specific star formation rate: sSFR=SFR/$M_*$. For the sSFR of galaxies with no recent SF we impose a fiducial value of $10^{-13}$ yr$^{-1}$.
    \item Stellar velocity dispersion ($\sigma_e$):  computed considering stellar particles within $R_{\rm hm}$.
    \item Dynamical mass: $M_{\rm dyn} = 5 \sigma_e^2 R_{\rm hm}/(3G)$. This estimation is performed assuming virialisation and an spherical distribution. $G$ the universal gravitational constant.
    \item Average surface density: $\Sigma_e = M_*/(2 \pi R_{\rm hm}^2)$. 
    \item Spin parameter ($\lambda$):  computed with the formula \linebreak $\lambda=\langle R |V| \rangle / \langle R \sqrt{V^2+\sigma^2} \rangle$ \citep[e.g.][]{Emsellem2011, Lagos2018} where $V$ and $\sigma$ are the tangential velocity and velocity dispersion along the line-of-sight, respectively, and $R$ the three-dimensional radius, for consistency with the calculation of other parameters in this paper. The brackets $\langle . \rangle $ correspond to a mass-weighted average. The calculations are done by considering all stars within $R_{\rm hm}$.
\end{itemize}

Hereafter, in figures involving data binning to estimate the medians and the 25 and 75 percentiles, we only consider bins with more than 10 galaxies in order to  obtain more robust and reliable trends.

It is important to mention that, despite the fact that these quantities clearly characterise the simulated galaxies, their detailed computations may differ from the methods applied to measure analogous properties in observations. Therefore, we must bear in mind that, in some cases, the comparison can be only done  qualitatively. 

\subsection{Definition and selection of ETGs}
\label{sec:sample}

We select central galaxies defined as the main system within a virial radius  which are resolved with more than 5000 stellar particles in both simulations. Therefore, satellite galaxies are not included in this work.
Taking into account the maximum mesh size resolution, galaxies with $R_{\rm hm}$ lower than 2 kpc are not considered.
We also remove galaxies with stellar masses greater than $10^{12.5}$ M$_{\odot}$ from the Horizon-noAGN simulation. These galaxies with unrealistic properties represent 0.3 per-cent of our sample at $z=0$ \citep{Beckmann2017}. 
The importance of AGN feedback in reproducing the high mass end of the galaxy stellar mass function was previously noted by \cite{Beckmann2017}
and is consistent with the fact that AGN feedback prevents excessive SF and in its absence, more massive galaxies can be formed.
These criteria are applied at all analysed redshifts.

In order to distinguish ETGs from late-type galaxies (LTGs), we follow \cite{Chisari2015} who adopt a threshold at  $V/\sigma=0.55$ (see Section \ref{sec:defi}). Taking this into account, galaxies with stellar $V/\sigma < 0.55$ are considered ETGs since they are dominated by dispersion velocity.
By applying this criterion, the number of members in our ETG samples are 4370 and 1560 for the Horizon-AGN  and  Horizon-noAGN run, respectively, at $z=0$. We remark that all analysed ETGs are central galaxies and hence, hereafter we will drop `central' in the text for the sake of simplicity, but will retain it in the figure and table captions.

In Table \ref{table:nfgal} we summarise the number of members in the resulting samples of ETGs for both simulations at each analysed redshift. As can be appreciated, the number of well-resolved ETGs decreases with increasing redshift in both runs, but there are fewer ETGs in Horizon-noAGN. Nevertheless, the fraction of ETGs over the total number of galaxies with more than 5000 stellar particles,  $R_{\rm hm}$ greater than 2 kpc and masses lower than $10^{12.5}$ M$_{\odot}$ remains approximately constant as a function of redshift in both simulations. It must be stressed that the only difference between both runs is the turn-on/turn-off of only the AGN feedback, the rest of the physics implemented, including SN feedback is always present.

The selection of galaxies and their morphological classification is performed by using the same definitions for both simulations at all analysed redshifts and hence, they are not followed along their merger trees. We follow this approach because  it is especially useful to compare statistically the properties of the simulated objects, taking into account the lack of tuning of Horizon-noAGN simulation.

\section{Characterization of Horizon ETGs at z=0}
\label{sec:char}

In this Section, we analyse the main properties of our subsamples of  ETGs from Horizon-AGN and Horizon-noAGN. A complete characterization of galaxies from these two simulations and a detailed comparison between them are performed by \cite{Dubois2016}. In that work, properties such as halo mass, $V/\sigma$ (and other kinematic parameters), and size are shown as a function of stellar mass at different redshifts. Hereafter, we restrict the analysis to our samples of ETGs obtained as described in the last section.

\begin{figure}
  \centering
\includegraphics[width=0.45\textwidth]{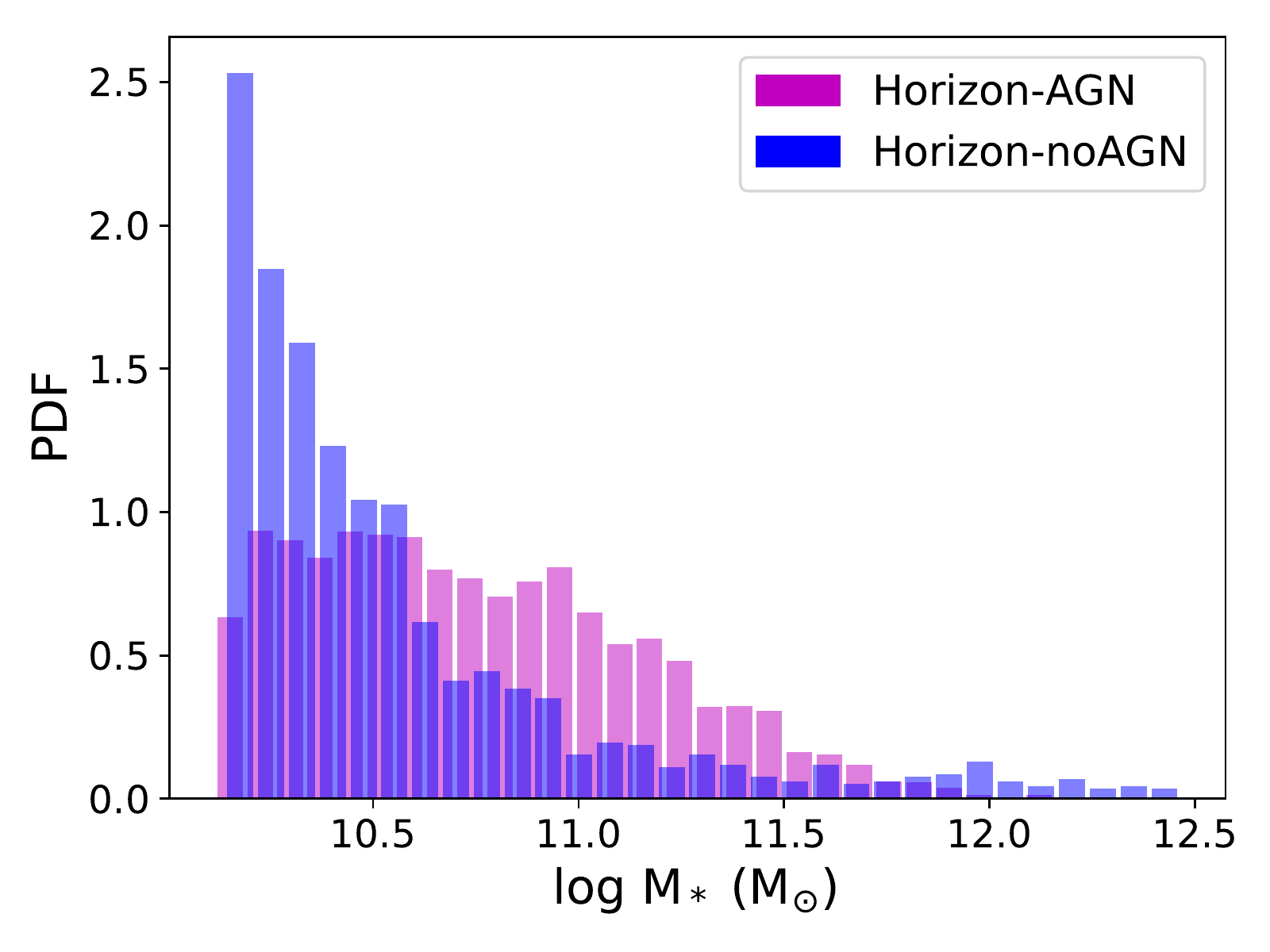}
  \caption{Stellar mass distributions for the selected samples at $z=0$ of central ETGs from Horizon-AGN (magenta) and Horizon-noAGN (blue). The differences show that in the absence of AGN feedback there are fewer massive central ETGs. We remark the existence of a small tail of massive galaxies in Horizon-noAGN, that may be a result of the lack of AGN feedback. }
   \label{fig:Mass_dist}
\end{figure}

In Fig.~\ref{fig:Mass_dist} we show the total stellar mass distributions for both galaxy samples. The histograms are, hereafter, normalised so that their enclosed area is 1 approximating, thus, a probability density function (PDF). This allows a direct comparison of the involved quantities. 
The number of well-resolved, and more massive, ETGs from Horizon-AGN is remarkably larger than those from Horizon-noAGN. 
The low number of massive ETGs from Horizon-noAGN is in agreement with the results of \cite{Dubois2016} as mentioned above and remarks the importance of modelling AGN feedback to achieve a better description of massive galaxies.
There is, however, a small tail of very massive ETGs (2.4 per-cent) in the stellar mass distribution of the Horizon-noAGN sample having total stellar masses greater than $10^{12}$ M$_{\odot}$.

\begin{table*}
\caption{Well-resolved central ETGs the Horizon simulations  at different redshifts.}    
\label{table:nfgal}      
\centering                          
\begin{tabular}{ccccc} 
\hline\hline   
$z$  & \multicolumn{2}{c}{Number}  & \multicolumn{2}{c}{Fraction over total number} \\
  & Horizon-AGN &  Horizon-noAGN &  Horizon-AGN  & Horizon-noAGN  \\
  \hline                        
0 &  4370 & 1560 & 0.25 & 0.08 \\
0.5 & 3641 & 745 & 0.23 & 0.04 \\
1 &  3034 & 441 & 0.24 & 0.03 \\ 
2 &  1507 & 274 & 0.26 & 0.03 \\
3 & 372 & 39 & 0.32 & 0.04 \\
  \hline 
\end{tabular} \\
\end{table*}

\subsection{Star formation activity}
\label{sec:sf}

We analyse the relation between sSFR and total stellar mass for our samples of ETGs, both defined in Section \ref{sec:defi}, from which we can verify the action of AGN. For comparison, we include observations of isolated elliptical galaxies from UNAM-KIAS catalog \citep{Hernandez-Toledo+2010}. We also consider the criterion reported by \citet{Lacerna2014} to distinguish active and passive galaxies.

\begin{figure}
  \centering
  \includegraphics[width=0.45\textwidth]{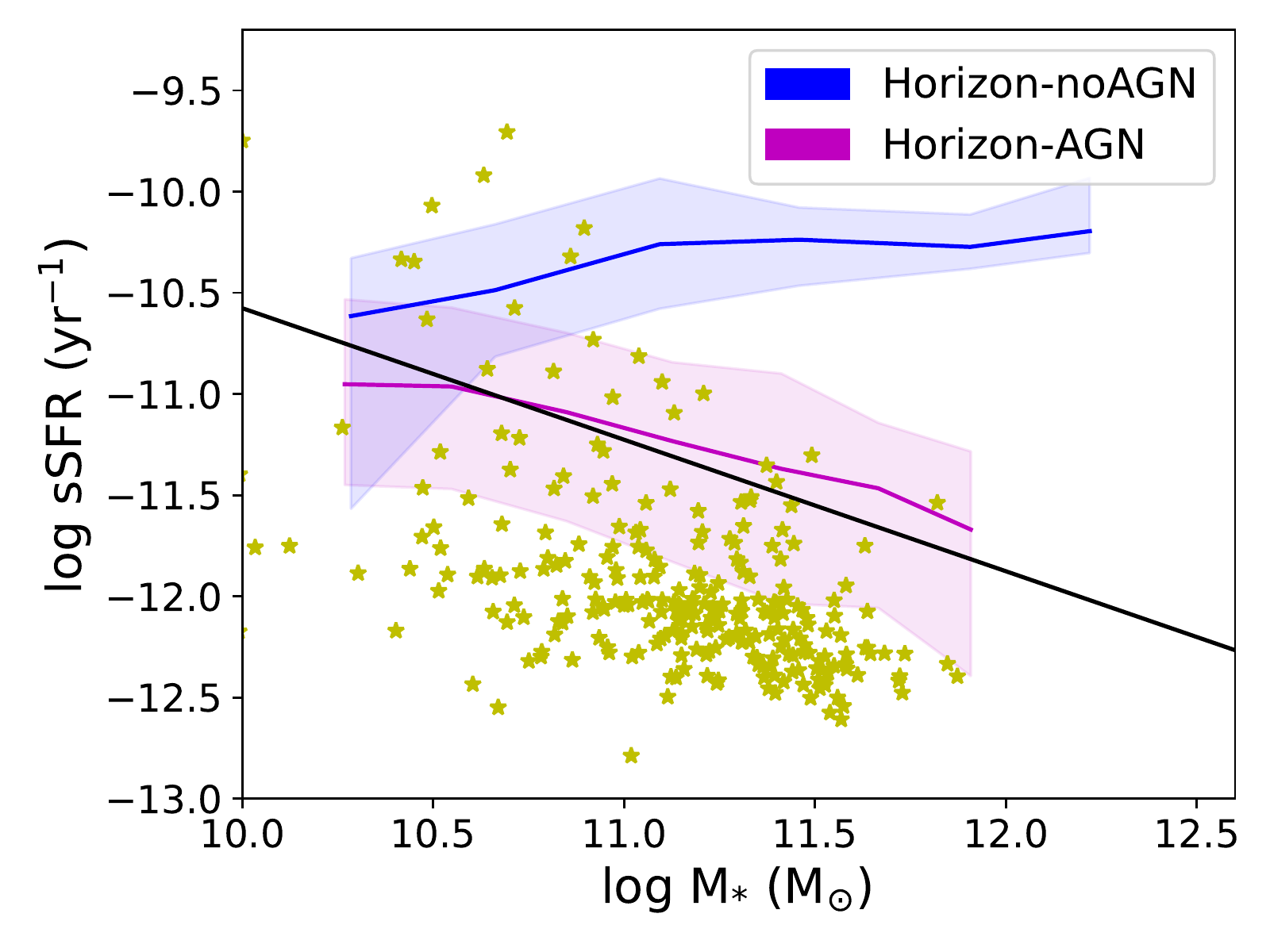}
  \caption{sSFR as a function of stellar mass for central ETGs from Horizon-AGN (magenta) and from Horizon-noAGN (blue). The 25 and 75 percentiles are depicted by the shaded region. We include the line separating active and passive galaxies (above and below the line, respectively) reported by \cite{Lacerna2014} and observations of isolated ETGs from UNAM-KIAS catalog \citep[yellow stars, ][]{Hernandez-Toledo+2010}.}
   \label{fig:sfr}
\end{figure}

As can be seen from Fig.~\ref{fig:sfr}, ETGs from Horizon-AGN  show a decreasing sSFR for increasing stellar mass, in agreement with observations albeit still more active.
A similar result is found for galaxies from Illustris simulation, for which feedback does not seem enough to suppress SF in some galaxies as reported in \cite{Genel2014}.
Other numerical works have already highlighted the impact of AGN feedback on the regulation of the SF activity \citep[e.g.][]{Crain2009, Schaye2010,Haas2013, Crain2015, Pillepich2018, Habouzit2019} as mentioned in the Introduction.
In a previous work, \cite{Chisari2018} conclude the need to modelling stronger AGN feedback in Horizon-AGN to decrease the amount of gas inside galactic haloes and match thus the observations.

On the other hand, ETGs from Horizon-noAGN are, as expected, significantly more star-forming with median sSFR which are almost independent of stellar mass, as can be seen in Fig.~\ref{fig:sfr}. 
It must be noticed, however, that in this simulation, SN feedback is not further tuned to act alone (i.e.the only difference with Horizon-AGN is the lack of black hole formation) and, hence, the differences with observations might be overestimated. 
In fact, galaxies from the Fenix simulation, which does not include AGN feedback but has been performed with a SN feedback model that reproduces the angular momentum content and the mass--size relation \citep{Pedrosa2015}, show a clear trend between sSFR and stellar mass in agreement with observations as reported by \cite{Rosito2018}.
These findings support previous claims that this scaling relation can be used to calibrate the feedback mechanisms in order to reproduce them as pointed out in previous results (see Introduction).

We check that imposing sSFR=$10^{-13}$ yr$^{-1}$ to galaxies with no young SPs does not affect significantly the results shown in Fig.~\ref{fig:sfr}. We compute the median log sSFR without these non-star-forming galaxies in the same mass bins considered in the figure. The differences between these medians and those plotted are negligible regarding the typical values of sSFR ($ < 0.2 $ dex). These galaxies with no recent SF represent the 9 per-cent and the 17 per-cent of the ETGs from Horizon-AGN and Horizon-noAGN.

\subsection{Mass--size relation}
\label{sec:ms}

In this Section, we investigate possible dependencies of the mass--size relation on age and level of rotation ($V/\sigma$) of the SPs. 
We compare the mass--size relation obtained from Horizon-AGN and Horizon-noAGN samples. 
We use an implementation in Python, coded by \cite{CappellariAtlasXX} of the two-dimensional Locally Weighted Regression method \citep{Cleveland88} so as to obtain smoothed distributions. 

\subsubsection{Galaxies formed under the action of AGN feedback}

\begin{figure*}
    \centering
    \includegraphics[width=0.45\textwidth]{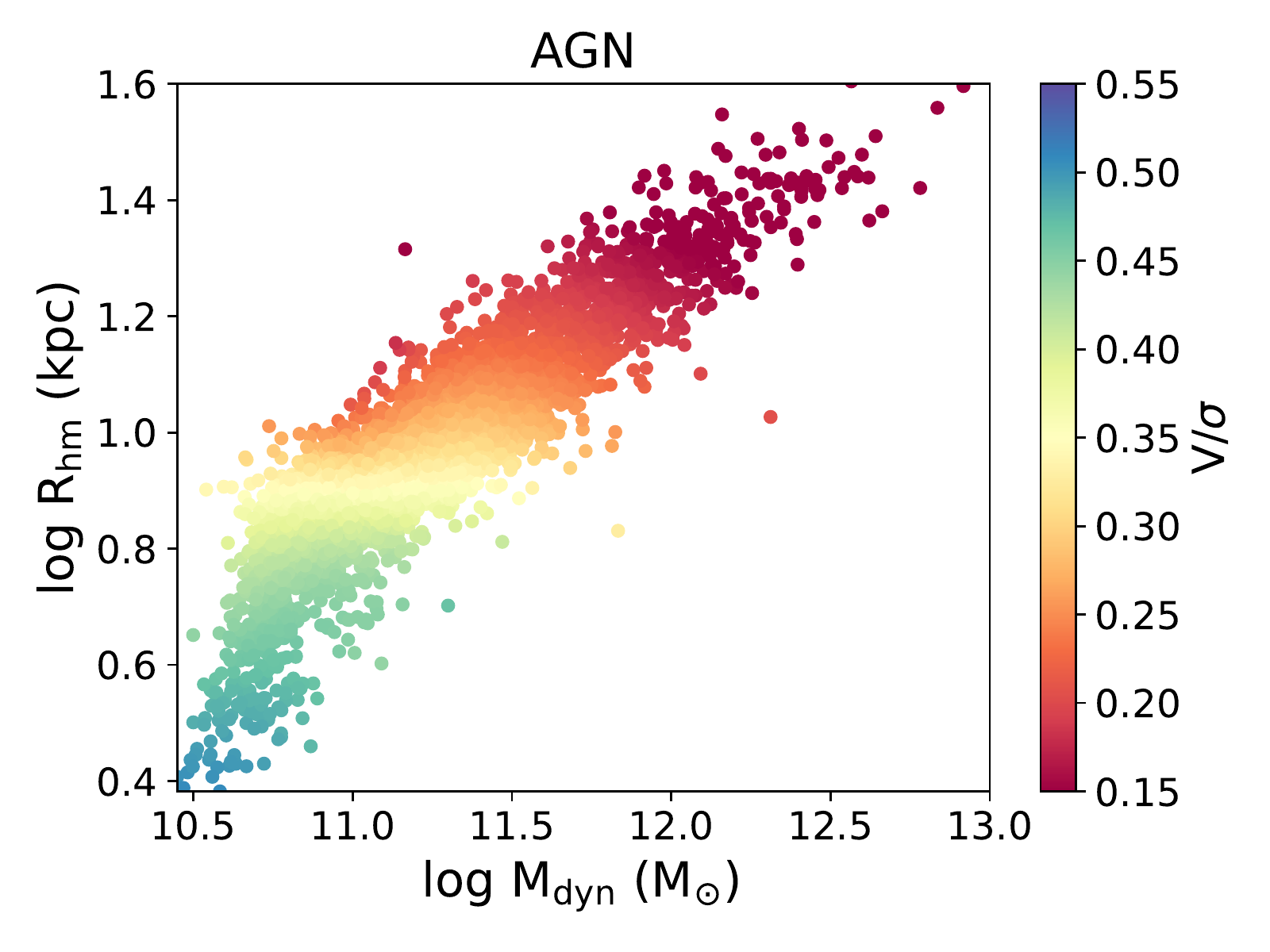}
    \includegraphics[width=0.45\textwidth]{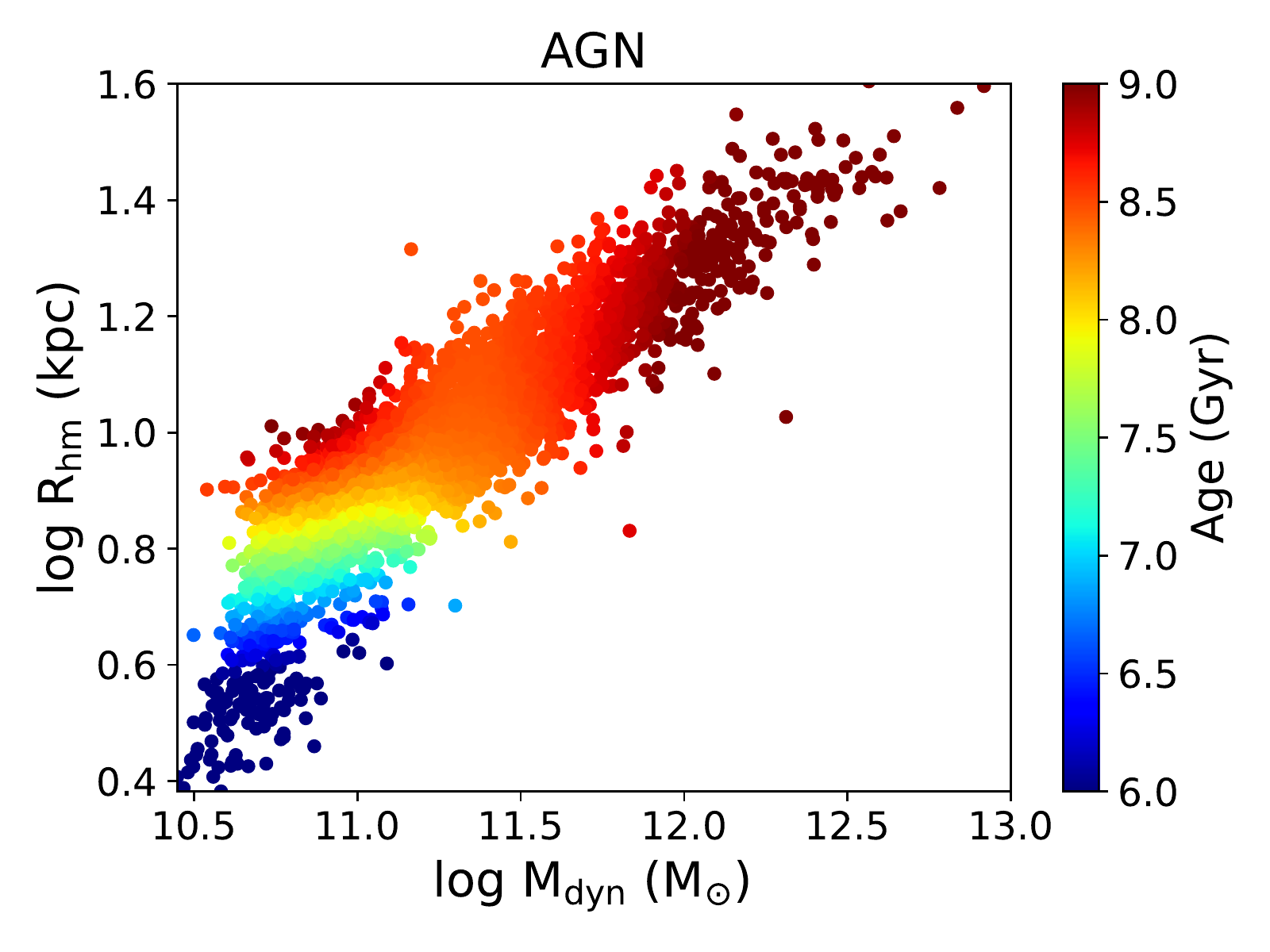} \\
    \includegraphics[width=0.45\textwidth]{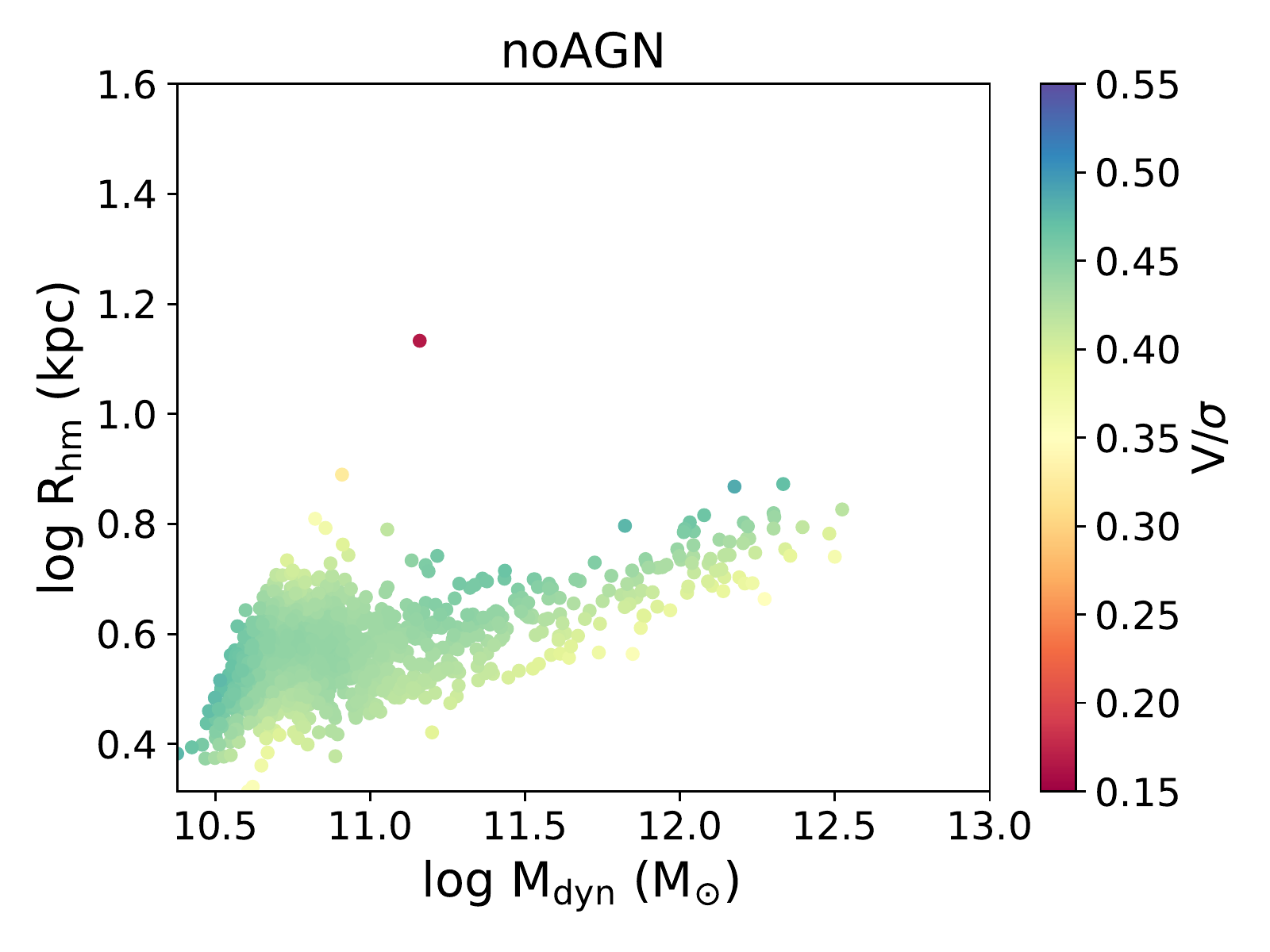}
    \includegraphics[width=0.45\textwidth]{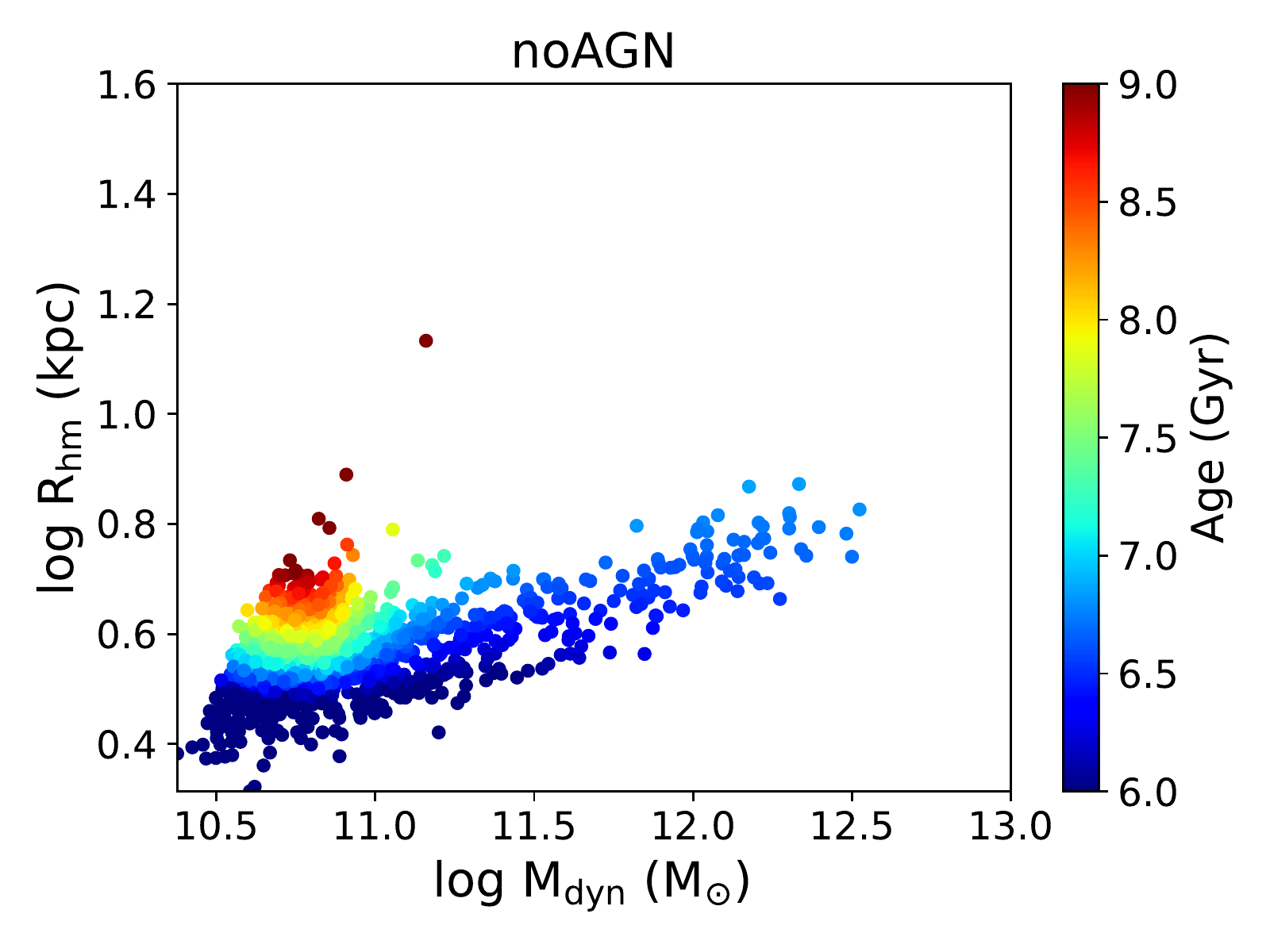} \\
    \caption{Mass--size relation as a function of stellar $V/\sigma$ (left panels) and mass-weighted average stellar age (right panels) for both samples of central ETGs at $z=0$.}
    \label{fig:mdv}
\end{figure*}

Let us first study the mass--size relation as a function of stellar $V/\sigma$ and stellar age. 
Here, following \citet{Rosito2019b} and \citet{Li2018}, we analyse the  mass--size relation using the dynamical mass, $M_{\rm dyn}$, estimated assuming virialisation.
\cite{Dubois2016} already report a strong correlation between stellar mass and the effective radius. Hence, the relation is also present between $M_{\rm dyn}$ and stellar $R_{\rm hm}$, as can be seen in the upper panels of Fig.~\ref{fig:mdv}.
We note that \cite{vdSande2019} find that galaxies from Horizon-AGN are more extended for all stellar masses than those observed, as mentioned in the Introduction.
 
In Fig.~\ref{fig:mdv} (left upper panel) a correlation between mass and $V/\sigma$ is found, such that more massive galaxies are more dispersion-dominated. 
This is the expected trend, considering that galaxies with a lower level of rotation are, generally, more massive, in agreement with  observational findings \citep[e.g.][]{Emsellem2011}. 
 \cite{Li2018} report that more massive galaxies present higher values of velocity dispersion. However, in their work, there is little dependence of velocity dispersion on radius. In our sample, since the increase of $R_{\rm hm}$ with dynamical mass is almost linear, and the correlation is, therefore, stronger, $V/\sigma$ decreases significantly with increasing size, as well as with dynamical mass. 

With respect to age variations across the mass--size relation, from the right upper panel of Fig.~\ref{fig:mdv}, it is clear that at higher masses, galaxies are dominated by older SPs. 
 A very low number of old, low-mass galaxies can be seen but it is not representative of the whole sample. This population is overemphasised by the smoothing method
\footnote{Although in some cases the smoothing methods can overemphasise features, we have carefully checked for possible spurious results between the smoothed and non-smoothed versions.}.
This small less massive old population cannot be seen in the non-smoothed distribution and we can therefore conclude that this trend is due to the smoothing method.
Our results are consistent with the ones from \cite{vdSande2019}.

\subsubsection{Galaxies formed in the absence of AGN}

The mass--size relation for galaxies formed in the absence of AGN feedback shows that they have lower radii at a given $M_{\rm dyn}$, as seen in lower panels of  Fig.~\ref{fig:mdv}.
As mentioned in the Introduction, it can be clearly seen that the presence of AGN feedback is needed to  produce more extended massive galaxies \citep{Dubois+2013}.
This is consistent with the fact that Horizon-noAGN ETGs also present higher values of the average surface density, defined in Section \ref{sec:defi}, with a maximum $\Sigma_e \sim 10^{10.2} $ M$_{\odot}$ kpc$^{-2}$. In contrast, the maximum $\Sigma_e$ for ETGs in Horizon-AGN is $\sim 10^9$ M$_{\odot}$ kpc$^{-2}$. This fact was previously reported by \cite{Peirani2019}, who conclude that without AGN feedback, simulated galaxies are more compact than observed ones, whereas the values of $\Sigma_e$ for massive ellipticals from Horizon-AGN match the observational data. Nevertheless, they remark that low-mass ETGs from that simulation are not compact enough to be in agreement with observations
(see the Introduction for comments on previous works such as \citealt{Crain2015}; \citealt{Rosas2015} and \citealt{Habouzit2019}).
We will return to this point in Section \ref{sec:empfp}. 

The trend of rotation across mass--size relation is less clear, but massive galaxies still show lower values of stellar $V/\sigma$ (Fig.~\ref{fig:mdv}), as expected.
For the stellar age, an old low-mass population can be found but it cannot be attributed to the smoothing method. 
This fact may be ascribed to the action of SN feedback, which is stronger for small galaxies \citep{Larson1974, White1978, Dekel86, White1991}, and therefore would quench SF in these objects at earlier times. When the AGN feedback is on, the regulation of the SF is clearly different across the whole mass range so a smoother variation of the mean age is achieved as a function of mass as shown in Fig.~\ref{fig:mdv}.

Considering that the SN feedback in this simulation has not been tuned to achieve better modeling of galaxy properties on its own (i.e. it is the same that is used when the AGN feedback is active), a good match with observations is not to be expected.

\subsection{Inner stellar discs}
\label{sec:inner}

As reported in \cite{Rosito2018} using the Fenix sample and \cite{Rosito2019a} the EAGLE sample, there is a rotational component embedded within the bulge, so-called inner stellar disc. In this Subsection, we explore the presence of this component in the Horizon galaxies.
We calculate $F_{\rm rot}$ that is a measure of the relevance of these inner discs defined in Section \ref{sec:defi}.
For this analysis, we discard galaxies with inner discs resolved with less than 250 star particles in order to mitigate numerical effects. With this condition, sub-samples comprising 1262 and 345 members are defined for Horizon-AGN and Horizon-noAGN, respectively.

\begin{figure*}
  \centering
\includegraphics[width=0.45\textwidth]{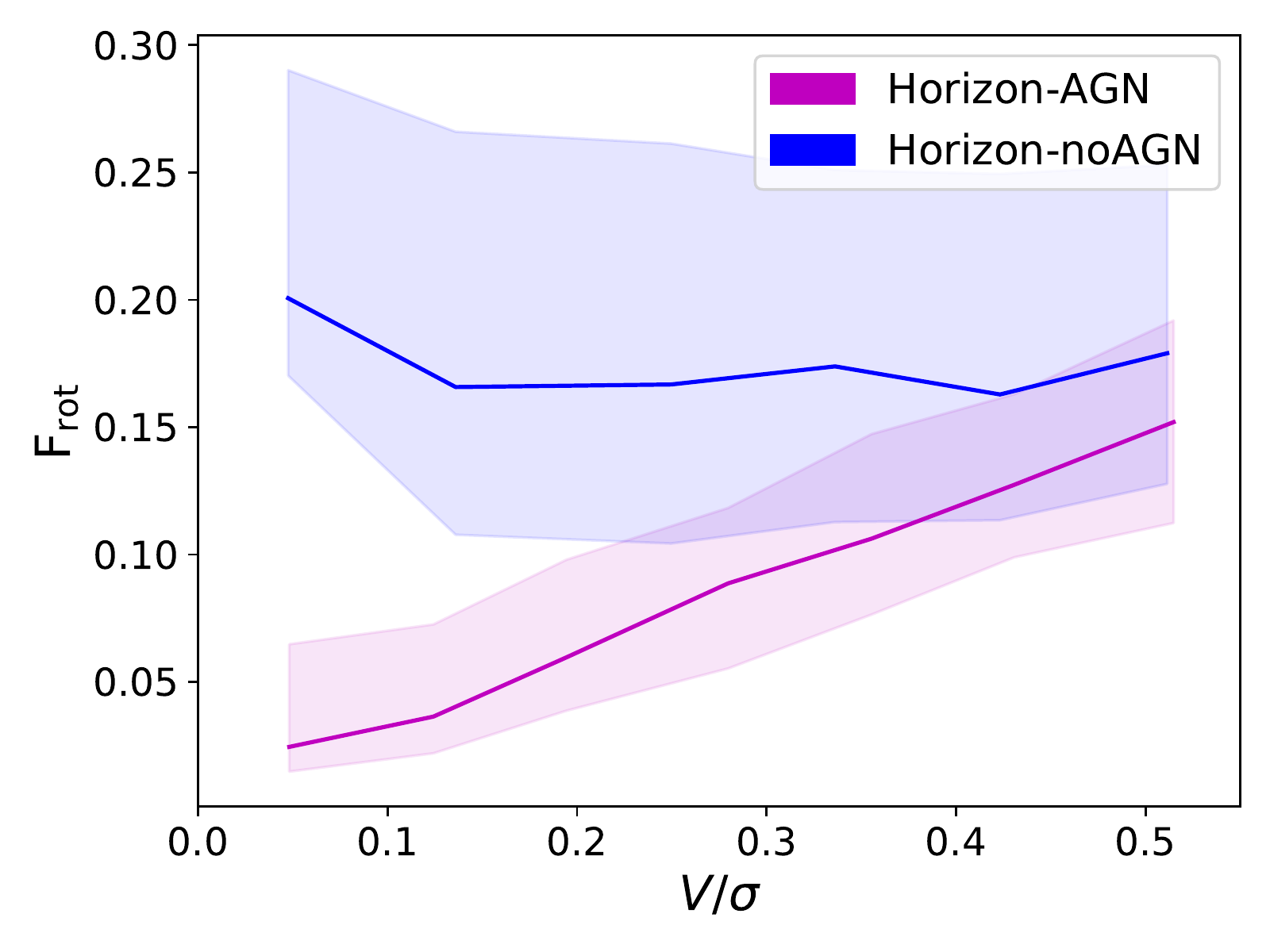}
\includegraphics[width=0.45\textwidth]{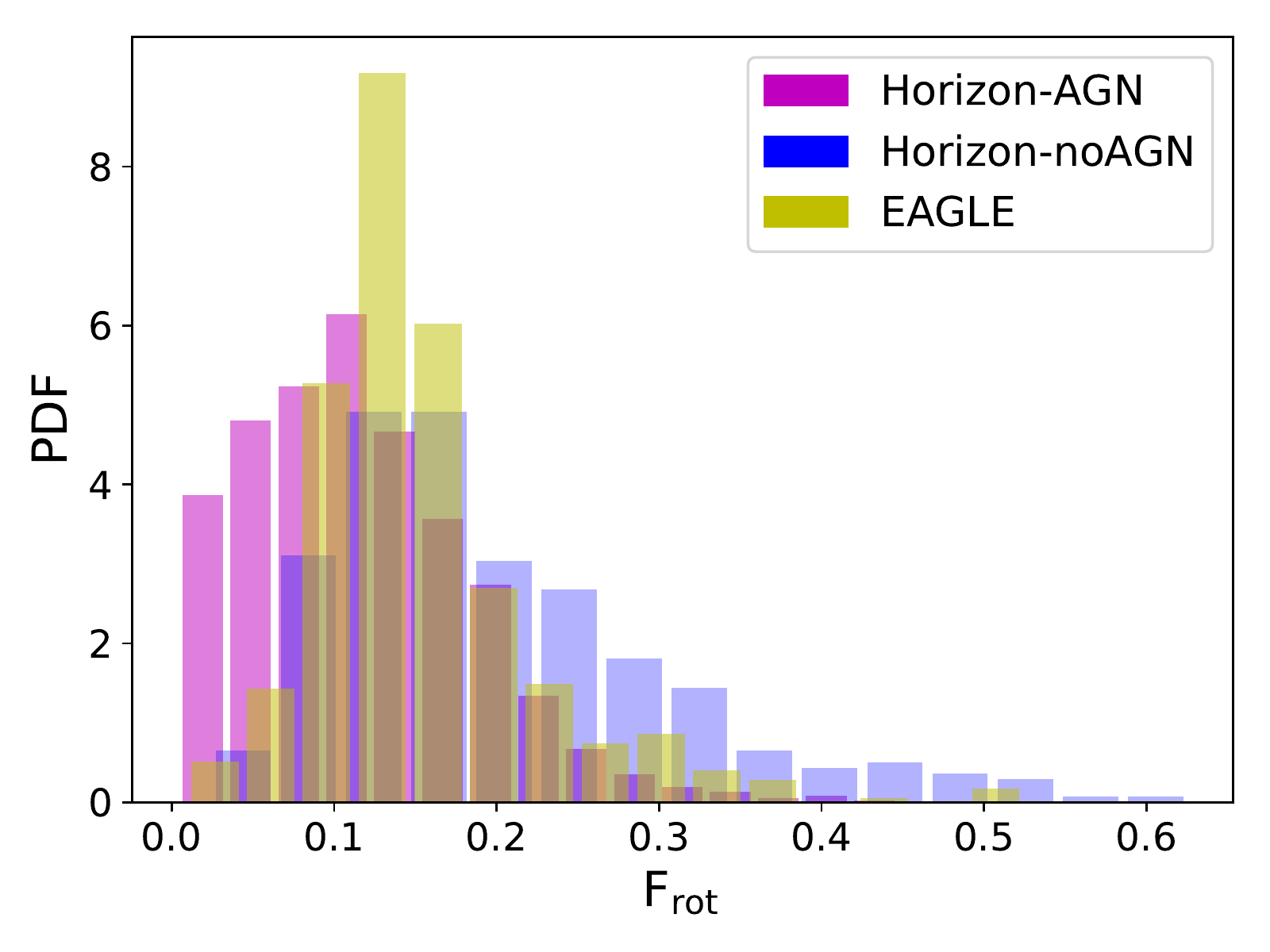}
  \caption{Left panel: $F_{\rm rot}$ as a function of $V/\sigma$ for central ETGs from Horizon-AGN (magenta) and Horizon-noAGN (blue) simulations. The 25 and 75 percentiles are depicted by the shaded regions. Right panel: Histograms of $F_{\rm rot}$ for central  ETGs selected from Horizon-AGN (magenta), Horizon-noAGN (blue) and EAGLE (yellow) at $z=0$.}
   \label{fig:id}
\end{figure*}

As can be seen from the left panel of Fig.~\ref{fig:id}, for ETGs in Horizon-AGN, there is a clear correlation between $F_{\rm rot}$ and the globally measured $V/\sigma$ \citep[as in][]{Dubois2014} with a p-value $\sim 0$. This is consistent with the results of \cite{Rosito2018} and \cite{Rosito2019a} where  anti-correlations between $F_{\rm rot}$ and $B/T$ are also reported. 
For galaxies in Horizon-noAGN, no clear correlation is found (Spearman correlation coefficient $\sim 0.06$ and p-value $\sim 0.30$).
 
In Fig.~\ref{fig:id} (right panel), normalised histograms of $F_{\rm rot}$ for our sub-samples (Horizon-AGN and Horizon-noAGN ETGs) are shown. For comparison, the distribution obtained by \cite{Rosito2019a} for the largest volume run of the EAGLE project simulations is included. It can be seen that the peaks of the distributions for Horizon-AGN and EAGLE ETGs are close to each other, but the latter is significantly higher while the former has a larger fraction of ETGs with negligible inner rotation. Horizon-noAGN galaxies present larger values of $F_{\rm rot}$, as can be seen in the same figure.
This suggests that AGN feedback can reduce the formation of inner discs and hence, the inner regions are more dispersion-dominated.

\subsection{Slow and fast rotators}

\begin{figure*}
  \centering
    \includegraphics[width=0.45\textwidth]{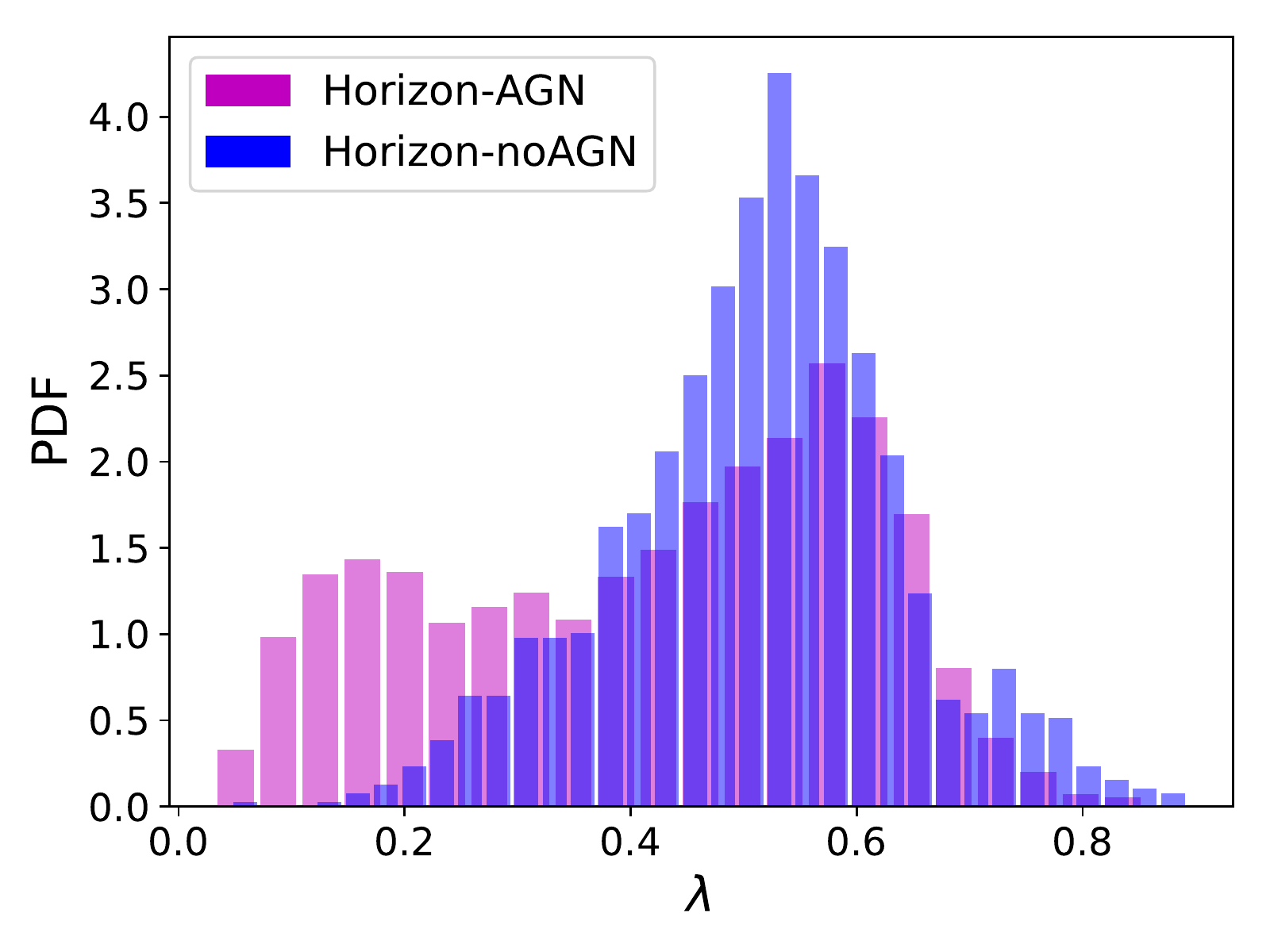}
    \includegraphics[width=0.45\textwidth]{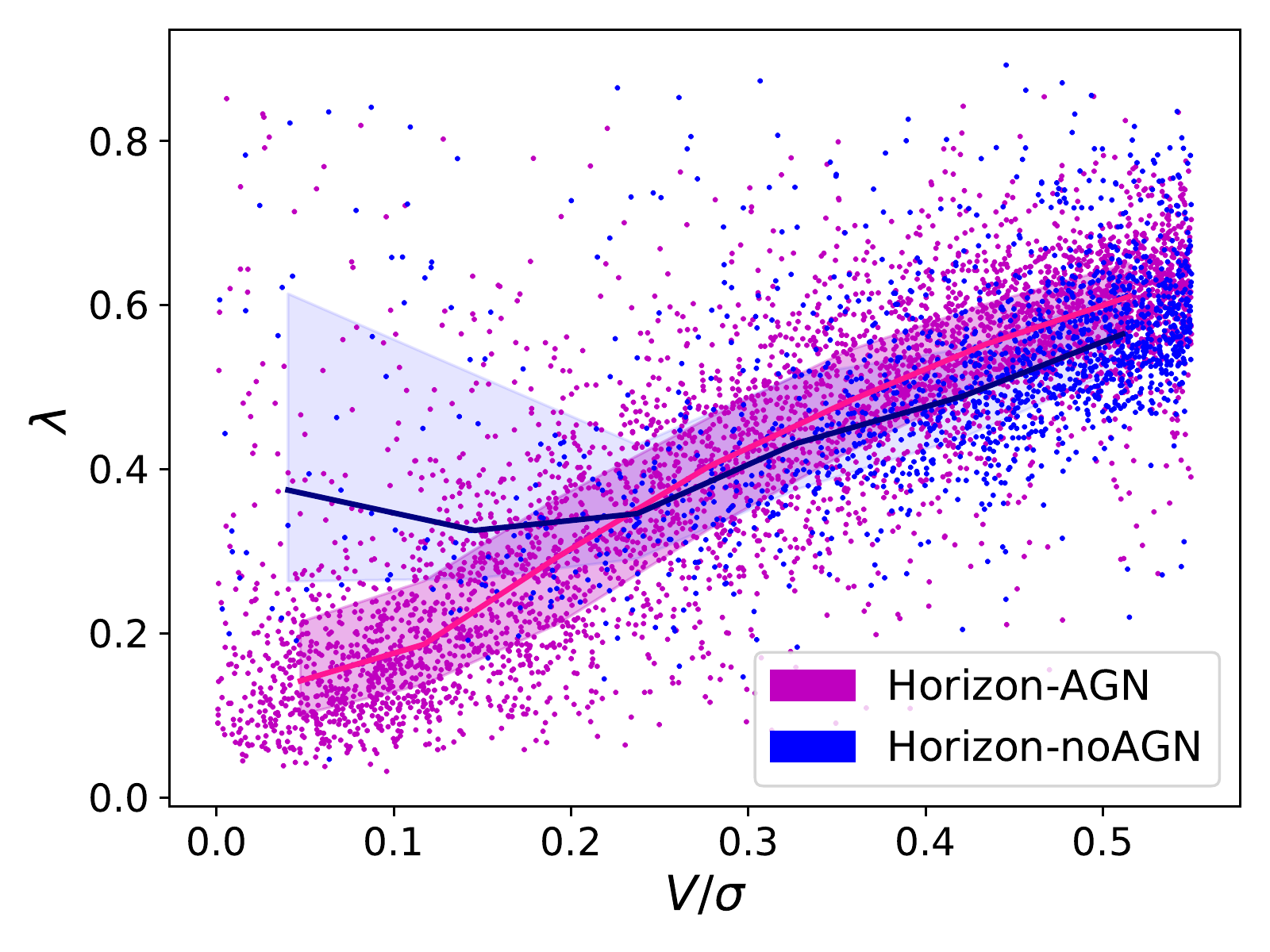}
  \caption{Left panel: Histograms of the spin parameter for central ETGs from Horizon-AGN (magenta) and Horizon-noAGN (blue). Right panel: Spin parameter as a function of $V/\sigma$ for Horizon-AGN (magenta) and Horizon-noAGN (blue) samples. Shaded regions depict the 25 and 75 percentiles. We include the scatter plots for both simulations.}
   \label{fig:spin}
\end{figure*}

As in many observational works \citep[e.g.][]{Emsellem2011}, we find that most Horizon-AGN ETGs are fast rotators \citep[see also][]{vdSande2019}. This is assessed using the spin parameter $\lambda$ within $R_{\rm hm}$ (see Section \ref{sec:defi}).
If we consider $\lambda \sim 0.2$ as a suitable threshold to distinguish fast and slow rotators as in \cite{Rosito2019b}, only 18 per-cent of the simulated ETGs are found to be slow rotators. In Fig.~\ref{fig:spin}, left panel, we show histograms of the spin parameters for these ETGs. For Horizon-AGN, we can observe a bimodality with peaks at $\lambda \sim 0.1$ (slow rotators) and  $\lambda \sim 0.6$. \cite{Graham2018} find a bimodal distribution of this parameter in observations, albeit the second peak is reported to be at $\lambda \sim 0.9$. The differences in the calculation of this parameter in that observational work must however be taken into account.
We remark that in a previous work by \cite{Choi2018}, where they use a relation between the spin parameter and the ellipticity to classify galaxies as slow/fast rotators, as in \cite{Emsellem2011}, they find a fraction of 11.2 per-cent of slow rotators in Horizon-AGN. In their work, they find a correlation of the amount of rotation with mass which is explained by the fact that massive galaxies have experienced a larger number of mergers regardless their environment and thus they rotate more slowly.

As expected, there is an even smaller fraction of slow rotators (defined above) in Horizon-noAGN (1 per-cent). This is consistent with the fact that in the absence of AGN, it is more difficult to reproduce massive ETGs \citep{Dubois2016}. For this simulation the distribution of $\lambda$ is unimodal with a peak at about $\lambda \sim 0.55$ (Fig.~\ref{fig:spin}, left panel). 
Therefore, we can conclude that AGN feedback is important for reproducing the bimodality present in the observations \citep{Graham2018} and hence, the angular momentum distribution of the SPs.

For the Horizon-AGN sample, the correlation between the spin parameter and $V/\sigma$ is excellent (right panel of Fig.~\ref{fig:spin}, magenta lines). 
For ETGs in Horizon-noAGN there is also a clear correlation when stellar $V/\sigma > 0.1$. The high values of $\lambda$ for galaxies with $V/\sigma \leq 0.1$ in the Horizon-noAGN run are due to a small number of outliers. Therefore, the existence of the correlation between morphology and spin parameter is not affected.
In fact, in Fig.~\ref{fig:id} it can be appreciated that ETGs with these low $V/\sigma$ have high inner stellar disc fractions, which are consistent with the high $\lambda$ parameters shown in Fig.~\ref{fig:spin}.

\section{The fundamental scaling relations and their redshift evolution}
\label{sec:empfp}

In this Section, we investigate the impact of the AGN feedback on the two fundamental relations: the mass--plane and the FP, their parameters compared to those obtained assuming virialization and their evolution with redshift.


\subsection{The mass--plane}
\label{sec:mpfp}

The mass--plane, which  can be understood as a consequence of the virial theorem \citep{Bolton2007}, relates the dynamical mass with the galaxy radius, $R$, and the velocity dispersion, $\sigma$ of galaxies:
\begin{equation}
    \label{eq:mp_def}
    M \propto \sigma^{\alpha}R^{\beta}/{G}.
\end{equation}
According to the virial theorem, the exponents should be $\alpha=2$ and $\beta=1$ \citep[see][]{Binney1987}.

\begin{figure*}
  \centering
\includegraphics[width=0.45\textwidth]{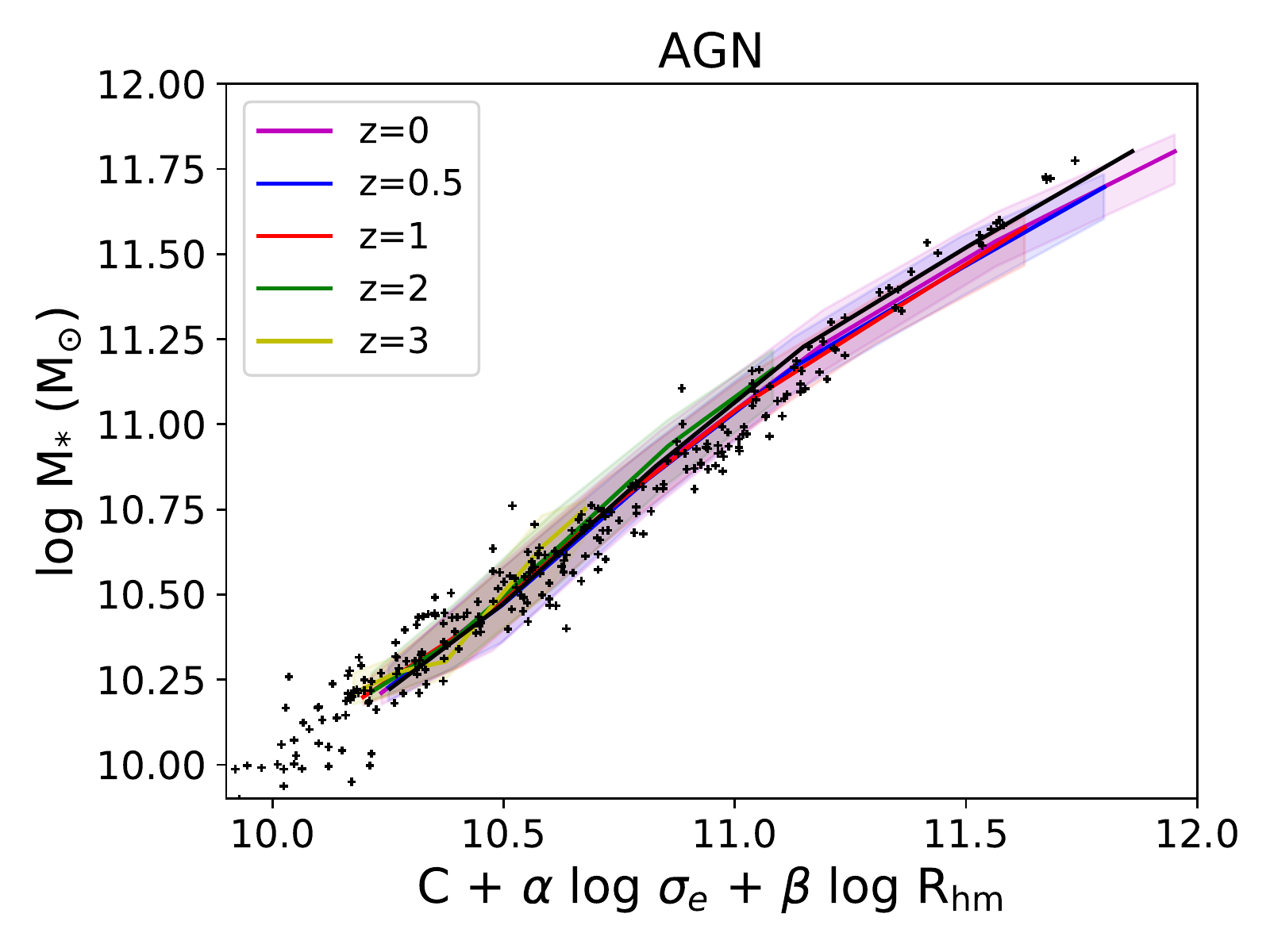}
\includegraphics[width=0.45\textwidth]{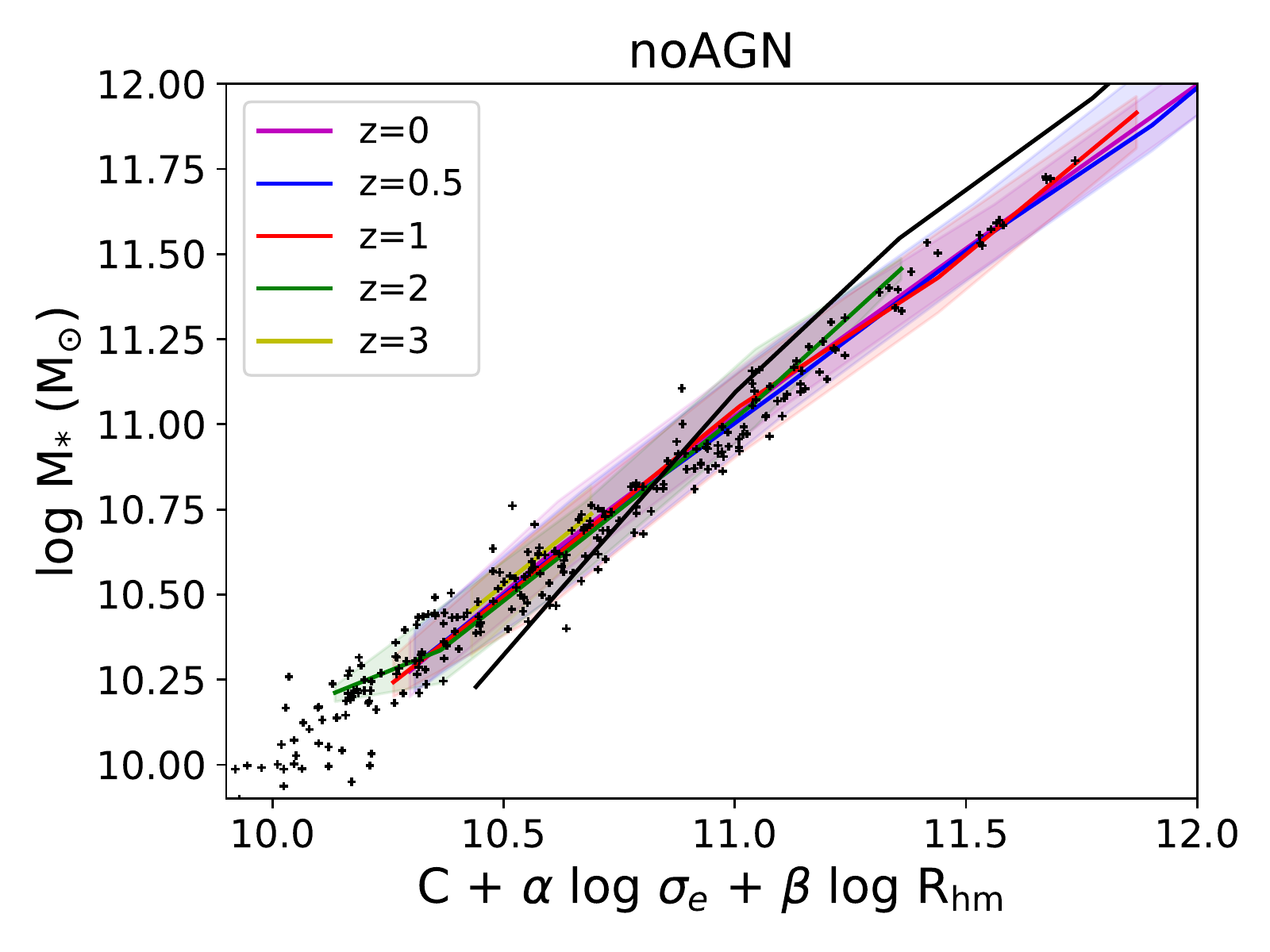}
  \caption{Mass--plane as a function of redshift for central ETGs from Horizon-AGN (left panel) and Horizon-noAGN (right panel). The magenta, blue, red, green and yellow lines depict the simulated relations at $z =0$, $z = 0.5$, $z= 1$, $z = 2$ and $z = 3$, respectively. The shaded regions delimit the 25 and 75 percentiles. We also show the virial relation for $z = 0$ in a black solid line. Observations from ATLAS$^{\rm 3D}$ project \citep{AtlasI, Cappellari2013, CappellariAtlasXX} at $z=0$ are included for comparison (black crosses).}
   \label{fig:mp-agnnagn}
\end{figure*}

Instead of working with dynamical masses, we use  stellar masses as in \cite{Onorbe2005} and \citet[][see Introduction]{Onorbe2006}. 
The galaxy properties used to estimate this plane are the total stellar mass, $R_{\rm hm}$ and $\sigma_e$ explained in Section \ref{sec:defi}.
In both Horizon simulations, we perform ordinary multiple linear regressions, being $M_*$ the dependent variable, to obtain the parameters in  Eq.~(\ref{eq:mp_def}) at each analysed redshift, as shown in Fig.~\ref{fig:mp-agnnagn}. For comparison,  we also include the virial relation ($\alpha=2$ and $\beta=1$) for $z=0$. The constant term in that relation at $z=0$ is set \textit{ad-hoc}. From this figure, we can appreciate that, in the presence of AGN, the mass--plane is in place since $z=3$ and is approximately consistent with the expectation from the virial theorem.  
On the other hand, the mass--plane for Horizon-noAGN ETGs is also in place since $z=3$, but it is clearly  tilted from the virial relation. 
This indicates that, given the same conditions, AGN feedback plays an important role in the regulation of  the SF, allowing the simulated galaxies to better reproduce the observed mass--plane. The larger impact is detected for $\beta$, which is associated with the galaxy size.
All the  parameters of the linear regression are summarized in Table~\ref{table:tableMP}.
The fact that we find different parameters from the linear regressions for Horizon-noAGN indicates that AGN feedback affects not only the stellar mass, the size, and the velocity dispersion but also the relation among them as clearly unveiled by the mass--plane. 

To compare with recent observations at $z=0$, we calculate the mass--plane for ETGs from ATLAS$^{\rm 3D}$ project \citep{AtlasI} by ordinary multiple linear regressions. 
We adopt the stellar mass estimated by the JAM method which provides a more robust estimation that does not largely depend on the IMF
\citep{CappellariAtlasXX} and which better compares with the stellar mass estimated from the simulations.
The effective radius (comparable with $R_{\rm hm}$) and the velocity dispersion within that radius are extracted from \cite{Cappellari2013}.
The resulting parameters for Eq.~(\ref{eq:mp_def}) are $\alpha_{\rm obs}=1.930 \pm 0.030$, $\beta_{\rm obs}=0.925 \pm 0.020$  and $C_{\rm obs}=6.241 \pm 0.064$. We calculate the errors by a bootstrap method.
\cite{Li2018} also calculate the mass--plane parameters for nearby elliptical and late-type galaxies (LTGs) classified by their S\'{e}rsic indexes \citep{Sersic1968}. They consider the major axis of the fitted ellipse of the half-light isophote, instead of the effective radius, and twice the dynamical mass, instead of the stellar mass.
By using least trimmed squares (LTS) regression \citep{Rousseeuw1987}, they obtain $\alpha_{\rm obs}=1.959 \pm 0.018$ and $\beta_{\rm obs}=0.964 \pm 0.013$ for the ETGs.
From Fig.~\ref{fig:mp-agnnagn}, it can be seen that both simulated mass--planes overlap the observations. However, the fitted parameters shown in Table~\ref{table:tableMP} differ significantly from the observed ones within their errors. At $z=0$, the parameters computed for Horizon-noAGN are noticeably farther away from the observed ones than those obtained from Horizon-AGN considering the estimated errors. This is particular clear for $\beta$ parameter.
Regarding this, in the case of Horizon-AGN, there is approximately the same level of scatter $\sigma \approx 0.10$ at all analysed redshifts while for Horizon-noAGN the scatters slightly increases with increasing redshift.

\begin{table*}
\small
\caption{Evolution of mass--plane parameters with and without AGN feedback. The parameters $\alpha$ and $\beta$ are given by  Eq.~(\ref{eq:mp_def}). The errors are estimated by applying a bootstrap technique. The standard deviations are also included ($\sigma$). } 
\label{table:tableMP}   
\centering              
\begin{tabular}{cccccccccc}   
\hline\hline 
 Redshift & \multicolumn{4}{c}{Horizon-AGN} & \multicolumn{4}{c}{Horizon-noAGN} \\
   &  $\alpha$ &   $\beta$ &  $C$  & $\sigma$ & $\alpha$ &  $\beta$ & $C$ & $\sigma$ \\
  \hline                   
0 & 2.553 $\pm$ 0.033 & 0.710 $\pm$ 0.017 & 4.103 $\pm$ 0.060 & 0.103 & 2.898 $\pm$ 0.013 & 0.249 $\pm$ 0.028 & 3.563 $\pm$ 0.032 & 0.066 \\
0.5 & 2.333 $\pm$ 0.030 & 0.794 $\pm$ 0.024 & 4.559 $\pm$ 0.054 & 0.099 & 2.725 $\pm$ 0.023 & 0.151 $\pm$ 0.055 & 4.093 $\pm$ 0.036 & 0.079 \\
1 & 2.190 $\pm$ 0.031 & 0.945 $\pm$ 0.027 & 4.734 $\pm$ 0.057 & 0.097 & 2.665 $\pm$ 0.036 & 0.233 $\pm$ 0.103 & 4.098 $\pm$ 0.062 & 0.108 \\
2 & 2.032 $\pm$ 0.044 & 0.722 $\pm$ 0.040 & 5.253 $\pm$ 0.087 & 0.115 & 2.460 $\pm$ 0.061 & -0.021 $\pm$ 0.134 & 4.628 $\pm$ 0.123 & 0.133 \\
3 & 1.573 $\pm$ 0.088 & 0.668 $\pm$ 0.078 & 6.266 $\pm$ 0.207 & 0.117 & 2.418 $\pm$ 0.167 & 0.706 $\pm$ 0.647 & 4.248 $\pm$ 0.492 & 0.186 \\
  \hline
\end{tabular} \\
\end{table*}

In order to evaluate the evolution of the mass--plane and to assess the effect of AGN feedback,  we use the parameters of Eq.~(\ref{eq:mp_def}) computed at $z =0$  
as a reference to estimate the relation defined by Eq.~(\ref{eq:mfit})  for ETGs  at different redshifts. 

\begin{equation}
    \log M_{\rm fit} = \alpha_{z=0} \log \sigma_e + \beta_{z=0}  \log R_{\rm hm}  + C_{z=0}
    \label{eq:mfit}
\end{equation}

\begin{figure*}
  \centering
\includegraphics[width=0.45\textwidth]{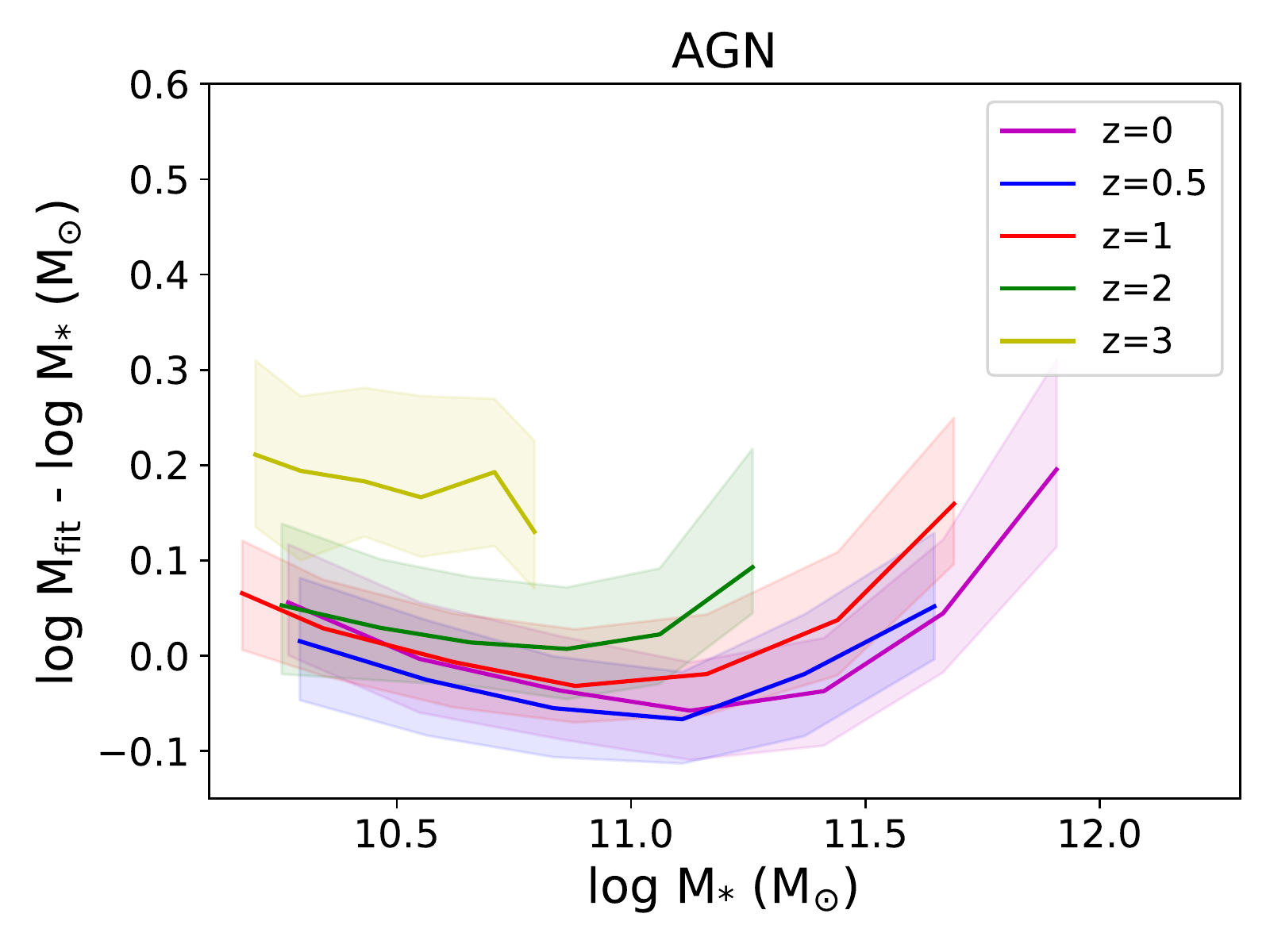}
\includegraphics[width=0.45\textwidth]{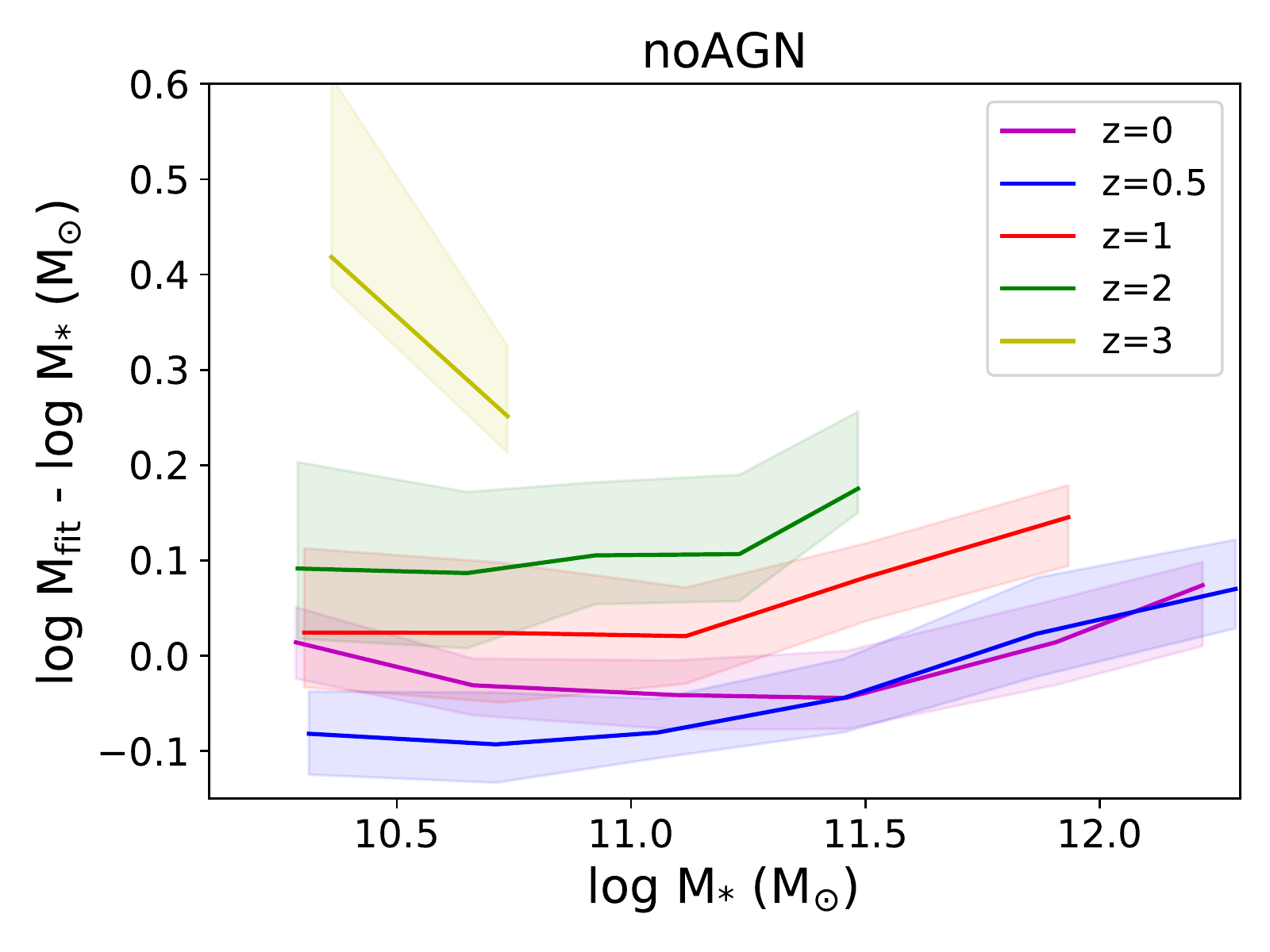}
  \caption{Differences between $\log M_{\rm fit}$ (defined in Eq.~\ref{eq:mfit})  computed with parameters at $z=0$ and stellar mass at different redshifts for central ETGs from Horizon-AGN (left panel) and Horizon-noAGN (right panel). The magenta, blue, red, green and yellow lines depict the simulated relations at $z = 0$, $z = 0.5$, $z=1$, $z = 2$ and $z = 3$, respectively. The shaded regions are delimited by the 25 and 75 percentiles.}
   \label{fig:displacement}
\end{figure*}

In Fig.~\ref{fig:displacement} we show the difference between $M_{\rm fit}$ and the galaxy stellar mass measured at different redshifts as a function of the stellar mass. This procedure was applied to both Horizon runs using the corresponding parameters for their mass--planes. As can be seen from this figure, a systematic displacement towards higher positive differences is detected for galaxies from both simulations.
When the AGN feedback is on, we note that there is no important evolution in the mass-plane up to $z =1$, so that the stellar masses are very close to $M_{\rm fit}$ within a wide  range.  However, for $ 1< z <3$ the differences are larger, indicating a stronger evolution of the mass-plane over this redshift interval (see Table \ref{table:tableMP}). Hence, although the mass--plane is set since $z \sim 3$, it converges to the $z=0$ plane by $z \sim 1$. We will come back to this point in Section \ref{sec:disc} in relation to the FP evolution.

For Horizon-noAGN ETGs, the evolution is more noticeable, with larger displacements for increasing  redshift. From Table~\ref{table:tableMP} we can see that the scatter of the regressions increases significantly for higher redshift and that the errors in the parameter $\beta$ are very large at $z = 2$ and $z = 3$.
In fact, $\beta$ is nearly zero for $z=2$ and, hence, there is almost no dependence on size in the mass--plane. For $z=3$ there is a small number of galaxies and the error for $\beta$ is also very high. Hence the large errors do not allow us to achieve robust conclusions with regards to the linear regressions at very high redshift.

\subsection{The fundamental plane}
\label{sec:fp}

The FP \citep{Faber1987, Dressler1987, Davis1987} is an alternative representation of the fundamental relation depicted in the mass--plane, assuming homology and a constant $M/L$ for all elliptical galaxies. Therefore, the functional form is similar:
\begin{equation}
    \label{eq:fp_def}
    R_{\rm eff} \propto \sigma_e^{\alpha} I_e^{\beta}.
\end{equation}
The subscript "e" indicates that the velocity dispersion is measured within $R_{\rm eff}$, which is replaced by $R_{\rm hm}$ in the computation of simulated data.
Here, $I_e$ is the average surface brightness  within the effective radius defined as $I_e=L/(2\pi R_{\rm eff}^2)$. By assuming virialisation and homology, $\alpha=2$ and $\beta=-1$ are obtained \citep[e.g.][]{Mo2010}. However, the observed FP shows a tilt with respect to the virial prediction, which could be attributed to a variation of $M/L$, the dark matter fraction, or possible non-universality of the IMF. It has been shown that a dynamical estimation of the stellar masses takes into account much of this tilt  \citep{cappellarireview2016}.
Since we are mainly interested in unveiling the role of AGN feedback on the FP, we work with stellar masses instead of luminosities (see Section \ref{sec:disc}
for a discussion on the latter).

We perform a similar analysis to that described in the last subsection for the mass--plane. For each analysed redshift, we estimate the parameters in the Eq.~(\ref{eq:fp_def}) for ETGs in the Horizon-AGN and Horizon-noAGN samples. An ordinary multiple linear regression where the dependent variable now is $R_{\rm hm}$ is applied.
We replace $I_e$ by $\Sigma_e$ and thus the  FP is defined by the total stellar mass
of the simulated galaxies. These stellar masses 
are comparable to the $M_{\rm JAM}$ estimated by
\citet{Cappellari2013}. 
To compare with the observations of these authors, a multiple linear regression is applied to the observed FP from which  the following values are obtained: $\alpha_{\rm obs}=1.678 \pm 0.035$ and $\beta_{\rm obs}=-0.848 \pm 0.019$.  The errors are calculated with a bootstrap method.

\begin{figure*}
  \centering
\includegraphics[width=0.45\textwidth]{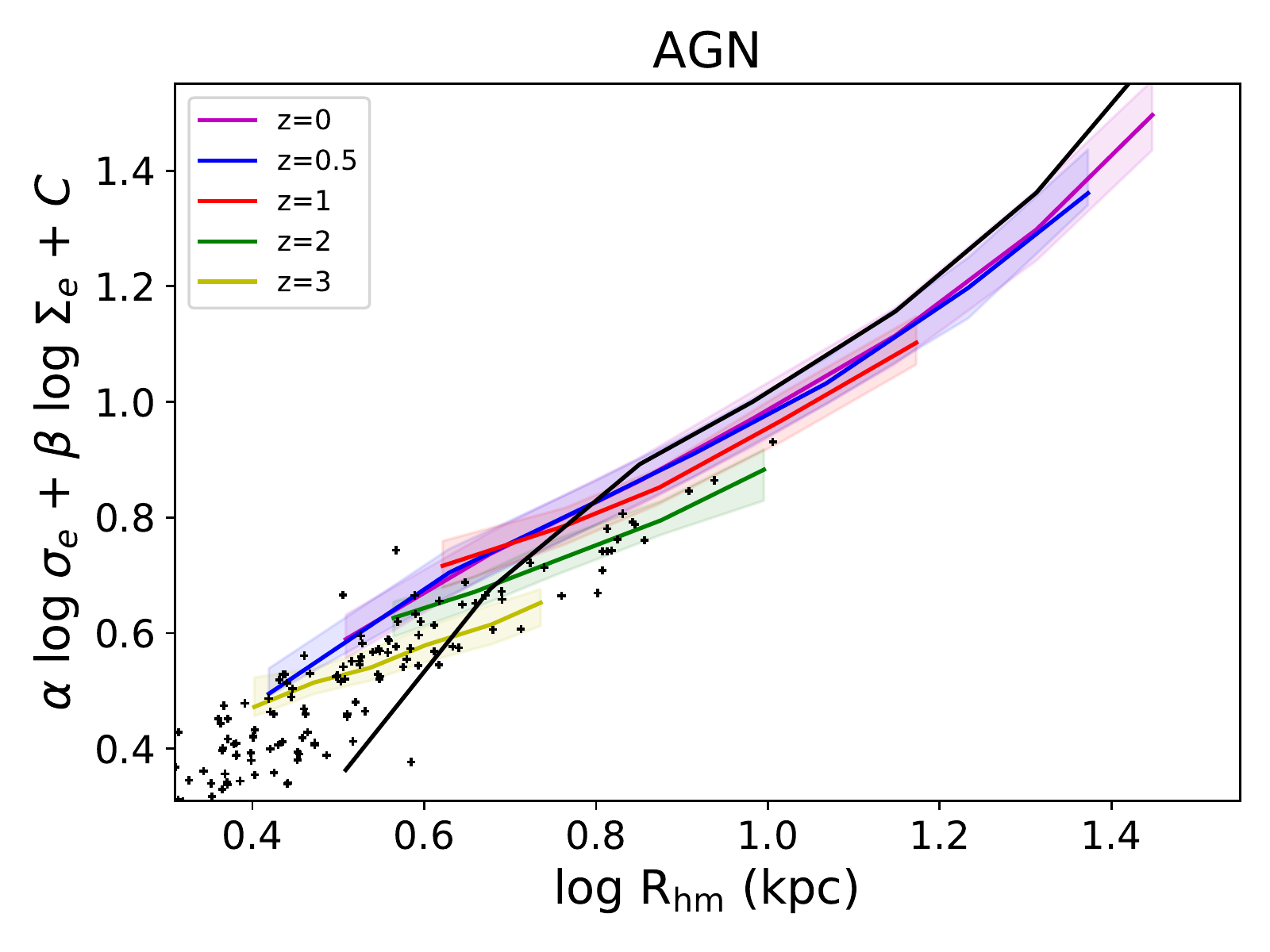}
\includegraphics[width=0.45\textwidth]{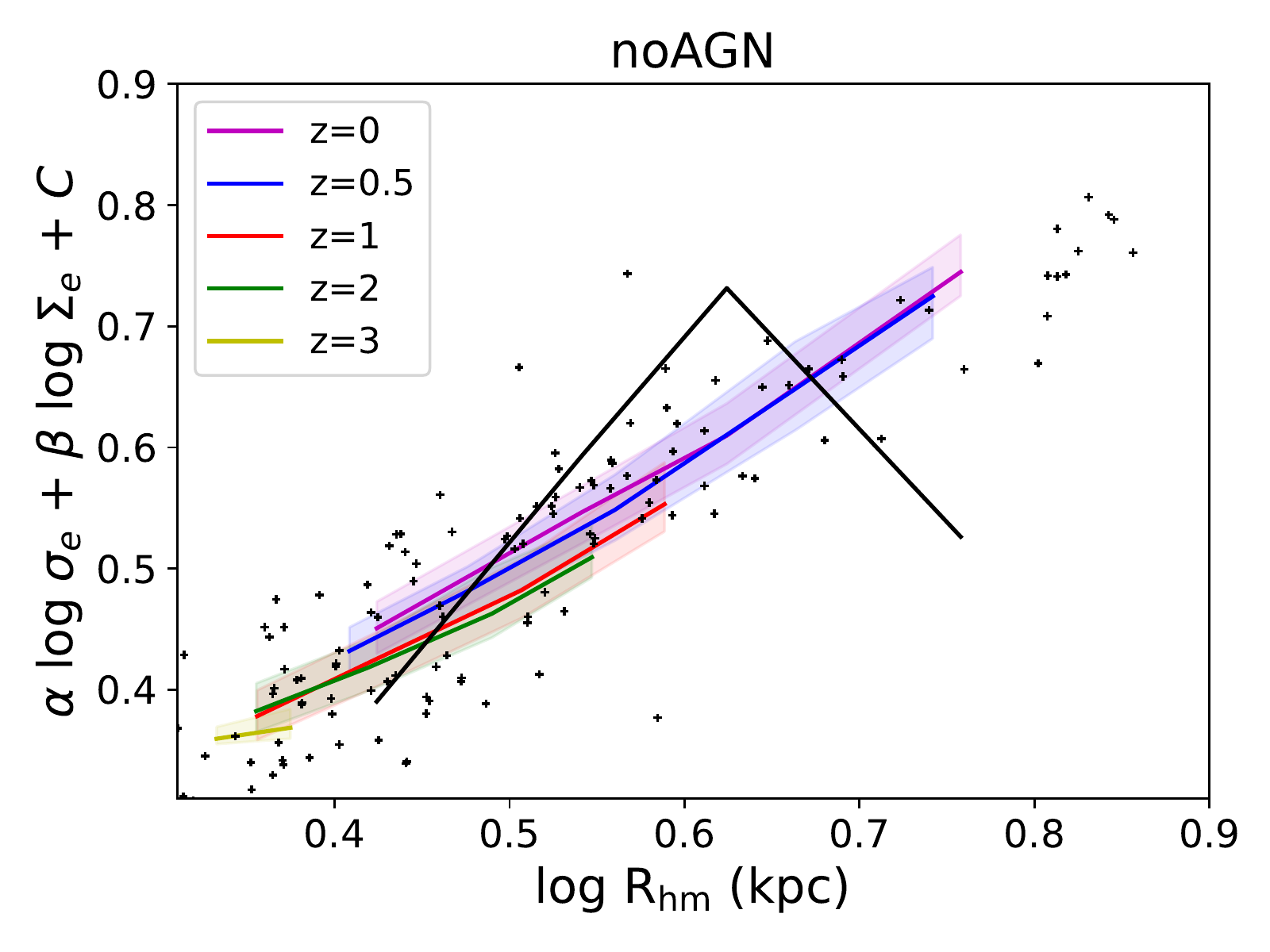}
  \caption{The FP as a function of redshift for central ETGs from Horizon-AGN (left panel) and Horizon-noAGN (right panel).
  The magenta, blue, red, green and yellow lines depict the simulated relations at $z =0$, $z = 0.5$, $z= 1$, $z = 2$ and $z = 3$, respectively. The regions between the 25 and 75 percentiles are shaded. We include the virial relation for galaxies at $z = 0$ (black solid line). For comparison, we also include observational results from ATLAS$^{\rm 3D}$  in black crosses.}
  \label{fig:fp_evol}
\end{figure*}

From Fig.~\ref{fig:fp_evol} it can be appreciated that the simulated  FPs are well-defined. For both runs, the planes show  low scatter in these regressions at all analysed redshifts.  Table~\ref{table:tableFP} summarizes the parameters obtained for both Horizon simulations.

For the Horizon-AGN sample (left panel of Fig.~\ref{fig:fp_evol}), the  FPs are slightly tilted from the virial relation for galaxies at $z = 0$ (see Table~\ref{table:tableFP} and Fig.~\ref{fig:fp_evol}).
This is consistent with many observational studies, as mentioned in the introduction.
More specifically, the $\beta$ parameter related with $\Sigma_e$ is the most affected one at $z=0$, while the $\alpha$ parameter, related with $\sigma_e$ is closer to the expected theoretical value. 
Regarding $\beta$, we note that $|\beta| \le |\beta_{\rm obs}|$. This can be related to the fact that $M_{\rm JAM}$ is a good approximation of the total stellar mass \citep{CappellariAtlasXX}, but there may be other  factors to take into account, such as variations of the IMF. Again, the zero-point of the virial relation was set \textit{ad-hoc}. 

On the other hand, the FP obtained from Horizon-noAGN galaxies by using the virial parameters  at $z = 0$ (right panel of Fig.~\ref{fig:fp_evol}) presents a remarkable break with respect to the slopes of the simulated FP at all the analysed redshifts. This means that, despite the fact that we can fit a plane at different redshifts with low scatter, in the absence of AGN feedback the observed and the theoretical relations cannot be reproduced, especially at radii larger than $ 4$ kpc (log $R_{\rm hm} = 0.6$). It is clear that without the regulation of the stellar mass and size by AGN feedback, the FP plane departures largely from the expectation of the virial theorem.

Another approach to assess the FP and its evolution is to use the
observed parameters at $z=0$ as reference values and estimate the changes with respect to them.
Following \citet{Rosito2018,Rosito2019a}, we estimate the FP  for the observational data of \citet{Cappellari2013}. The $M_{\rm JAM}$ is used to compute $\Sigma_e$ at $z=0$.  
Then, the observed parameters are applied to estimate the FP (Eq.~(\ref{eq:fp_obs})) as a function of redshift. The constant term is set \textit{ad-hoc} to better compare with observations.

\begin{equation}
\label{eq:fp_obs}
 \mathrm{FP}=\alpha_{\rm obs} \log{\sigma_e} + \beta_{\rm obs} \log{\Sigma_e} + C
\end{equation}

\begin{figure*}
  \centering
\includegraphics[width=0.45\textwidth]{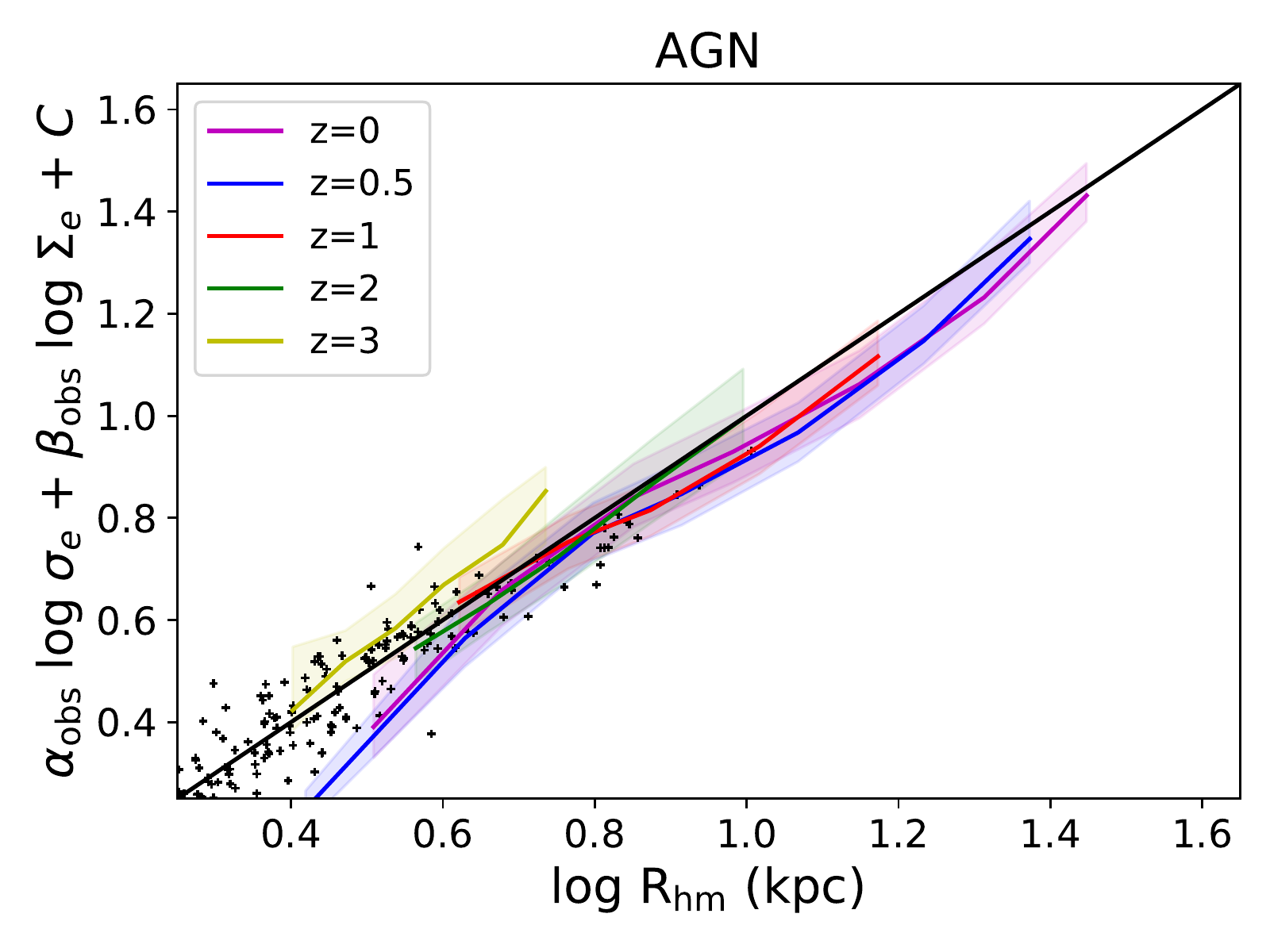}
\includegraphics[width=0.45\textwidth]{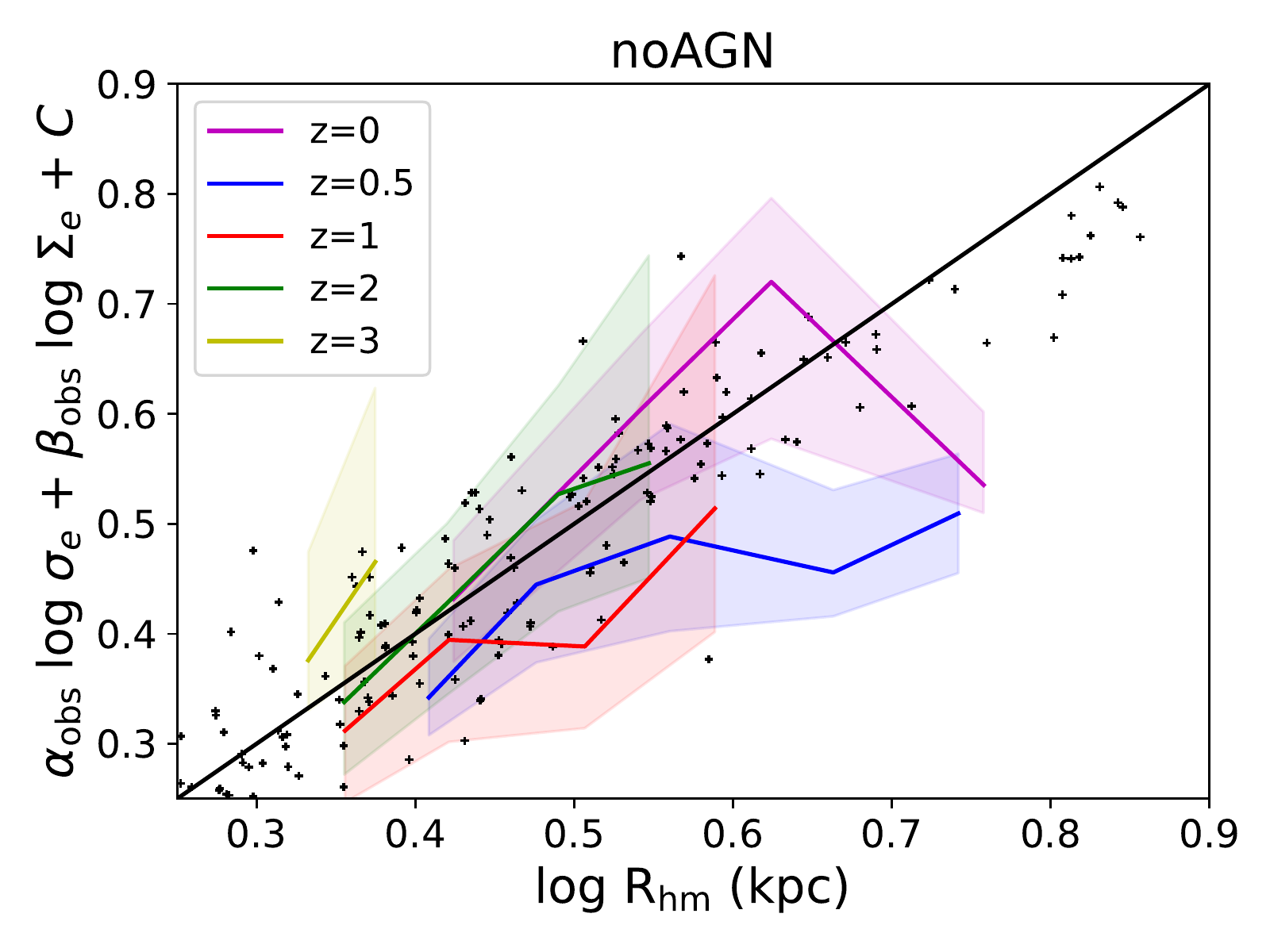}\\
\includegraphics[width=0.45\textwidth]{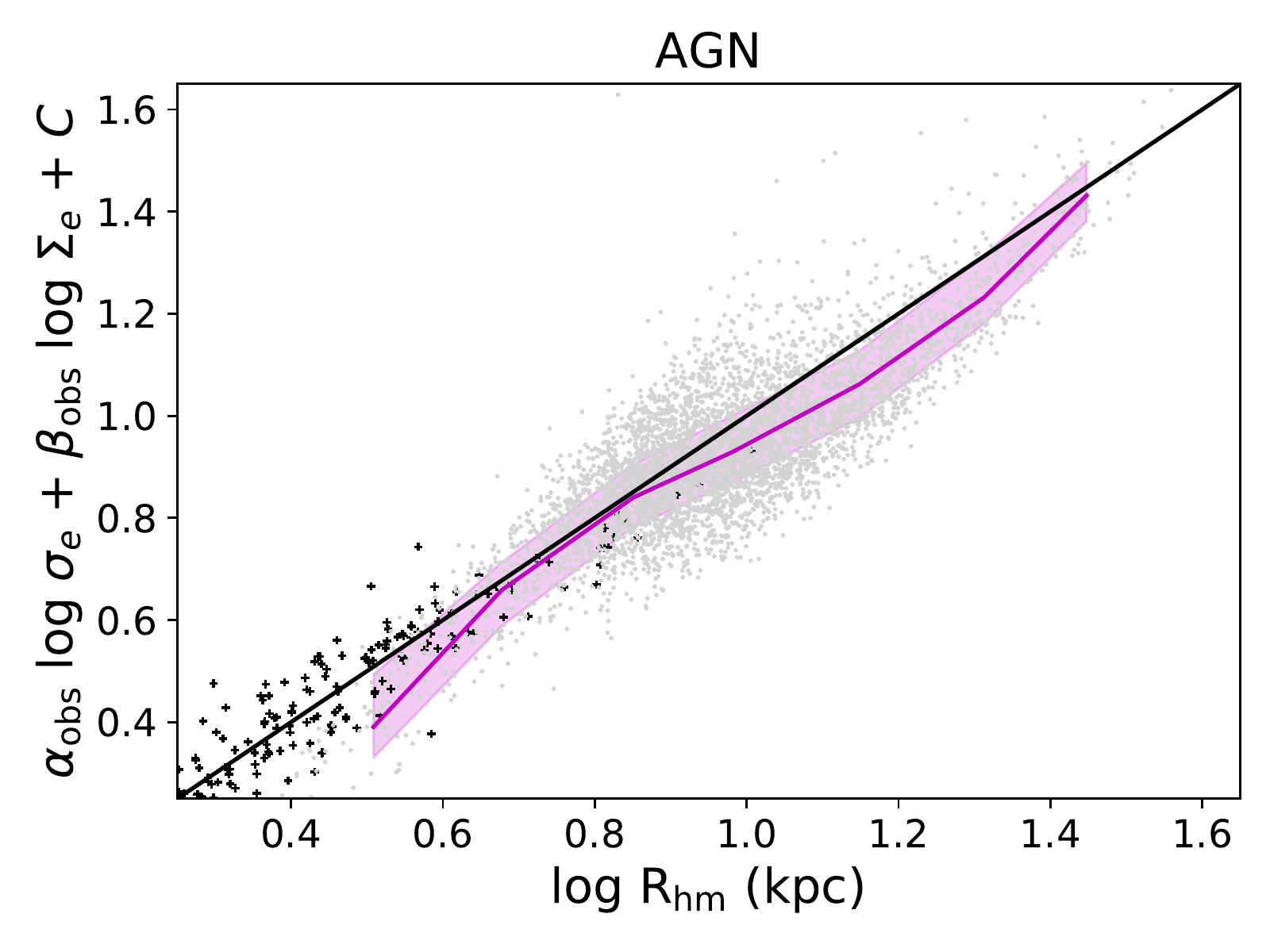}
\includegraphics[width=0.45\textwidth]{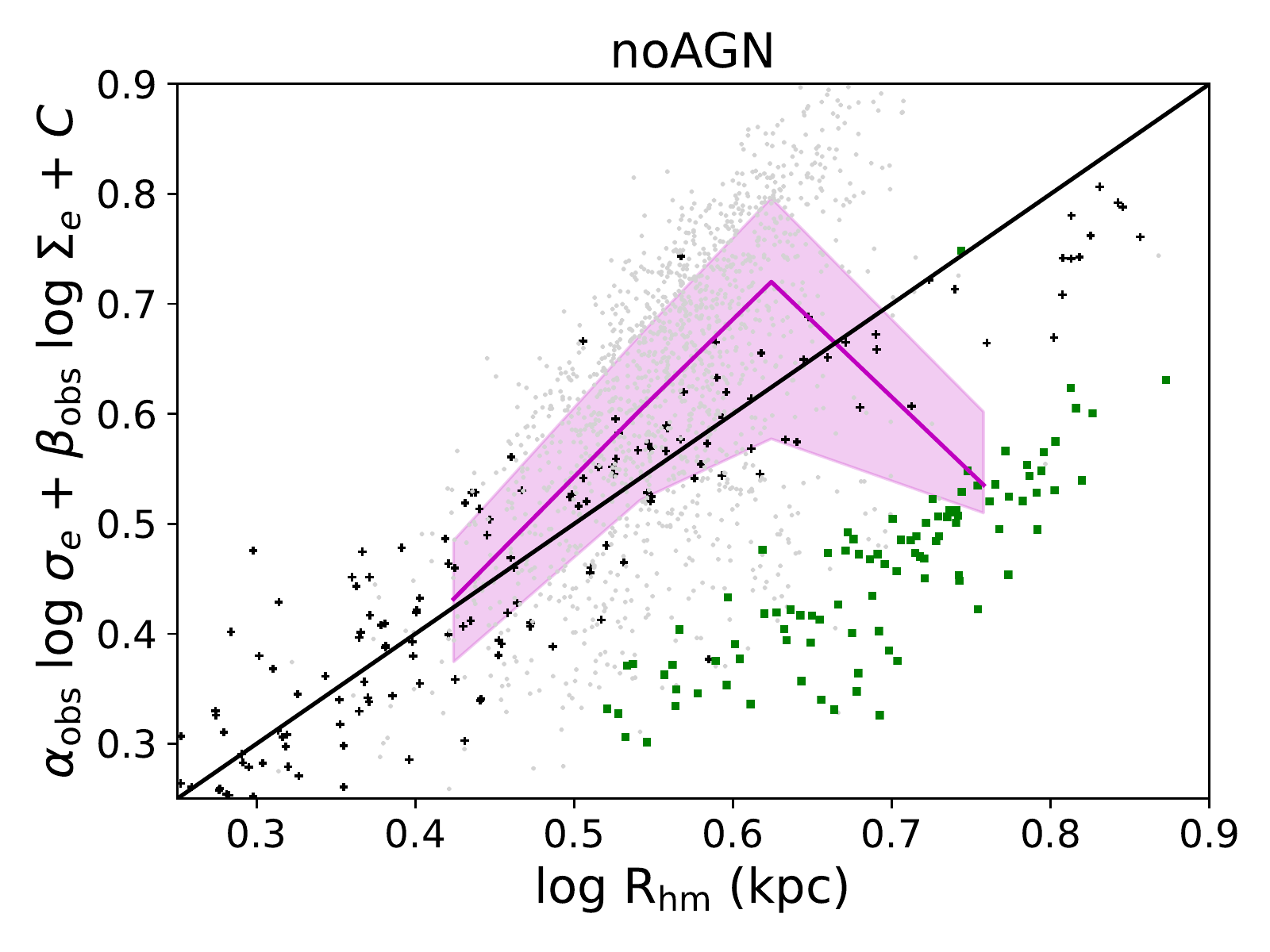}
  \caption{Upper panels: The FP for Horizon-AGN (left panel) and Horizon-noAGN (right panel) central ETGs obtained with the parameters calculated with data from \cite{Cappellari2013} estimated at $z = 0$ (magenta lines), $z = 0.5$ (blue lines), $z= 1$(red lines), $z = 2$ (green lines) and $z = 3$ (yellow lines).  The regions between the 25 and 75 percentiles are shaded. For comparison, we also include the observational results from ATLAS$^{\rm 3D}$(black crosses). Lower panels: Similar relations for $z=0$ including the individual galaxies.
  The scatter plot of the simulated galaxies is depicted in gray circles in both panels. Horizon-noAGN central ETGs with average surface density greater than $10^{9.5}$ M$_{\odot}$ kpc$^{-2}$  are highlighted (green squares). There are not such dense galaxies in Horizon-AGN sample.
  The 1:1 line is  depicted in black.}
  \label{fig:fp_evol_Cappe}
\end{figure*}

Hence, we obtained the expected $R_{\rm hm}$  in case of no evolution of the FP since $z=0$. These values are compared to the simulated $R_{\rm hm}$ at a given $z$.
The results are shown in the upper panels of  Fig.~\ref{fig:fp_evol_Cappe}. From this figure, we can see a consistency of the simulated FP for the Horizon-AGN sample with these parameters. However, in the absence of AGN feedback, the FP departures from the observed relation at high radii. 
This break from the FP in the Horizon-noAGN ETGs is produced by an excess of stellar mass at given $R_{\rm hm}$, which produces the high $\Sigma_e$ as previously mentioned.
This sub-population of compact galaxies starts to be noticeable for  $z < \sim 1$. 
To illustrate this, in the lower panels of Fig.~\ref{fig:fp_evol_Cappe} we show the relation for $z=0$ together with the individual simulated galaxies. It can be appreciated that the deviation from the observed FP at large radii in Horizon-noAGN is caused by galaxies with high average surface density ($\Sigma_{e}>10^{9.5}$ M$_{\odot}$ kpc$^{-2}$) at a given $R_{\rm hm}$ (right, lower panel).
Hence, AGN feedback plays an important role in regulating the SF and prevents the formation of very dense galaxies, which otherwise depart from the FP.  This trend is in agreement with \cite{Peirani2019} who already reported  higher values of $\Sigma_e$ for galaxies formed in the  absence of AGN in the Horizon simulation.
These galaxies clearly do not follow the FP, however, they are still consistent with the mass--plane.
For comparison, we include the scatter plot for Horizon-AGN at $z = 0$, where we do not find such dense galaxies.

\begin{figure}
  \centering
\includegraphics[width=0.4\textwidth]{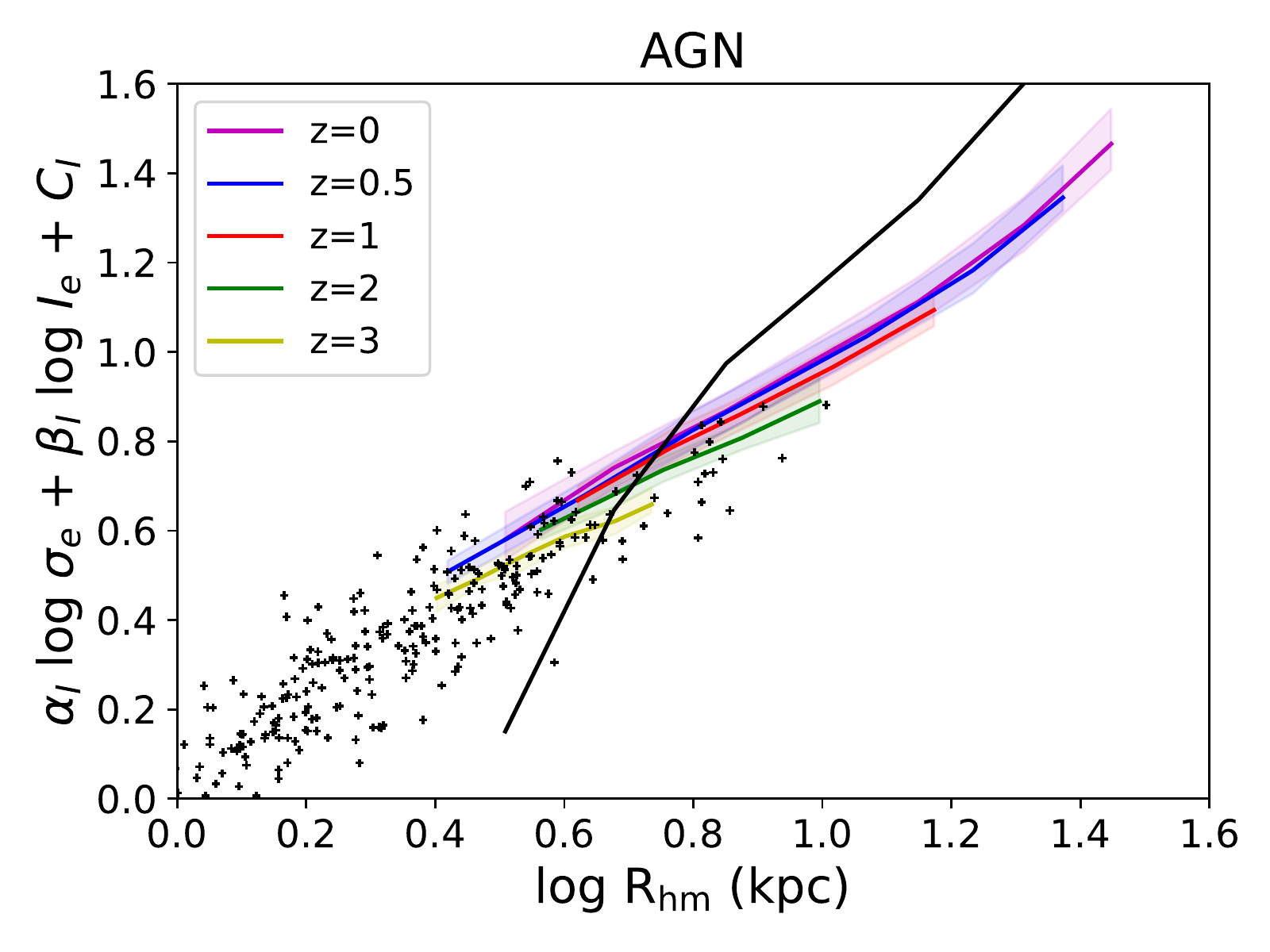}
  \caption{The L-FP for Horizon-AGN central ETGs sample computing the average surface brightness for both observations \citep{Cappellari2013} and simulation, using the luminosity in the r-band. }
  \label{fig:l_fp}
\end{figure}

Finally, in order to analyse the Luminosity FP (i.e. the FP obtained using $I_e$ instead of $\Sigma_e$, L-PF\footnote{Hereafter, the subscript I will be used to denote the fitting parameters for the L-PFs.}) we used the luminosities in $r$-band\footnote{Simulated AB magnitudes are obtained by means of SP models from \cite{BruzualCharlot2003} and a Salpeter IMF. The total flux for each frequency is passed through the different SDSS filters, in particular, the $r$-band (see more details in \cite{Chisari2015})} to estimate $I_e$.
In Fig.~\ref{fig:l_fp}, we display the L-FP as a function of redshift for the Horizon-AGN.
For comparison, the expected relation from the virial theorem is also included. As can be seen, the tilt is strong at all redshifts (Table~\ref{table:tableFPlum}) whereas, when using the stellar mass, the deviation from the virial expectations is weaker for $z=0$ (Fig.~\ref{fig:fp_evol}). 
Our results {suggest} that an important factor causing the tilt in the L-FP  is  the variation of the $M/L$ from galaxy to galaxy and as a function of redshift. We will further address this point in the Discussion.

\begin{table*}
\small
\caption{Evolution of FP parameters in Horizon simulations using $\Sigma_e$. The parameters $\alpha$ and $\beta$ are given by  Eq.(\ref{eq:fp_def}). The errors are estimated by applying a bootstrap technique. The linear regression scatter is included ($\sigma$). } 
\label{table:tableFP}   
\centering              
\begin{tabular}{cccccccccc}   
\hline\hline   
 Redshift & \multicolumn{4}{c}{Horizon-AGN} & \multicolumn{4}{c}{Horizon-noAGN} \\
   &  $\alpha$ &   $\beta$ &  $C$  & $\sigma$ & $\alpha$ &  $\beta$ & $C$ & $\sigma$ \\
  \hline                   
0 & 1.700 $\pm$ 0.017 & -0.542 $\pm$ 0.007 & 1.345 $\pm$ 0.040 & 0.066 & 1.317 $\pm$ 0.027 & -0.440 $\pm$ 0.010 & 1.251 $\pm$ 0.027 & 0.033 \\
0.5 & 1.471 $\pm$ 0.017 & -0.475 $\pm$ 0.009 & 1.331 $\pm$ 0.047 & 0.062 & 1.069 $\pm$ 0.030 & -0.355 $\pm$ 0.014 & 1.101 $\pm$ 0.051 & 0.035 \\
1 & 1.360 $\pm$ 0.024 & -0.468 $\pm$ 0.012 & 1.490 $\pm$ 0.057 & 0.064 & 0.921 $\pm$ 0.041 & -0.299 $\pm$ 0.017 & 0.884 $\pm$ 0.060 & 0.045 \\
2 & 0.990 $\pm$ 0.033 & -0.375 $\pm$ 0.013 & 1.503 $\pm$ 0.068 & 0.062 & 0.726 $\pm$ 0.051 & -0.261 $\pm$ 0.022 & 0.999 $\pm$ 0.104 & 0.048 \\
3 & 0.695 $\pm$ 0.051 & -0.366 $\pm$ 0.020 & 2.001 $\pm$ 0.152 & 0.061 & 0.168 $\pm$ 0.134 & -0.084 $\pm$ 0.050 & 0.714 $\pm$ 0.192 & 0.047 \\
  \hline
\end{tabular} \\
\end{table*}

\begin{table*}
\small
\caption{Evolution of L-FP parameters for Horizon-AGN central ETGs and for the subsample obtained with the restriction $V/\sigma< 0.3$. The parameters $\alpha_I$ and $\beta_I$ are given by  Eq.(\ref{eq:fp_def}). The errors are estimated by applying a bootstrap technique.  The linear regression scatter is included ($\sigma$). } 
\label{table:tableFPlum}   
\centering              
\begin{tabular}{cccccccccc}   
\hline\hline   
 Redshift & \multicolumn{4}{c}{Horizon-AGN ETGs} & \multicolumn{4}{c}{Horizon-AGN ETGs with $V/\sigma < 0.3$} \\
   &  $\alpha_I$ &   $\beta_I$ &  $C_I$  & $\sigma$ & $\alpha_I$ &  $\beta_I$ & $C_I$ & $\sigma$ \\
  \hline                        
0 & 1.388 $\pm$ 0.013 & -0.379 $\pm$ 0.004 & 0.605 $\pm$ 0.033 & 0.067 & 1.313 $\pm$ 0.018 & -0.363 $\pm$ 0.008 & 0.658 $\pm$ 0.048 & 0.067 \\
0.5 & 1.181 $\pm$ 0.011 & -0.324 $\pm$ 0.006 & 0.720 $\pm$ 0.037 & 0.063 & 1.194 $\pm$ 0.016 & -0.321 $\pm$ 0.008 & 0.666 $\pm$ 0.056 & 0.063 \\
1 & 1.108 $\pm$ 0.015 & -0.341 $\pm$ 0.007 & 1.037 $\pm$ 0.047 & 0.062 & 1.188 $\pm$ 0.021 & -0.372 $\pm$ 0.011 & 1.098 $\pm$ 0.069 &  0.065 \\
2 & 1.028 $\pm$ 0.028 & -0.412 $\pm$ 0.013 & 1.813 $\pm$ 0.072 & 0.056  & 1.100 $\pm$ 0.042 & -0.420 $\pm$ 0.019 & 1.710 $\pm$ 0.106 &  0.055 \\
3 & 0.796 $\pm$ 0.056 & -0.373 $\pm$ 0.018 & 1.981 $\pm$ 0.140 & 0.053 & 0.804 $\pm$ 0.103 & -0.381 $\pm$ 0.036 & 2.036 $\pm$ 0.246 & 0.059 \\
  \hline
\end{tabular} \\
\end{table*}

\section{Discussion on the FP}
\label{sec:disc}

In the previous section, we have shown that the impact of AGN feedback on ETGs allows the FP  to be reproduced in close agreement with observations. Here we explore in more detail the correlation between different galaxy properties and the FP as a function of redshift with the aim at discovering how AGN feedback works at different mass scales and redshifts.

\subsection{sSFR, age and morphology dependence}
\label{disc1}

\begin{figure*}
  \centering
\includegraphics[width=0.45\textwidth]{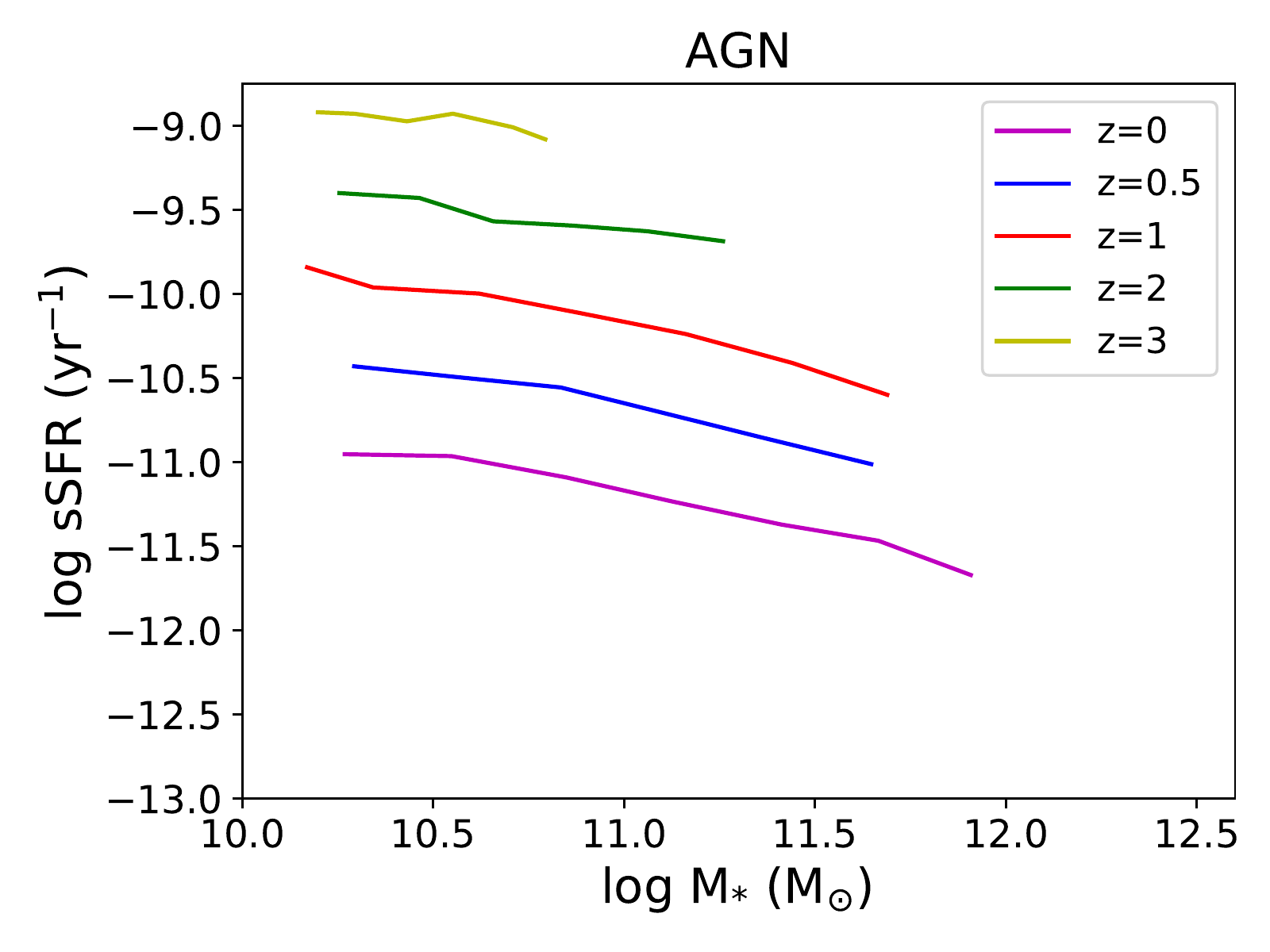}
\includegraphics[width=0.45\textwidth]{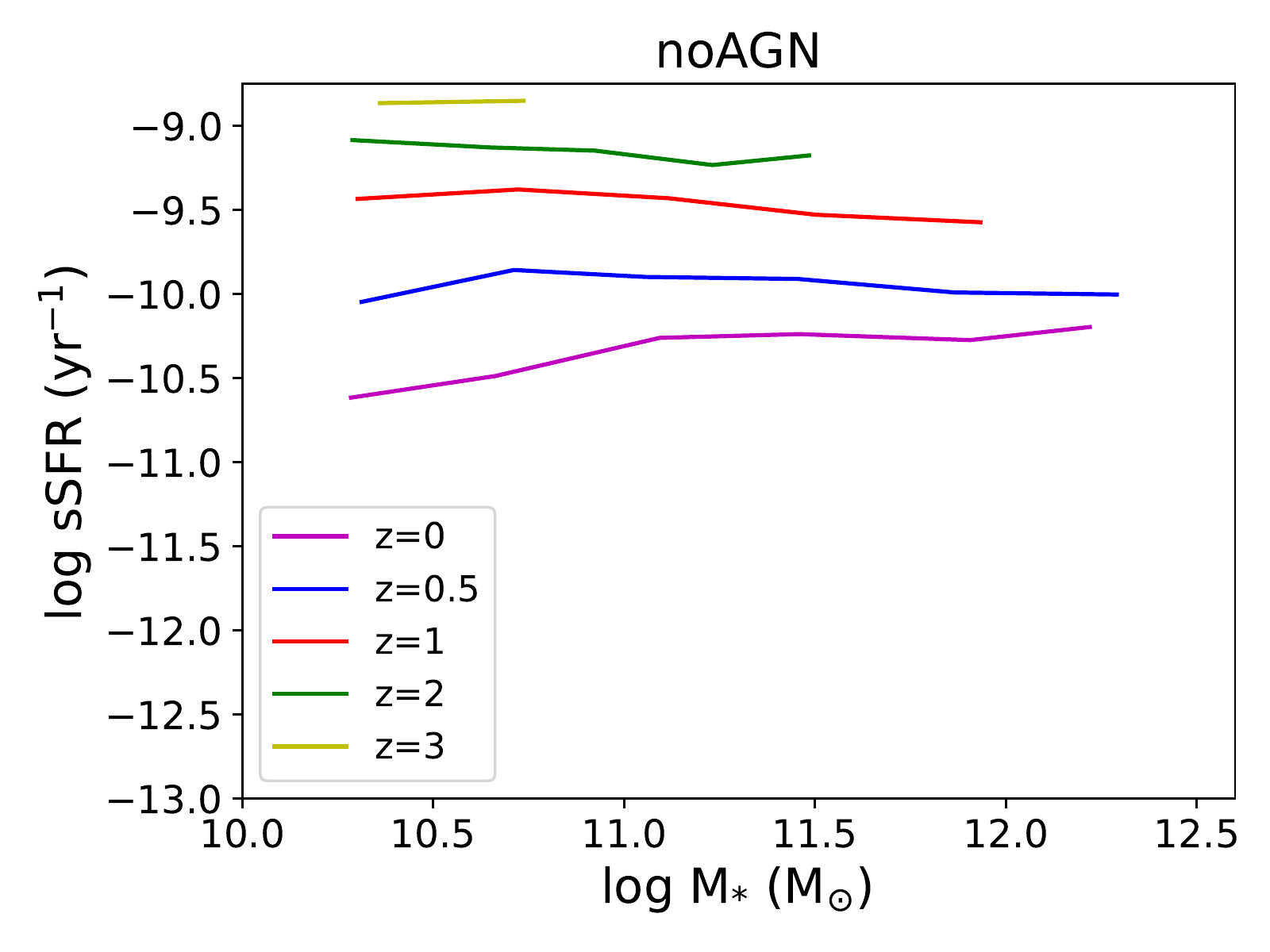}
  \caption{sSFR as a function of stellar mass for Horizon-AGN (left panel) and Horizon-noAGN (right panel) central ETGs at different redshifts. At $z = 1$ there is a change in the slope for the Horizon-AGN samples while in the absence of AGN feedback, the sSFR does not depend on stellar mass. }
   \label{fig:evol_sf}
\end{figure*}

One of the main roles of AGN feedback is the regulation of the SF activity. 
This can be clearly appreciated from Fig.~\ref{fig:evol_sf} where the median sSFR as a function of the stellar mass for the different analysed redshifts are shown.  For the noAGN run, the sSFR decreases with decreasing redshift as expected. However, it remains approximately constant as a function of stellar mass (right panel). The activation of the AGN feedback modulates the SF activity across time with the larger impact at  $z\sim 1$ from where it starts to be quenched, principally for high-mass galaxies \citep[see also][]{Beckmann2017}. The overall effects are a change in the slope of this relation and a larger suppression  of the SF activity than in the noAGN run for lower redsfhits. Both of them contribute to  better reproduce the observed trend.

\begin{figure*}
  \centering
\includegraphics[width=\textwidth, height=0.95\textheight]{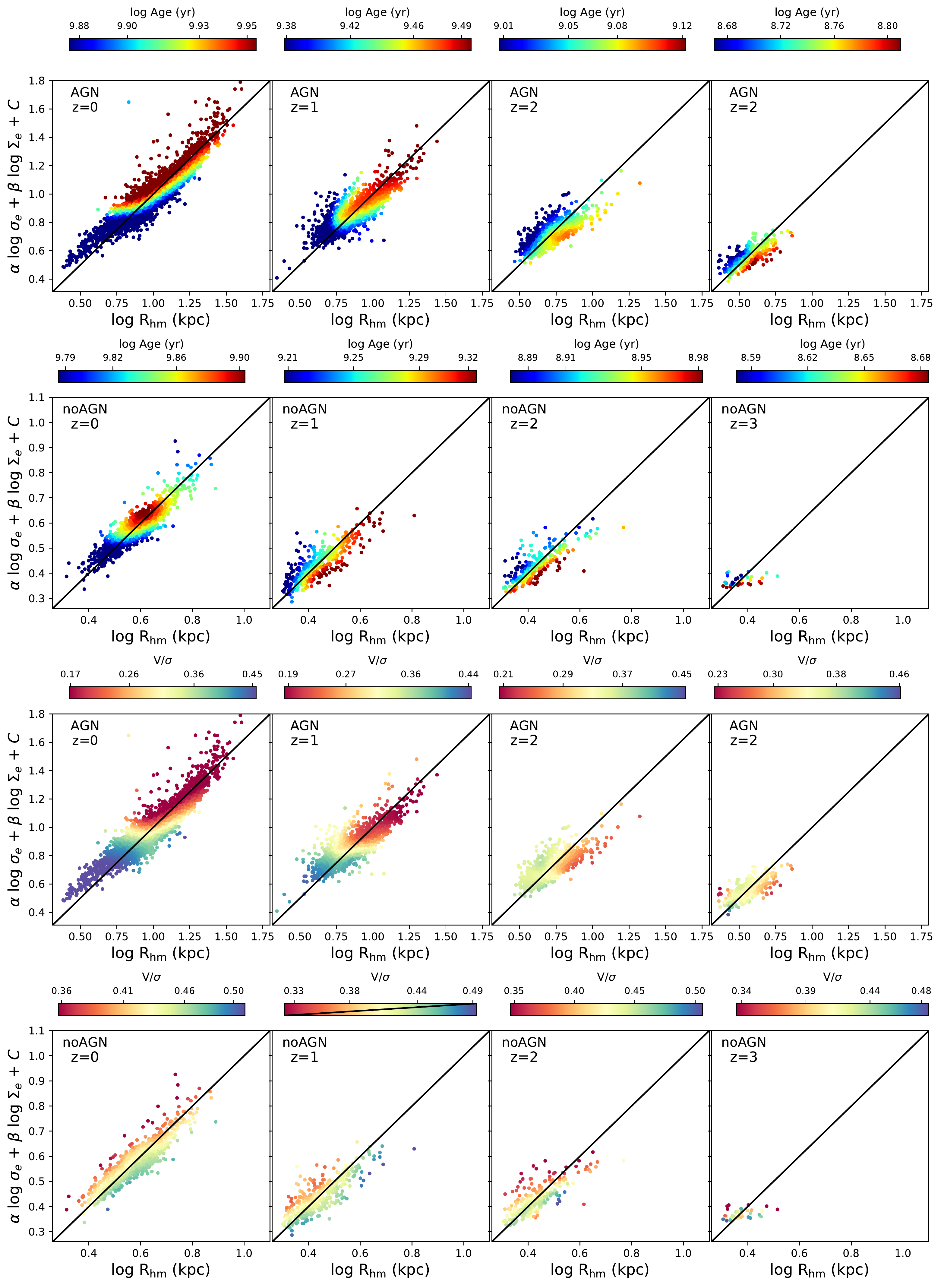}
  \caption{The FP for Horizon-AGN and Horizon-noAGN central ETGs fitted at different redshifts. The symbols are coloured according to mass-weighted average stellar age (upper panels) and the stellar $V/\sigma$ (lower panels).
  The 1:1 line is depicted in black.}
  \label{fig:FPcolor}
\end{figure*}

The quenching of the SF activity is reflected on the evolution of mean ages of the SPs. In Fig.~\ref{fig:FPcolor}  we show the FP for both simulations as a function of mass-weighted average stellar age\footnote{Hereafter,  we fix the colour-bar limits to the first and third quartiles of the corresponding values.}.
At $z=0$,  galaxies dominated by old stars lie above the 1:1 line, particularly for large radii. The opposite trend can be observed for higher redshifts ($z = 2$ and $z = 3$), while the transition between these behaviours occurs at  $z \sim 1$ in agreement with the SF trends of Fig.~\ref{fig:evol_sf}.

We also find that  $V/\sigma$ varies across that plane as can be seen from Fig~\ref{fig:FPcolor} (left panels) similar to the behaviour shown by the ages. This suggests a link between  star formation histories and  galaxy morphologies (quantified by $V/\sigma$, see Fig~\ref{fig:DT_vsigma}). Those galaxies which are faster rotators tend to have younger SPs, on average.  At $z\sim 1$, galaxies dominated by velocity dispersion present larger sizes compared to those more rotation dominated. Those with $V/\sigma$ within the range
[0.30-0.35] depict a U-shape in the distribution across the FP at a given galaxy size. In the absence of AGN feedback, galaxies dominated by velocity dispersion are systematically located above the 1:1 relation. 

\begin{figure*}
  \centering
\includegraphics[width=0.3\textwidth]{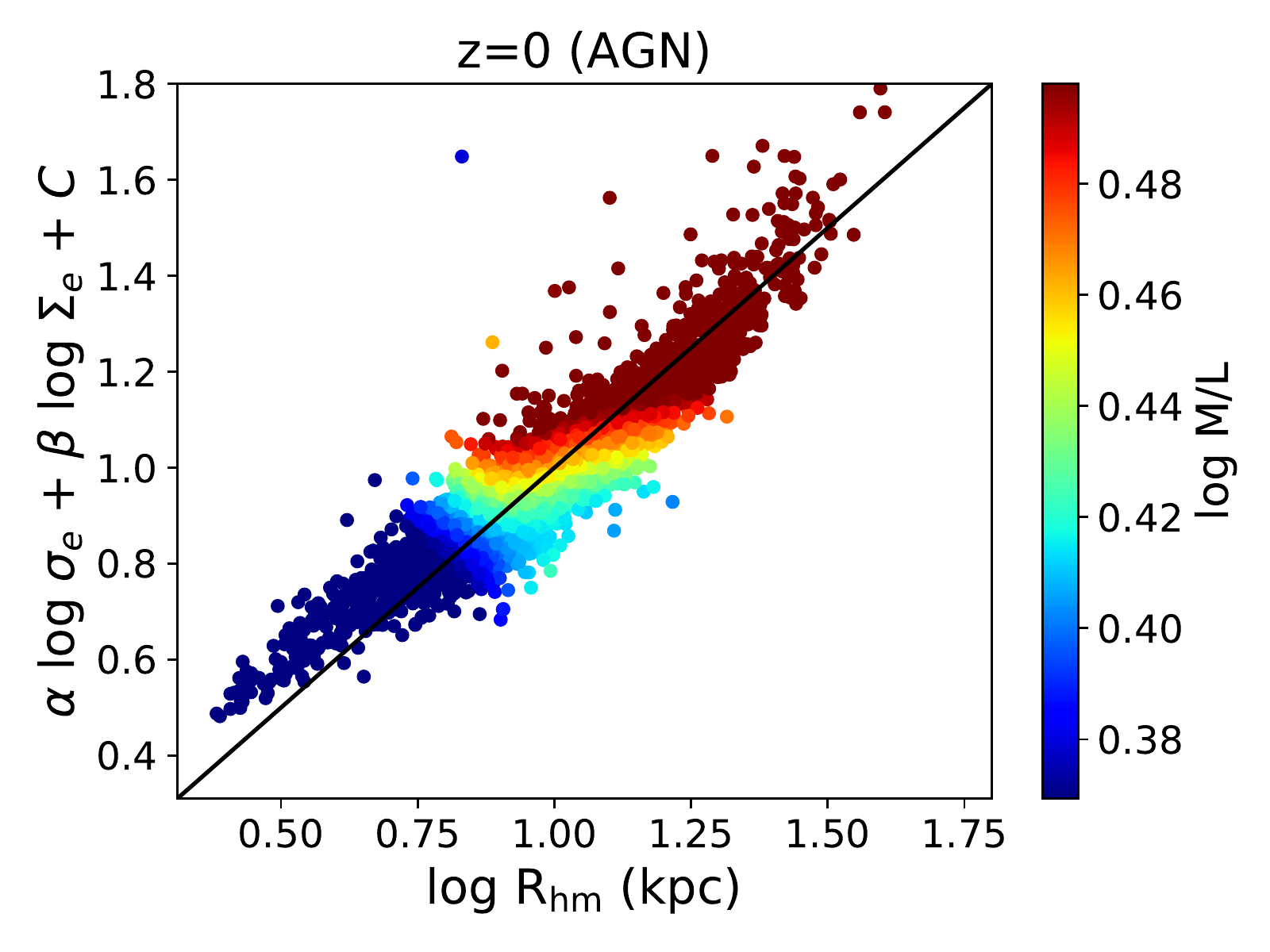} \
\includegraphics[width=0.3\textwidth]{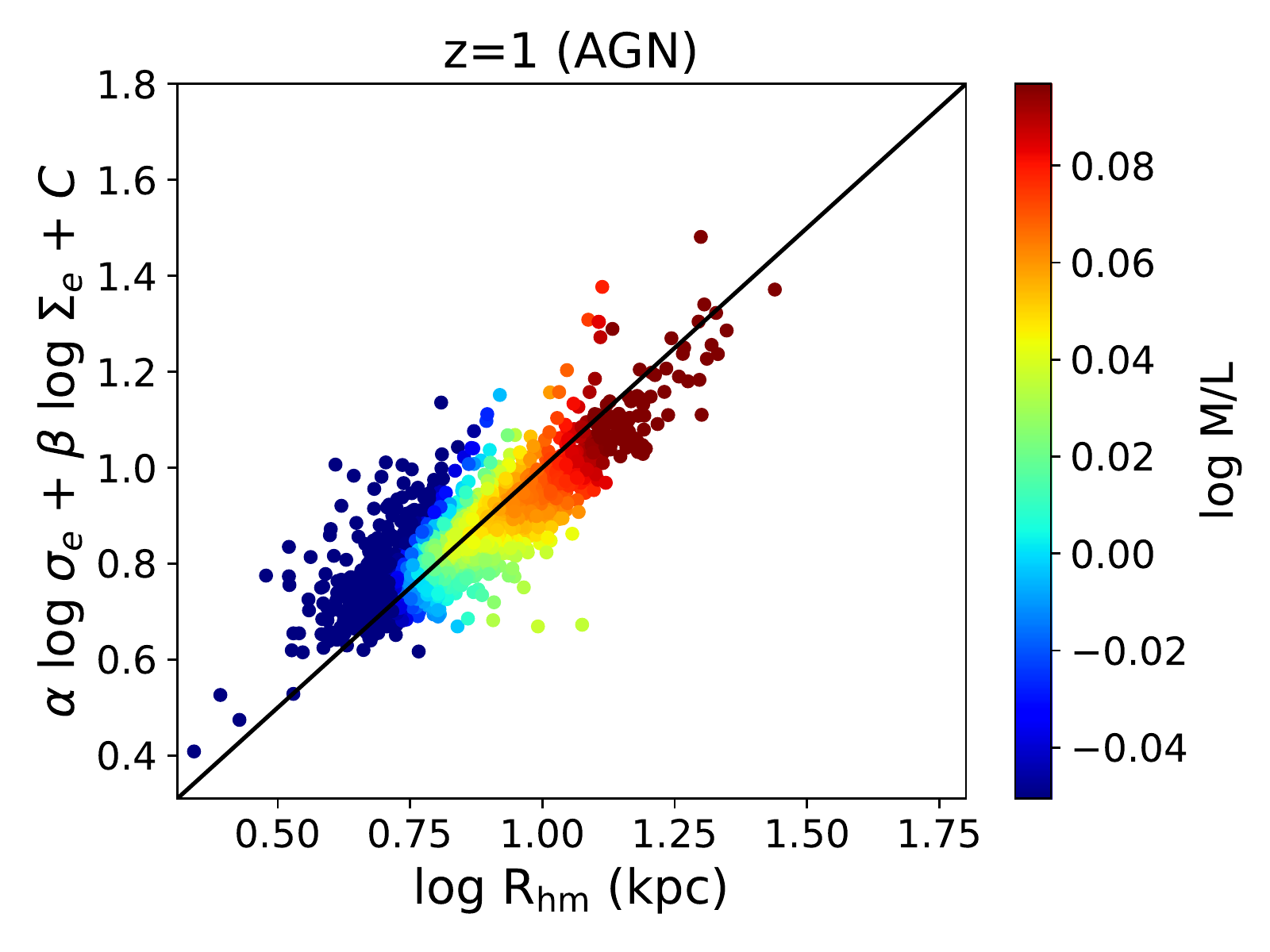}
\includegraphics[width=0.3\textwidth]{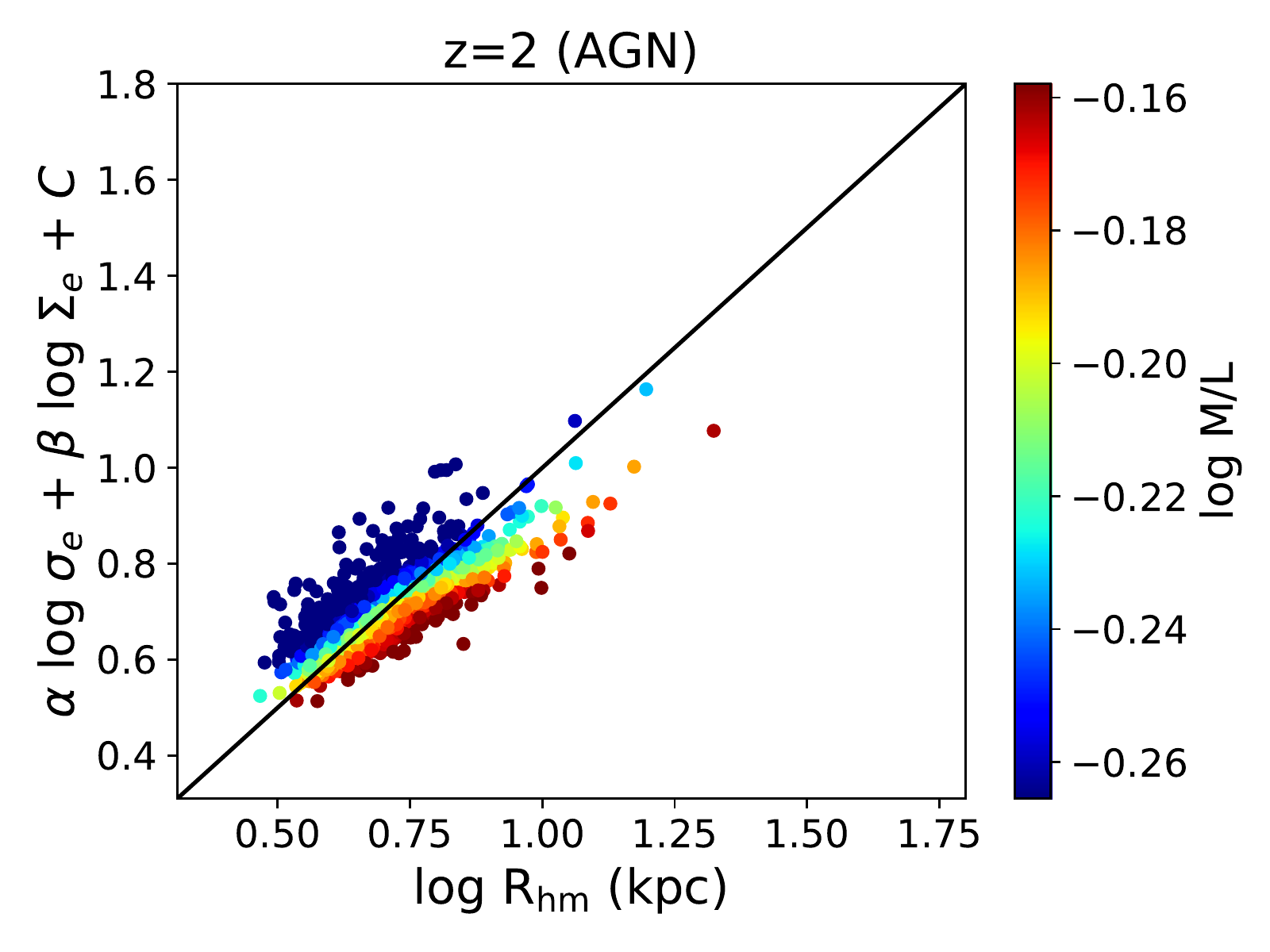}\\
\includegraphics[width=0.3\textwidth]{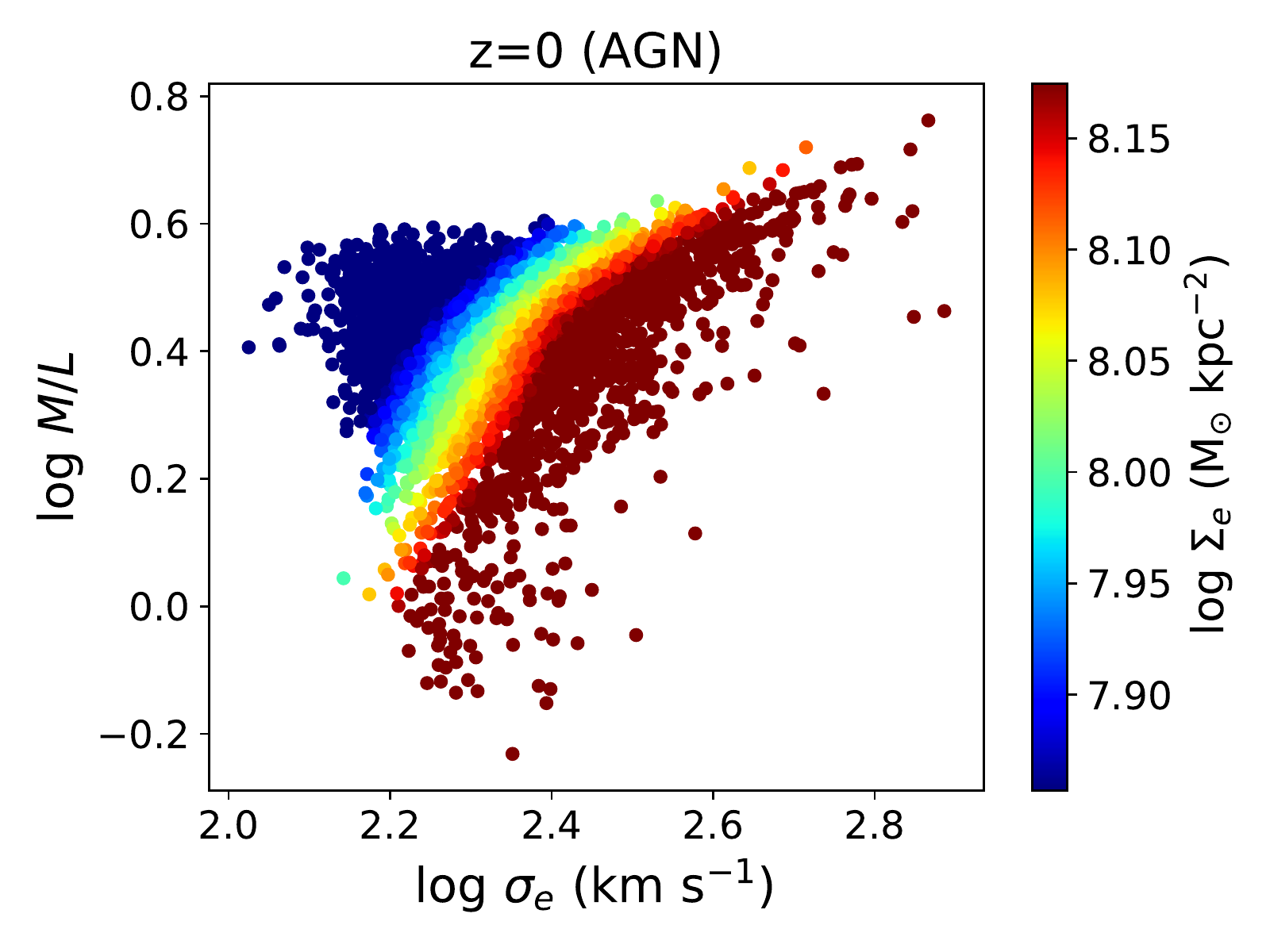} \
\includegraphics[width=0.3\textwidth]{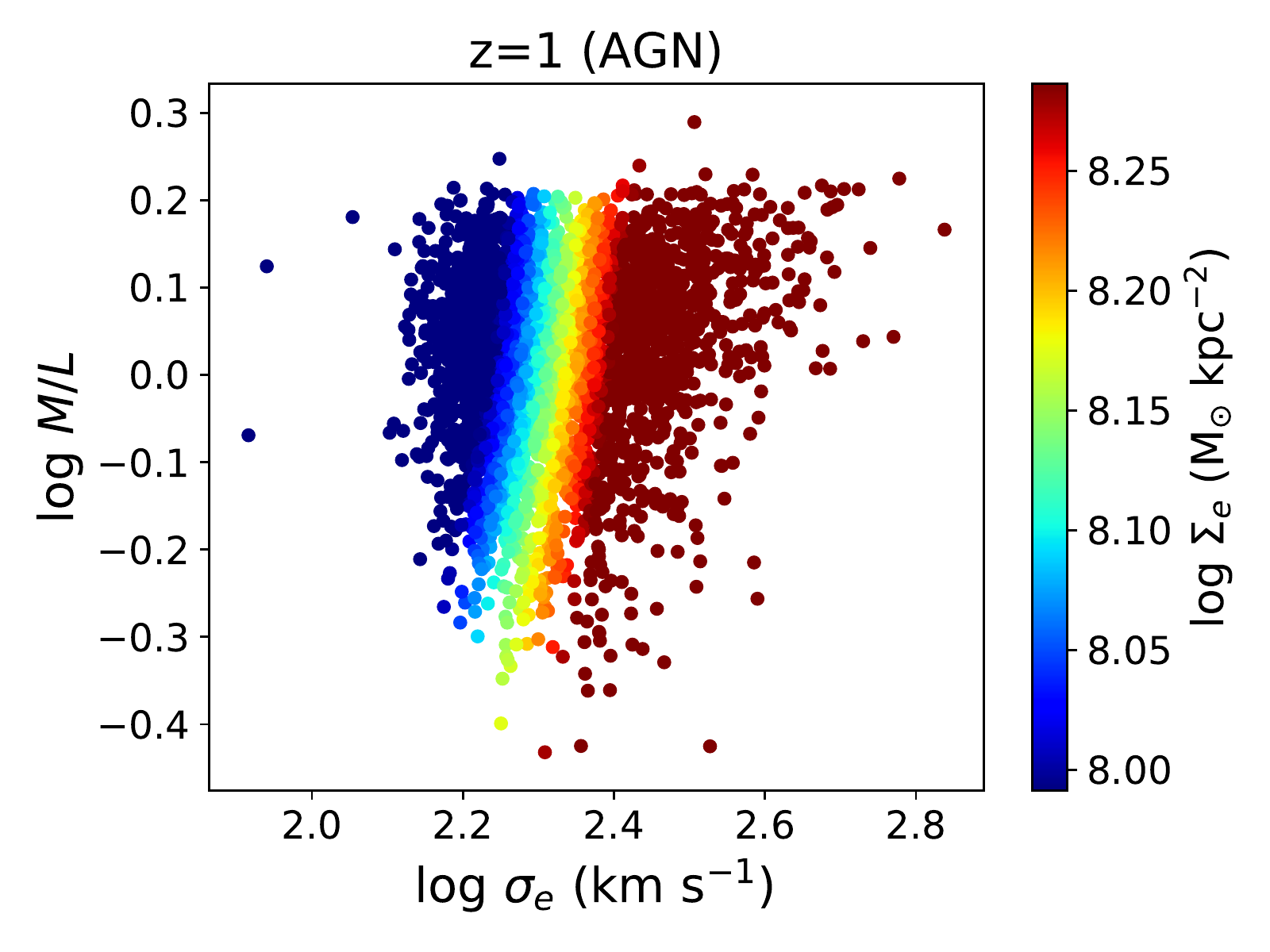}
\includegraphics[width=0.3\textwidth]{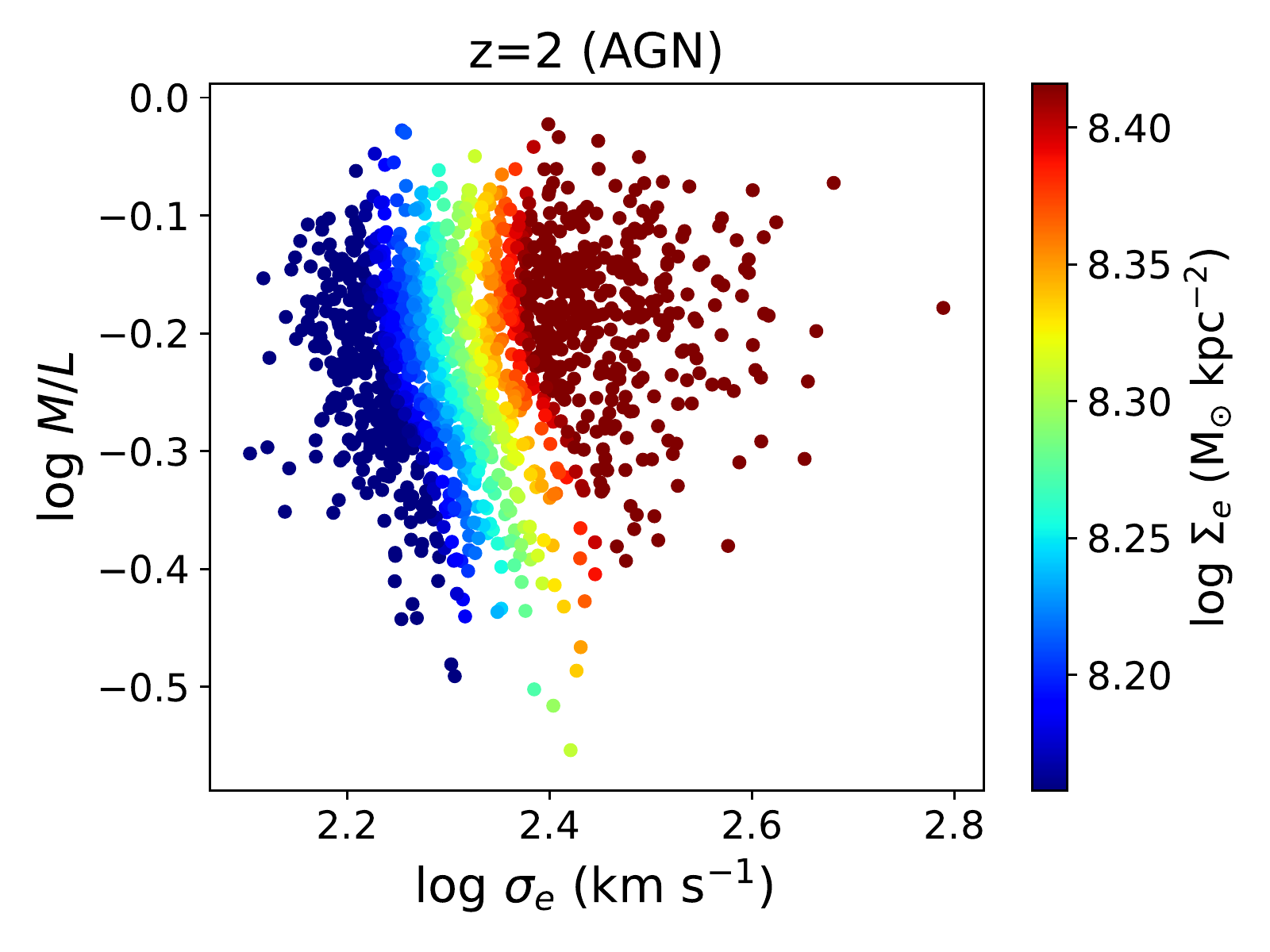}
  
  \caption{Upper panels: The FP for Horizon-AGN central ETGs as a function of $M/L$ at $z=0, 1$ and $z=2$. Symbols are coloured by $M/L$. Lower panels: Relation between $M/L$ ratio and velocity dispersion for the same galaxies. Symbols are coloured by the average surface density $\Sigma_e$.}
  
   \label{fig:fp_ML}
\end{figure*}

\subsection{The mass-to-light ratio variation across the FP}
\label{sec:disc2}

Hereafter, we will only focus on the Horizon-AGN run, which better reproduces the observations.
The regulation of the SF activity by AGN feedback also modulates the luminosities and, consequently, the $M/L$ ratios. As can be seen from the upper panel of Fig.~\ref{fig:fp_ML}, the FP as a function of $M/L$ in the $r$-band shows a clear change from a sequence of galaxies populated by young stars with low $M/L$ and dominated by rotation which populate the region above the 1:1 relation for $z>1$ to  galaxies with older populations and larger $M/L$ at $z \sim 0$.
As previously stated, at $z \sim  1$ the action of AGN feedback can be linked to a quenching of the SF and the progressive aging of the SPs. 

In summary, the action of AGN feedback in the Horizon simulations regulates the SF activity so that it is significantly quenched by $z \sim 1$ and mainly for  massive galaxies. This also has an impact on the $M/L$ ratios which get larger with decreasing redshift as the SPs get older (Fig.~\ref{fig:fp_ML}). However, this is not the only effect. Some galaxies have lower $\Sigma_e$ at similar $M/L$, as shown in lower panel of Fig.~\ref{fig:fp_ML}. Additionally, we also find that $V/\sigma$ varies so that more massive spheroidal-dominated galaxies are able to form. 
Hence, the regulation of the star formation activity by the AGN feedback shapes the $M/L$ ratio for galaxies of intermediate and large $\sigma_e$.

\begin{figure*}
  \centering
\includegraphics[width=0.3\textwidth]{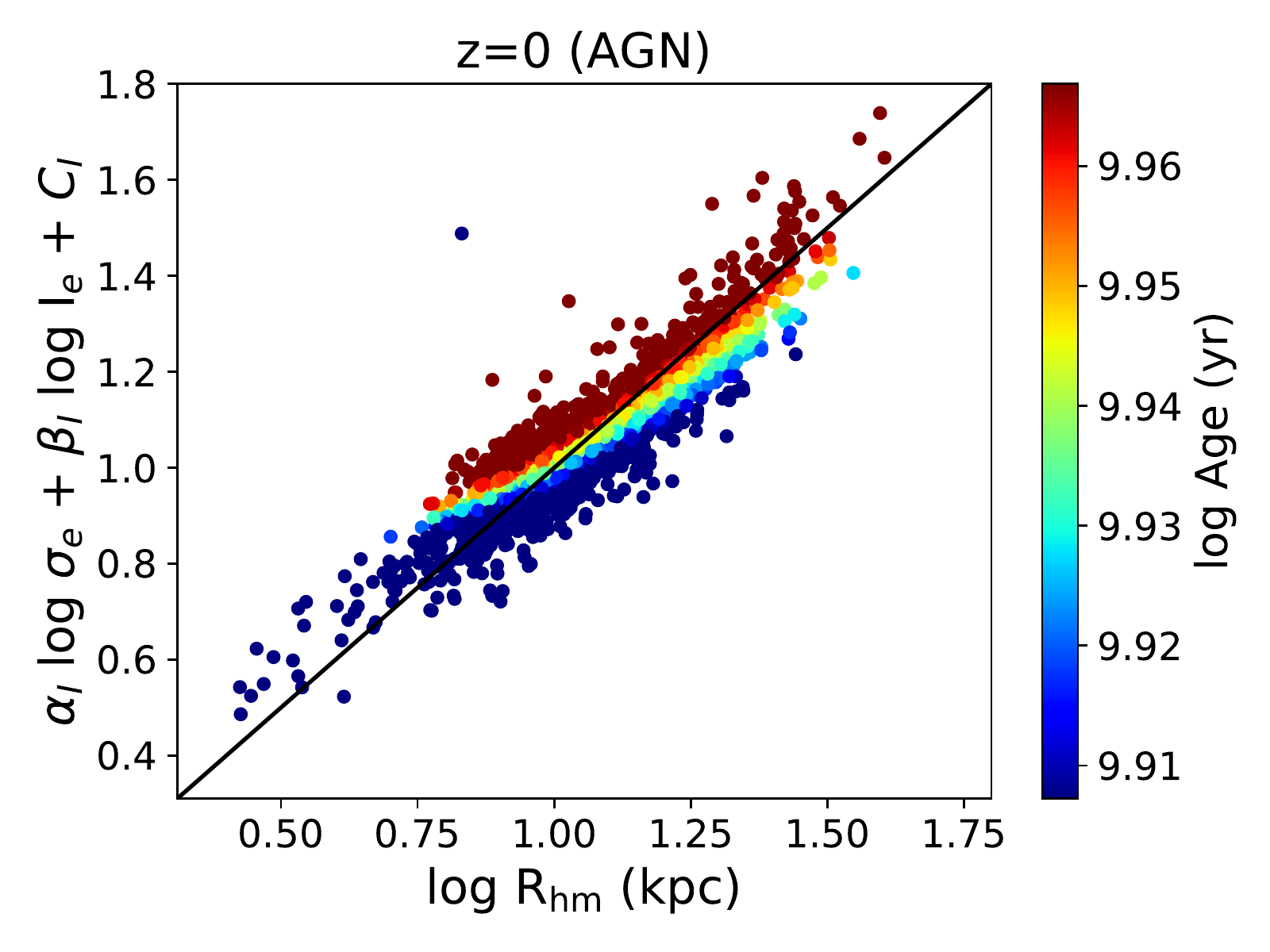}
\includegraphics[width=0.3\textwidth]{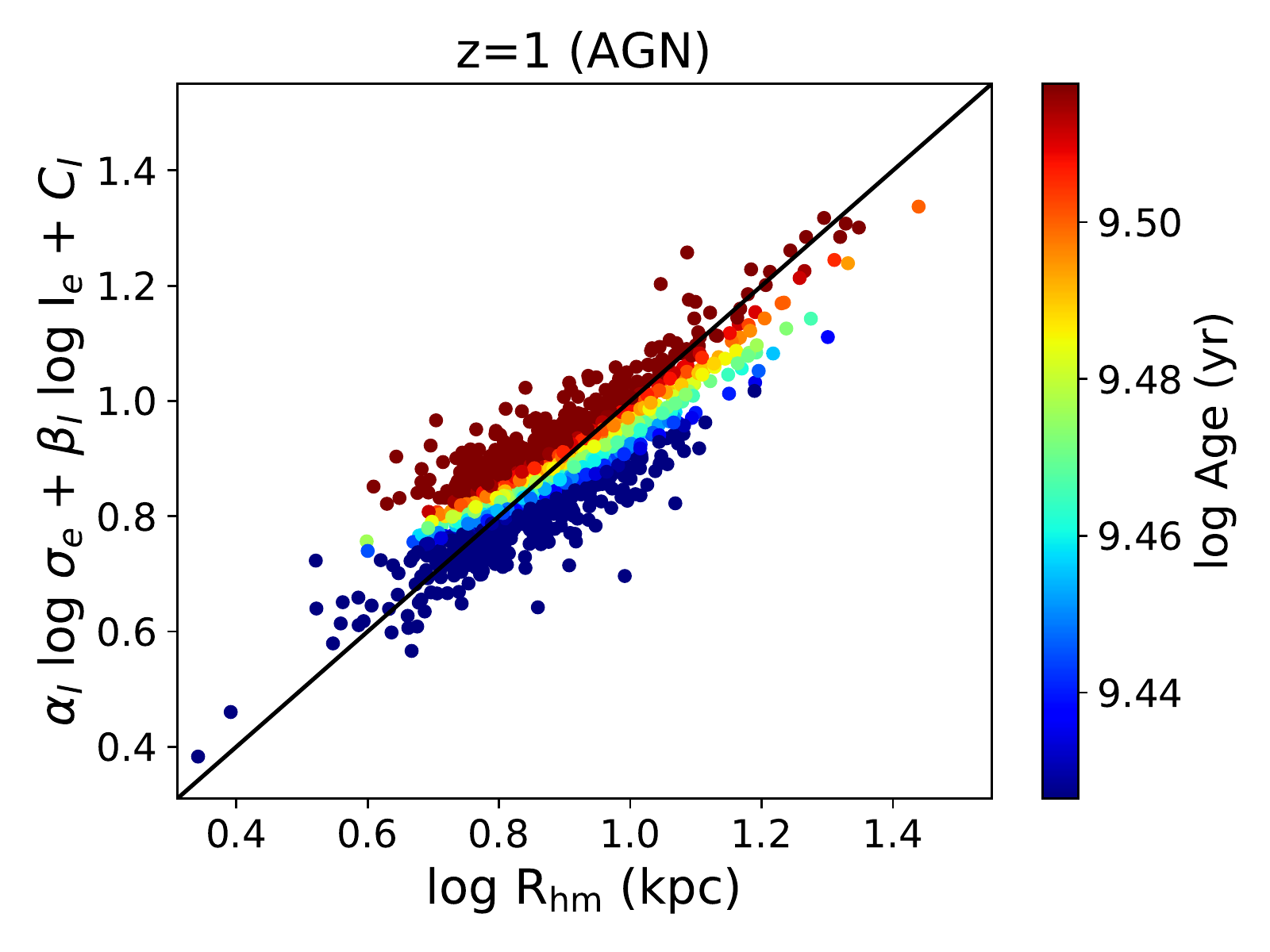}
\includegraphics[width=0.3\textwidth]{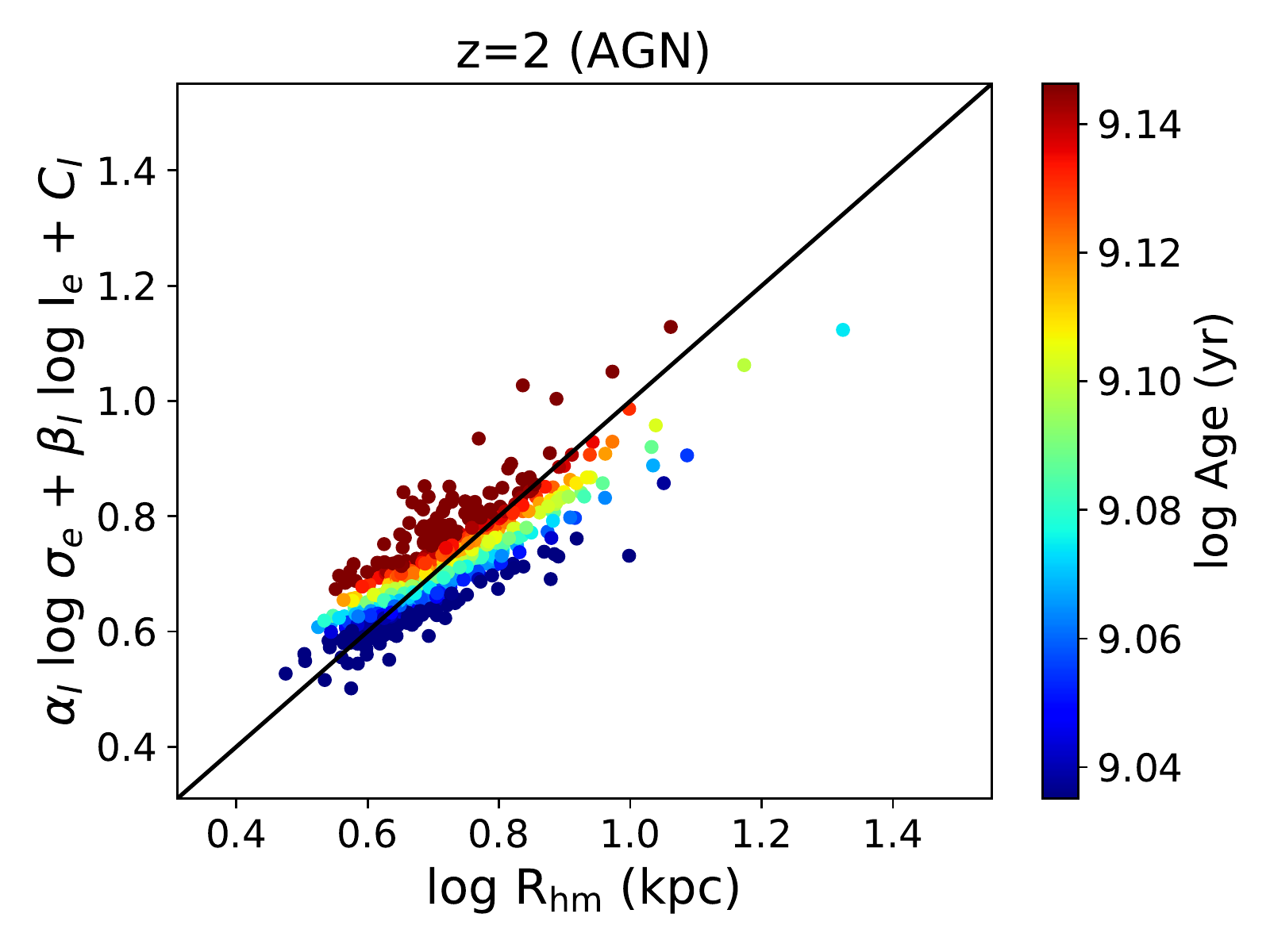}
  \caption{The L-FP for Horizon-AGN central ETGs fitted at different redshifts where ETGs are defined as the ones with $V/\sigma < 0.3$. The symbols are coloured according to mass-weighted stellar ages.
  The 1:1 line is depicted in black.}
   \label{fig:fp_age_lumVS}
\end{figure*}

\begin{figure}
  \centering
\includegraphics[width=0.3\textwidth]{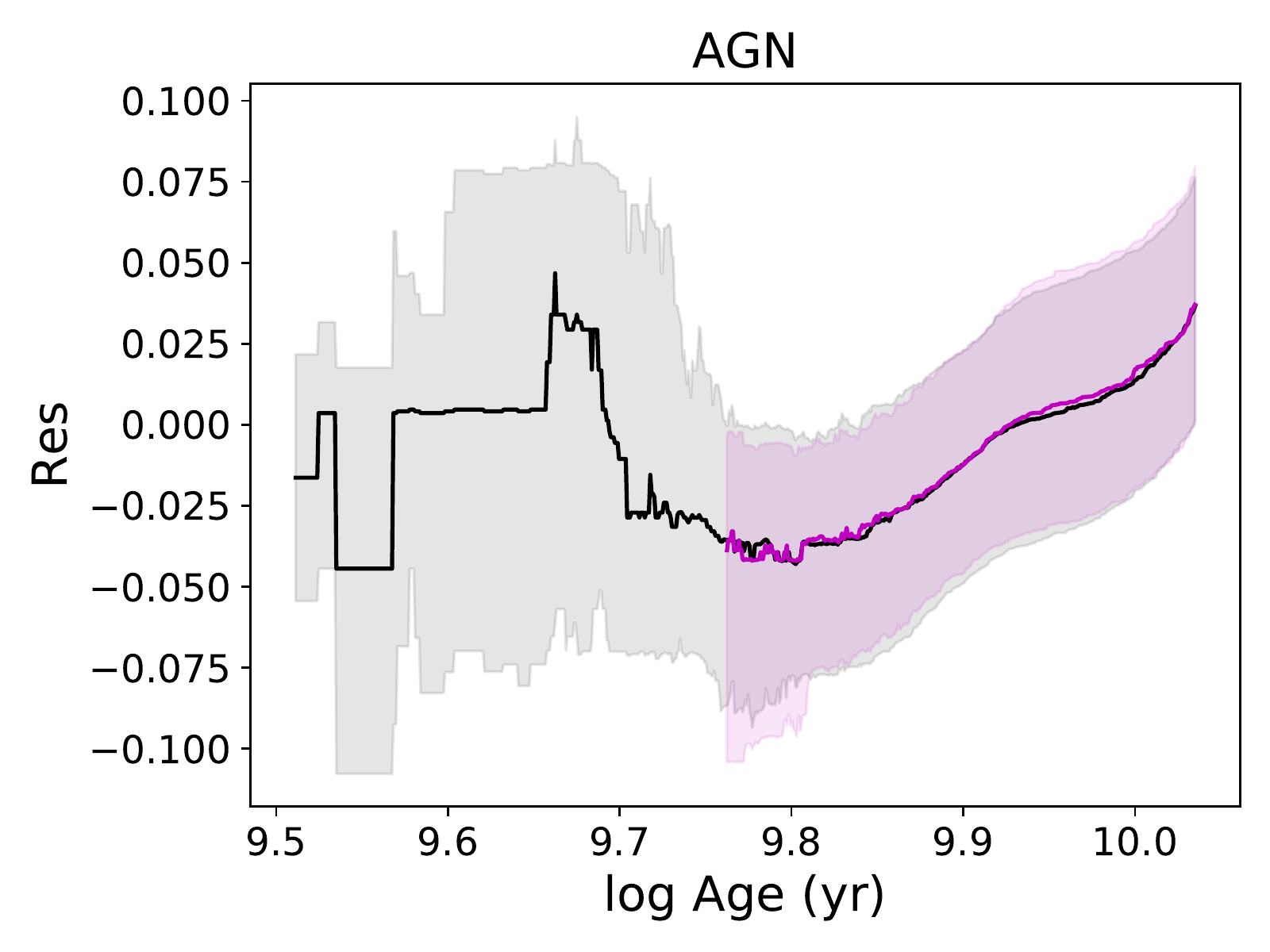}
  \caption{The residuals of the L-FP at $z=0$ with the restriction $V/\sigma < 0.3$ as a function of mass weighted average age (black line). We depict the residuals of the subsample of passive galaxies \citep[defined by][]{Lacerna2014} in magenta lines. The regions between the 25 and 75 percentiles are shaded.}
   \label{fig:res_fp}
\end{figure}

In a recent observational work, \cite{Bernardi2020} analyse the FP of nearby galaxies from MaNGA DR15 \citep{Aguado2019}. They found a tight plane, especially for slow rotator ETGs. In agreement with \cite{Saglia1993}, who found a dependence of the FP on morphology, \citet{Bernardi2020} remark that the FP parameters are linked to the morphological mix of ETGs. 
Our results are in global agreement with this trend at $z=0$ although the simulated L-FP is more tilted for the coefficient of the surface brightness compared to the results reported by  \citet{Bernardi2020}
($\alpha_{\rm obs}=1.275 \pm 0.025$ and $\beta_{\rm obs}=-0.723 \pm 0.013$). The scatter found in observed L-FP (0.077 dex) is similar to that measured in the Horizon-AGN (0.067 dex).

\cite{Lu2020} study the evolution of the FP from the IllustrisTNG-100 simulation, finding  a tight plane since $z=2$, in general agreement with our findings.  
There are, however, some differences in the fitted parameters as can be seen from the comparison of  their Table 1 and our {Table~\ref{table:tableFPlum}} regarding the whole ETG sample. Our $\alpha_I$ parameters are slightly different to theirs taking into account the errors at all redshifts (with the exception of $z=2$ in which both calculations are consistent). On the other hand, $\beta_I$ values are always more negative than those reported by \cite{Lu2020}. In both works, the parameters evolve with time with similar scatter. 
However, we note that \cite{Lu2020} select their sample by applying different criteria. In order to improve the comparison between theirs and our results, we restrict our sample to the most dispersion-dominated ETGs by imposing $V/\sigma < 0.3$, {hereafter, the restricted sample}.  This restricted sample excludes  faster rotator ETGs in Horizon-AGN. 
In Table~\ref{table:tableFPlum} we show the PF parameters for this subsample. The $\alpha_I$ parameters  are consistent to those reported by \cite{Lu2020} within the errors.

As can be seen from Fig.~\ref{fig:fp_age_lumVS}, there is a clear age dependency of the L-FP, which is similar at all analysed redshifts. Those ETGs dominated by older stars are located above the 1:1 relation, having more positive residuals (defined by  $\mathrm{Res} = \alpha_I \log{\sigma_e} + \beta_I \log{I_e} + C - \log R_{\rm hm}$ where the coefficients are computed at each redshift and summarised in Table \ref{table:tableFPlum}). However, we find that ETGs with similar mean ages are located on planes that are slightly tilted with respect to the overall L-FP.

These trends are different from what we found for the stellar mass FP where, for $z>1$,  ETGs dominated by younger  stellar populations tend to have positive $\mathrm{Res}$ while at $z=1$ there is an inversion in the relation which drives the final age distribution at $z=0$. 
The variation of stellar age across the FP is related to the  variation of the $M/L$ for galaxies with different $\Sigma_e$. From the lower panel of Fig.~\ref{fig:fp_ML}, we can see that high $\Sigma_e$ galaxies have high $\sigma_e$ and low $M/L$ at high redshift, but at a given $\sigma_e$ some galaxies become denser as the $M/L$ get higher for lower redshift. 
For $z <1$  the dependence on $\Sigma_e$ has changed slope with some galaxies becoming less dense for a given $\sigma_e$ and $M/L$. 

In Fig.~\ref{fig:res_fp} we show the residuals of the L-FP as a function of the median stellar age at $z=0$ for the restricted sample (black line and gray shaded regions). Additionally, we also highlight the residuals considering an extra condition to select only passive galaxies \citep{Lacerna2014} from the restricted sample so that  $\log({\rm sSFR}) \leq -0.65 (\log(M_* - 10.56)) -10.94$ ($M_*$ is in units of M$_{\odot}$ and sSFR is in units of yr$^{-1}$,  pink line and shaded regions).

As can be seen from Fig.~\ref{fig:res_fp}, the relation between   $\mathrm{Res}$ and stellar age obtained for passive, dispersion-dominated Horizon-AGN ETGs depicts a monotonous increasing function of stellar age\footnote{We applied the moving median to estimate this relation.}, in good agreement with \citet{Lu2020}. However, there are 22 ETGs from the restricted sample (1.11 per-cent) with ages lower than $10^{9.75}$ yr. Their residuals present a different behaviour as a function of stellar age, being 14 of them positive (0.07 per-cent of the restricted sample). These slow rotator ETGs have been rejuvenated and hence determined a turn in the residual relation. 
The confrontation of these results with recent reported observation of ETGs could shed light on the efficiency of feedback mechanisms.

\section{Conclusions}
\label{sec:conclusions}

We have performed a comprehensive study of ETGs from the Horizon-AGN and Horizon-noAGN simulations, which allows us to analyse the role of AGN feedback in reproducing the main scaling relations for this type of galaxies.
We select central galaxies resolved with more than 5000 stellar particles, $R_{\rm hm}$ larger than 2 kpc and total stellar masses lower than $10^{12.5}$ M$_{\odot}$.
Galaxies are classified as ETGs if $V/\sigma < 0.55$ following  in \cite{Chisari2015}.

Our main conclusions are:

\begin{itemize}

    \item ETGs albeit dispersion-dominated by definition show a variety of morphologies, ranging from less to more rotational supported systems. They show a rotational component within their inner region (defined in Sec.~\ref{sec:inner}). The
    stellar mass fraction of this rotational component to the dispersion-dominated component correlates with $V/\sigma$ for ETGs in Horizon-AGN in agreement with previous results \citep{Rosito2018,Rosito2019a}, as seen in Fig.~\ref{fig:id}.
    ETGs in Horizon-noAGN show higher fractions ($\sim 0.20$) which are independent of stellar mass. This implies that 
    ETGs are more dominated by rotation in the Horizon-noAGN and that AGN feedback contributes to building up more dispersion-dominated systems leading to a better agreement with observations.
    
     \item The presence of AGN feedback is required to i) reproduce the bimodality in the spin parameter found in \cite{Graham2018} and, hence, has an impact on the angular momentum retained by ETGs consistent with the previous item \citep[Fig.~\ref{fig:spin}, see also][]{Sande2020} and ii) to regulate the star formation activity and to reproduce the decreasing trend with stellar mass (Fig.~\ref{fig:sfr}) in  better agreement with observations \citep[e.g.][]{Wuyts2011}.
    
     \item In both simulations, there is a tight relation between mass and radius,  as shown in Fig.~\ref{fig:mdv}. Horizon-AGN galaxies present clearer trends of rotation and age (more massive galaxies are older and slower rotators) across the mass--size relation than those found in the absence of AGN. Moreover, a few low-mass galaxies dominated by old SPs are found in  Horizon-noAGN, which may be attributed to the fact that these systems are more affected by SN feedback. We note that this process has not been recalibrated in  Horizon-noAGN. Hence, its role might be not well-described. In Horizon-AGN, the transition to ETGs dominated by younger stars and more rotationally supported occurs at $\sim 6$ kpc. Around this radius, the mean stellar mass is $\sim 2.2 \times 10^{10}$ M$_{\odot}$ and the mean dynamical mass is $\sim 7.0 \times 10^{10}$ M$_{\odot}$.     

    \item  As can be seen from Fig.~\ref{fig:displacement}, in Horizon-AGN, we find that the mass--plane evolves for $z >1$. This agrees with the redshift at which AGN feedback starts to have an impact on the  slope of the relation between sSFR and stellar mass (Fig.~\ref{fig:evol_sf}). In Horizon-noAGN, the mass--plane evolves strongly for $z>0$ (Fig.~\ref{fig:displacement}). The tilt from the virial relation is more important in the absence of AGN feedback  (Fig.~\ref{fig:mp-agnnagn}). In the Horizon-AGN simulation, the coefficient of the velocity dispersion remains closer to that expected from the virial theorem as a function of redshift, whereas the coefficient of $R_{\rm hm}$ is slightly smaller than the theoretical one.
    
    \item From Fig.~\ref{fig:fp_evol_Cappe}, it can be seen that the FP of ETGs with AGN follows a similar relation to the observed one at $z = 0$ \citep{Cappellari2013}.  The evolution of the Horizon-AGN FP with redshift is very mild from $z\sim 2$.  Conversely, the   Horizon-noAGN FP departs from the observed one  for radii larger than $\sim 4$ kpc at all analyzed redshifts. The main difference is caused by the presence of extended galaxies with high surface density  \citep[see also][]{Peirani2019}.

    \item It can be seen from Fig.~\ref{fig:sfr} that the SF activity  is regulated by AGN feedback so that the observed anti-correlation between the sSFR and the stellar mass is reproduced.  We note that such regulation is not enough to reproduce \cite{Hernandez-Toledo+2010} lower values of sSFR. However, recent observations report the existence of rejuvenated AGNs  galaxies \citep{Yates2014,lacerna2020} which should be taken into account when revising the models. Horizon-noAGN ETGs present an excess of sSFR with similar values for all analysed stellar masses. The evolution of the sSFR as a function of time shows a similar flat trend contrary to the relations obtained when AGN feedback is turned on (Fig.~\ref{fig:evol_sf}). AGN feedback regulates the level of star formation activity as a function of redshift and consequently, it also modulates its dependence on stellar mass,  producing relations in better agreement with observations.  
    The impact of AGN feedback is very clear from $z\sim1$.
    
    \item The strong impact of AGN feedback from $z\sim 1$ is impressed in the evolution of the mass--plane and the FP. Both relations show a stronger evolution since this redshift.
    The imprint of AGN feedback on the power spectrum and subsequent regulation of the SF at galaxy scales leave features in the dependence of the scatter of the FP on stellar age and $V/\sigma$ (Fig.~\ref{fig:FPcolor}). Hence, the analysis of the FP and its scatter as a function of redshift opens the possibility to underpin the variation of the   AGN feedback efficiency across time. 
    
    \item The action of AGN feedback  has also an impact on the $M/L$ ratios which get larger with decreasing redshift as the stellar population ages, as shown in Fig.~\ref{fig:fp_age_lumVS}. The variation of the $M/L$ as a function of stellar mass surface density explains the differences (at least in part) found in L-FP which is strongly tilted with respect to the theoretical prediction compared to the weaker one found when using surface stellar density and has a different dependence on stellar age.
    
    \item Fig.~\ref{fig:res_fp} shows that when restricting the analysis of the FP to passive  ETGs, we found that those with  median ages older than $10^{9.75}$ yr have residuals that increase for increasing age in agreement with \cite{Lu2020} for the IllustrisTNG simulation and  with observations \citep[e.g.][]{Dokkum1996, Terlevich2002}. However, there is a  0.70 per-cent of Horizon ETGs with $V/\sigma < 0.3$ which have been slightly rejuvenated and have median positive residuals for which there seems to be an anticorrelation with age but there are not enough data to draw a robust conclusion. The presence of these ETGs suggests the need for further suppression of star formation activity .

\end{itemize}

Overall, 
our  findings show that the FP is more sensitive to the action of the AGN feedback than the mass--plane since it affects the stellar surface density and the $M/L$ significantly. The evolution of the FP as well as its dependence on stellar age and galaxy morphology can thus provide detailed insight into the action of AGN feedback as a function of time.

\begin{acknowledgements}
We thank the referee for useful comments, which helped to improve this paper. PBT acknowledges support from  CONICYT project Basal AFB-170002 (Chile) and Fondecyt Regular 1200703-2020.
This work was partially supported through MINECO/FEDER (Spain) PGC2018-
094975-C21 grant. This project has received funding from the European Union’s
Horizon 2020 Research and Innovation Programme under the Marie Skłodowska-Curie
grant agreement No 734374- LACEGAL. MSR and PBT acknowledge funding from the
same Horizon 2020 grant for a secondment at the Astrophysics group of Univ.
Aut\'onoma de Madrid (Madrid, Spain).

This work has made use of the HPC resources
of CINES (Jade and Occigen supercomputer) under the time
allocations 2013047012, 2014047012 and 2015047012 made
by GENCI. This work is partially supported by the Spin(e)
grants ANR-13-BS05-0005 (\url{http://cosmicorigin.org}) of
the French Agence Nationale de la Recherche and by the ILP
LABEX (under reference ANR-10-LABX-63 and ANR-11-
IDEX-0004-02). We thank S. Rouberol for running smoothly
the Horizon cluster for us. Part of the analysis of the simulation was performed on the DiRAC facility jointly funded
by STFC, BIS and the University of Oxford.
This work is part of the Delta ITP consortium, a program of the Netherlands Organisation for Scientific Research (NWO) that is funded by the Dutch Ministry of Education, Culture and Science (OCW). NEC was partly supported by a Royal Astronomical Society Research fellowship during the preparation of this work.

\end{acknowledgements}

\bibliographystyle{aa}

\def\apj{ApJ}
\def\apjl{ApJ}
\def\aj{AJ}
\def\mnras{MNRAS}
\def\aa{A\&A}
\def\nat{Nature}
\def\araa{ARA\&A}
\def\aap{A\&A}

\bibliography{Rosito_horizon}

\begin{appendix}

\section{Morphological classification}
\label{app:morp}

In this work, we apply the criterion in \cite{Chisari2015} to morphologically classify galaxies in ETGs and LTGs.
In this Appendix, we verify the consistency of this criterion with that used by \cite{Rosito2018, Rosito2019b} and \cite{Rosito2019a}, in which galaxies are classified by using the $B/T$ ratio following the method described in \cite{tissera2012}, being $B/T=0.5$ the adopted threshold for a galaxy to be considered spheroid dominated.

\begin{figure}
  \centering
\includegraphics[width=0.45\textwidth]{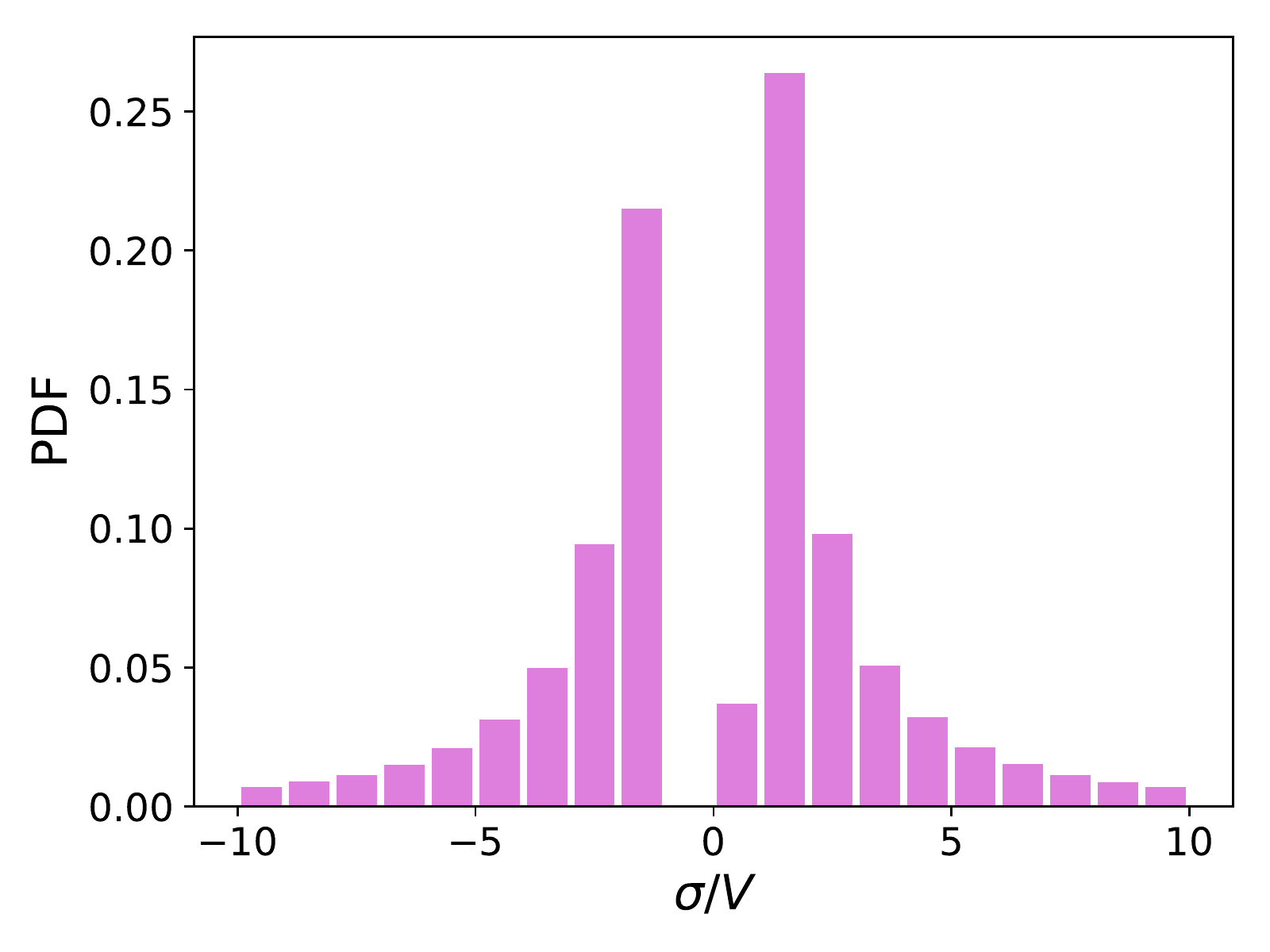}
  \caption{Distribution of $\sigma/V$ estimated for each stellar particles for a typical ETG selected from the Horizon-AGN simulation.}
   \label{fig:sigmav_gal1}
\end{figure}

We measure quantitatively the amount of rotation in the analysed samples. For this purpose, we calculate a parameter $\sigma/V$ for each stellar particle within $R_{\rm opt}$. We define the dispersion locally by taking radial bins around the centre of a galaxy in which we calculate the average of each velocity component. Thus, $\sigma$ for each particle is calculated relative to that averages and $V$ is the particle tangential velocity. If $0 < \sigma/V < 1$, that particle is considered part of the disc and thus, we are able to calculate disc-to-total stellar mass ratio ($D/T$), which may play the role of $B/T$ in the works mentioned above.
On the other hand, the bulge is defined by the stellar particles with $|\sigma/V| <1$ located within 0.5 $R_{\rm hm}$\footnote{ We note that here the definition of the bulge is different from that employed by \cite{Rosito2018} and \cite{Rosito2019a}. In those works, the bulge particles are defined as the ones without ordered rotation that are most gravitationally bound.}. 
In Fig.~\ref{fig:sigmav_gal1} we show the distribution of $\sigma/V$ for an individual ETGs from Horizon-AGN simulation. It is clear that there is no counter-disc and that the galaxy is dominated by velocity dispersion.

\begin{figure}
  \centering
\includegraphics[width=0.45\textwidth]{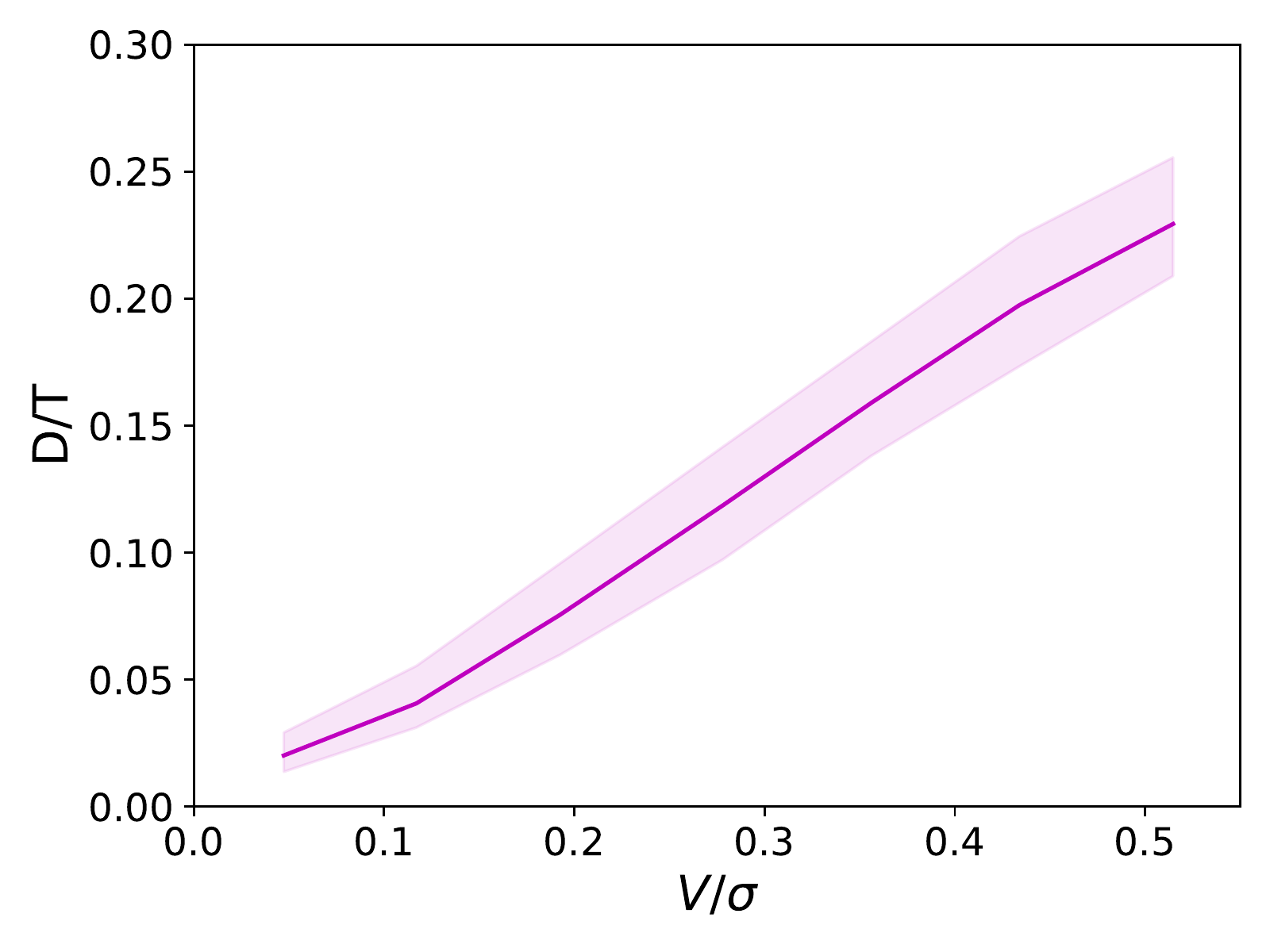}
  \caption{$D/T$ as a function of $V/\sigma$ for central ETGs from Horizon-AGN simulation. The remarkable correlation (p-value $\sim$ 0) between them verify the consistency of our morphological classification and the one presented in \cite{Rosito2018}.}
   \label{fig:DT_vsigma}
\end{figure}

As regards the comparison between $D/T$ calculated as described above and the global parameter $V/\sigma$ \citep{Dubois2014} utilised for the classification in this work, we find an excellent correlation between them, with a p-value of the Spearman coefficient near 0 in the selected sample of Horizon-AGN (see Fig~\ref{fig:DT_vsigma}).
Therefore, we conclude the consistency between both classifications.

\end{appendix}

\end{document}